\documentclass[
aps,%
12pt,%
final,%
notitlepage,%
oneside,%
onecolumn,%
nobibnotes,%
nofootinbib,%
superscriptaddress,%
noshowpacs,%
centertags]%
{revtex4}

\begin{document}


\title
{\small Multi-frequency Studies of Massive Cores with Complex Spatial and Kinematic Structures}

\author{\firstname{L.~E.}~\surname{Pirogov}},
\email{pirogov@appl.sci-nnov.ru}
\affiliation{%
Institute of Applied Physics, Russian Academy of Sciences, Nizhny Novgorod, Russia}
\author{\firstname{V.~M.}~\surname{Shul'ga}},
\affiliation{Institute of Radio Astronomy, National Academy of Sciences of Ukraine, Kharkov, Ukraine}
\author{\firstname{I.~I.}~\surname{Zinchenko}},
\affiliation{%
Institute of Applied Physics, Russian Academy of Sciences, Nizhny Novgorod, Russia}
\author{\firstname{P.~M.}~\surname{Zemlyanukha}},
\affiliation{%
Institute of Applied Physics, Russian Academy of Sciences, Nizhny Novgorod, Russia}
\affiliation{%
Lobachevsky State University, Nizhnyi Novgorod, Russia}
\author{\firstname{O.~N.}~\surname{Patoka}},
\affiliation{Institute of Radio Astronomy, National Academy of Sciences of Ukraine, Kharkov, Ukraine}
\author{\firstname{M.}~\surname{Thomasson}}
\affiliation{%
Chalmers University of Technology, Onsala Space Observatory, Sweden}

\begin{abstract}

Five regions of massive star formation have been observed in various molecular 
lines in the frequency range $\sim 85-89$ GHz. The studied regions possess 
dense cores, which host young stellar objects. The physical parameters of the 
cores are estimated, including kinetic temperatures ($\sim 20-40$ K), sizes 
of the emitting regions ($\sim 0.1-0.6$ pc), and virial masses ($\sim 40-500 M_{\odot}$). 
Column densities and abundances of various molecules are calculated 
in the local thermodynamical equilibrium approximation. 
The core in 99.982+4.17, associated with the weakest IRAS source, is characterized 
by reduced molecular abundances. Molecular line widths decrease with increasing 
distance from the core centers ($b$). For $b\ga 0.1$~pc, the dependences $\Delta V(b)$ 
are close to power laws ($\propto b^{-p}$), where $p$ varies from $\sim 0.2$ to $\sim 0.5$, 
depending on the object. In four cores, the asymmetries of the optically thick 
HCN(1--0) and HCO$^+$(1--0) lines indicate systematic motions along the line of sight: 
collapse in two cores and expansion in two others. Approximate estimates of the 
accretion rates in the collapsing cores indicate that the forming stars have masses 
exceeding the solar mass.

\end{abstract}

\maketitle

\section{Introduction}

The theory of massive star formation is by no means complete, even in general 
terms (see e.g. \cite{Tan}). Massive ($\ga 8~M_{\odot}$) stars are born in clusters, 
which are located at larger distances than regions of low-mass star formation, 
and evolve much faster than low-mass stars. Massive stars at early stages of their 
evolution actively interact with their parent dense cores, increasing their turbulence
and temperatures, accelerating shocks and massive outflows, and leading to fragmentation, 
further compression and, under certain conditions, to a new phase of star formation. 
The gas-phase molecular composition in massive star-forming regions is fairly rich due to the evaporation of molecules from grain surfaces, and the observed line profiles are significantly broadened as a result of turbulent and systematic motions. Individual compact regions may give rise to maser emission (e.g., in the water vapor, hydroxyl, or methanol lines). As a result of inhomogeneities and systematic motions in the cores, optically thick lines often exhibit non-Gaussian profiles. Detailed analyses of these profiles may enable estimation of the parameters of the core structures and velocity fields.

Among molecular lines, which trace dense gas, the HCN(1--0) line is one of the most sensitive to the core structure and kinematics. Observations of this line performed in recent years with the 22-m radio telescope of the Crimean Astrophysical Observatory (CrAO) toward $\sim 150$ methanol masers showed that the HCN(1--0) lines often have fairly complex profiles. Analysis of the HCN(1--0) spectra is hindered by the fact that, even when the profiles of hyperfine components are close to Gaussian, the ratios of their intensities deviate from those expected for local thermodynamical equilibrium (LTE) (see, e.g. \cite{Sandell83,Harju89,Pir99}). Interpretation of the spectra requires modeling of the HCN excitation taking into account deviations from LTE and overlapping of closely spaced hyperfine components, which requires many independent input parameters.

The goal of this study is to determine the physical parameters of five selected massive star formation regions associated with methanol masers in order to use them for subsequent modeling. Data obtained with the CrAO 22-m radio telescope showed that our objects demonstrate complex HCN(1--0) profiles, and one more a nearly Gaussian profile. The presence of dips and asymmetric peaks in the HCN(1--0) profiles toward four objects suggests inhomogeneous structure and complex kinematics. This paper presents physical parameters such as kinetic temperatures, sizes, virial masses, and velocity dispersions and their radial profiles of these objects measured using multi-frequency spectral observations in various molecular lines. The parameters of the contracting gas are presented for two objects. The results of modeling of the line profiles will be given in subsequent publications.

\section{Sources}

We selected five regions associated with methanol masers as our targets; according to CrAO 22-m results, four of these exhibit complex HCN(1--0) profiles, while the fifth region demonstrates a nearly Gaussian profile. The source list is presented in Table 1. Source names and coordinates were taken from the catalog of methanol masers  \cite{Bayandina}. The coordinates for 37.427+1.518 were taken from \cite{Wu14}. Table 1 also presents the source distances and associations with other objects. We adopted the distances for all sources except for 99.982+4.17 to be those determined using parallax measurements of the maser sources, which are the most reliable. For three objects (34.403+0.233, 37.427+1.518, and 77.462+1.759), these distances are appreciably different from previously measured kinematic distances. The selected objects are well-known regions of massive star formation. In addition to methanol and water vapor masers, they also host IRAS point sources, near-IR and mid-IR sources, submillimeter and radio sources, and molecular outflows. A review of available observational data on four regions (excluding 121.28+0.65) is presented in Appendices A23, A24, A37, and A45 of \cite{Varricatt}. Reviews of observations of 121.28+0.65 can be found, e.g., in \cite{Xu06,Quanz}. The bolometric luminosities of the IRAS sources, calculated by integrating over ``greybody''
 curves fitted to the flux-frequency dependences \cite{Pir07} and taking into account the distances from Table 1, fall in the range $3\cdot 10^2-7\cdot10^3~L_{\odot}$. The luminosity of IRAS 21391+5802, which is associated with 99.982+4.17, is approximately an order of magnitude lower than those of the other sources.

\section{Observations}

The observations were carried out in April--May 2015 with the 20-m radio telescope of the Onsala
Space Observatory (Sweden). We used a new 3-mm dual-polarization receiver with an SiS mixer at the input, which has a noise temperature of $\sim 50-60$~K at 85--116 GHz \cite{Belitsky}. The system temperature during most of the observations varied in the range: $\sim 150-250$~K, depending on the frequency and source elevation. A new Fourier spectrum analyser (FFTS) with a 2.5-GHz bandwidth and a 76-kHz frequency resolution (32\,768 channels) was used for the spectral analysis. These broadband observations made it possible to record several molecular lines simultaneously. The central frequencies of the observed bands were 86.6 and 88.9 GHz. To study line profiles in more detail, we observed some source positions using a Fourier spectrum analyser with a 625-MHz bandwidth and a 19-kHz frequency resolution. The signals from two polarizations were added during the reduction, increasing the sensitivity by a factor
of $\sqrt{2}$. The full width at half maximum (FWHM) of the main beam of the Onsala telescope at 86 GHz is
44$''$ \footnote{$http://www.chalmers.se/en/centres/oso/radioastronomy/20m/Documents/OSOman\_v1p9.pdf$}. 
The efficiency of the main beam depends on the source elevation, and varies in the range $\sim 0.4-0.58$
at 86 GHz for elevations of $\sim 30-70$. This parameter was used to convert antenna temperatures to main-beam brightness temperatures ($T_{\rm MB}$). An example of the spectrum for 121.28+0.65, observed at the positon offset (20$''$,20$''$) is shown in Fig. 1.

A list of detected molecular lines, including transitions, frequencies, and upper level energies
(in temperature units), is presented in Table 2.

\section{Results of observations}

All the observed sources were mapped in the lines c-C$_3$H$_2$(2$_{1,2}$--1$_{0,1}$), CH$_3$C$_2$H(5--4), SO(2$_2$--$1_1$), H$^{13}$CN(1--0), H$^{13}$CO$^+$(1--0), SiO(2--1), HN$^{13}$C(1--0), C$_2$H(1--0~3/2--1/2), HCN(1--0), HCO$^+$(1--0), HC$^{15}$N(1--0) (except 37.427) and NH$_2$D(1$_{1,1}$--$1_{0,1}$) (except 37.427).
We detected the HNCO(4$_{0,4}$--$3_{0,3}$),
HCS$^+$(2--1), C$_4$H(9--8~19/2--17/2), CCS(7$_6$--$6_5$) and H$^{15}$NC(1--0) lines in several positions.
34.403+0.233 was mapped in the HNCO(4$_{0,4}$--$3_{0,3}$) line.

\subsection{Source maps}
\label{maps}

Maps of the integrated intensities of specified molecular lines are presented in Figs. 2-5. Only the central part of 37.427+1.518 was mapped. The maps show the IRAS and maser positions, as well as the positions of infrared, submillimeter, and radio sources.

In most cases, the maps in different molecular lines display spatially correlated compact regions of enhanced intensity, which are obviously related to regions of enhanced density (dense cores). IRAS sources, as well as maser and other sources, are usually concentrated near the centers of the cores, testifying that the process of star formation has started there. The NH$_2$D emission is less well correlated with the emission of other molecules.

In four objects, the core shapes are nearly spherically symmetric. The maps of molecular emission in 34.403+0.233 (the dark infrared cloud G034.43+00.24) are elongated in the north-south direction. Some differences between the images of this source in different molecular lines may be related to the fact that the cloud contains two compact clumps with different molecular composition \cite{Rathborne08}, which are unresolved in our observations. The core in 99.982+4.17 is located in the southern part of the studied region. We observed a fairly extended and probably less dense region to the north of the core.

To estimate the sizes of the regions, we fitted the maps of the integrated intensities using convolutions of two-dimensional elliptical Gaussians with circular Gaussians having extents equal to the antenna main beamwidth \cite{Pir03}. In this fitting procedure, we used the Marquardt-Levenberg iteration algorithm (the ``neighborhood of maximum'' method; see, e.g. \cite{ST,Press}), which yields the parameters and their errors. The core sizes were determined as the geometrical mean of the extents of the best-fit two-dimensional elliptical Gaussians. The offsets of the core centers, axial ratios of the best-fit ellipses, and full widths at half power of the regions emitting in the different molecular lines with their errors are presented in the corresponding table in Section 5.3.

\subsection{Line parameters}
\label{parlines}

The spectral data were reduced using both standard methods using the GILDAS package
 \footnote{http://iram.fr/IRAMFR/GILDAS} and our own original codes. After subtracting baselines (polynomials of an order not higher than three) from spectral regions with lines, we fitted the observed lines using one or more Gaussians, in order to determine line intensities, radial velocities of the line centers, and line widths. The parameters of the observed molecular lines toward emission peaks are presented in Table 3. When reducing the CH$_3$C$_2$H(5--4) spectra, we fixed the offsets between the lines with different K values and specified the line widths to be equal. The same approach was used when reducing the H$^{13}$CN(1--0) spectra, which consist of three close hyperfine components. The NH$_2$D(1$_{1,1}$--$1_{0,1}$) lines were reduced assuming LTE. The component intensity ratios were used to determine  optical depths, which were then used to estimate column densities (see Section 5.2). The profiles of optically thick lines (HCN, HCO$^+$) in all objects except 37.427+1.518 exhibit complex structures, and deviate appreciably from Gaussians as a result of self-absorption, systematic motions, and possibly inhomogeneous gas distribution along the line of sight. Only the integrated intensities of such lines are presented in Table 3. The uncertainties in the integrated intensities were calculated from $\Delta T_{\rm MB} \sqrt{N_{\rm ch}}\,\delta V_{\rm ch}$ where $\Delta T_{\rm MB}$ is the noise level in a channel without any line (calculated after the baseline is removed),$N_{\rm ch}$ is the number of channels occupied by the line, and $\delta V_{\rm ch}$  is the velocity resolution (channel width).

\section{Physical parameters of the cores}

This section presents calculated physical parameters of the cores. 
Kinetic temperatures are given in Section 5.1, column densities and molecular abundances are given in Section 5.2, and the parameters of the emitting regions, including their sizes, mean velocity dispersions, and virial masses, are given in Section 5.3. Section 5.4 presents the results of calculations of the radial profiles of the velocity dispersions in the cores. In Section 5.5, we analyze the line profiles for two objects and estimate their collapse velocities and accretion rates.

\subsection{Kinetic temperatures}
\label{Tkin}

Methyl acetylene (CH$_3$C$_2$H) lines can be used to determine kinetic temperatures in dense cores.
The  $J=5-4$ CH$_3$C$_2$H spectrum consists of a group of lines with different values of the quantum number $K$. As the populations of levels with the same $J$ values but different $K$ values are governed only by
collisions, they are close to their LTE values. Observations of these lines in dense regions enable estimation
of the corresponding kinetic temperatures (see, e.g. \cite{Mal}). The slope of the best-fit line describing the dependences of the integrated intensities of lines with different $K$ values on the level energies can be used to
determine the rotation temperature, which is usually adopted as an estimate of the kinetic temperature.
We used this method to calculate kinetic temperatures toward positions where two or more CH$_3$C$_2$H lines were reliably detected. As the integrated intensities of lines with different $K$ values, we adopted the product of $T_{\rm MB}$ and $\Delta V$, which were determined from the fitting of the CH$_3$C$_2$H(5--4)
spectra. The uncertainties in the kinetic temperatures were calculated from the uncertainties of the fits, using standard error propagation. The temperatures determined toward selected positions in the sources are presented in Table 4. The region 121.28+0.65 demonstrates a tendency for the kinetic temperature to decrease from the center to the edge. We could not find a spatial gradient of the temperature in any other source given the temperature uncertainties. The calculated temperatures lie in the range: $\sim 20-40$~K.
The temperatures in the 34.403+0.233 core are higher than those in other cores. We were not able to derive significant estimate of kinetic temperature (above $3\sigma$) for 37.427+1.518. The kinetic temperature in this core calculated using the ammonia lines \cite{Molinari} is 29 K which falls in the range of temperatures obtained for the other objects.

\subsection{Molecular column densities}
\label{sec:Nmol}

Integrated intensities of optically thin lines can be used to find molecular column densities in the LTE approximation. In general, molecular column density can be found in this approximation using the relation (see \cite{MangumShirley}, eq. (79)):

\begin{equation}
N_{\rm MOL}=\Bigl(\frac{3\,k}{8\pi^3\,\nu\,S\,\mu^2}\Bigr)\,
\Bigl(\frac{Q_{rot}}{g_J\,g_K\,g_I}\Bigr)\,
\frac{\exp({\frac{E_u}{k\,T_{\rm EX}}})}{1-\frac{J(T_{\rm BG})}{J(T_{\rm EX})}}\,
\frac{\int\,T_R\,dv}{R_i\,f} \hspace{2mm},
\end{equation}

\noindent{where $k$ is the Boltzmann constant, $\nu$ is the transition frequency, $\mu$ is the dipole moment, $S=(J^2-K^2)/J/(2J+1)$ is the line strength, $J$ and $K$ are the quantum numbers, $Q_{rot}$ is the partition function over all states, $ g_J$, $g_K$, $g_I$ are degeneracies, $E_u$ is the energy of the upper state of the transition, $T_{\rm EX}$ is the excitation temperature, which is the same for all levels (LTE), $T_{\rm BG}$=2.73~K is the cosmic background temperature, $J(T)=\frac{h\nu}{k}/(\exp(\frac{h\nu}{kT})-1)$, $h$ is the Planck constant, $\int\,T_R\,dv$ is the integral over the line profile, $R_i$ is the relative intensity of the hyperfine component, $f$ is the main beam filling factor.

We used this formula to calculate molecular column densities toward the intensity peaks. The filling factors $f$ were always taken to be unity, leading to underestimation of the column density when the source size is smaller than the main-beam size. The relative intensities $R_i$ were taken to be unity in all cases except for C$_2$H and HCO; we calculated the column densities of each of these molecules from the integrated intensity of one hyperfine component. We calculated $N$(C$_2$H) using the integated intensities of the $F=1-1$ ($J$=3/2--1/2) component, which is probably optically thin and for which $R_i=0.17/4$ \cite{Reitblat}. We calculated $N$(HCO) using the $F=2-1$ ($J$=3/2--1/2) component, whose relative intensity $R_i$ was taken to
be 1.67/4. The excitation temperatures were taken to be 10 K for all lines except CH$_3$C$_2$H(5--4),
for which we adopted the kinetic temperature estimates (see Section 5.1) as the excitation temperatures.
For $T_{\rm EX}$=10~K, the calculated  $N_{\rm MOL}$ values are close to the minimum ones. The NH$_2$D column densities were calculated from line optical depths and widths using the method of \cite{Busquet}. The optical depths of the NH$_2$D hyperfine component $F=2-2$ (1$_{1,1}$--$1_{0,1}$) were determined taking into account the relative intensities of the hyperfine components of this transition \cite{Bester}, and for our sources proved to be $\sim 0.5-2$. The calculated molecular column densities are presented in Table 5. However, these values may be underestimated by a factor of a few if the excitation temperatures differ from 10 K.

The ranges of molecular column densities observed in either all of the objects or in four of them (excluding
37.427+1.518) are as follows:
$N$(C$_3$H$_2)\simeq (4-11)\,10^{12}$~á¬$^{-2}$,
$N$(CH$_3$C$_2$H$)\simeq (0.3-1.9)\,10^{14}$~á¬$^{-2}$,
$N$(C$_2$H$)\simeq (3.0-5.5)\,10^{14}$~á¬$^{-2}$,
$N$(NH$_2$D$)\simeq (2.5-12)\,10^{14}$~á¬$^{-2}$,
$N$(H$^{13}$CN$)\simeq (0.5-1.8)\,10^{13}$~á¬$^{-2}$,
$N$(H$^{13}$CO$^+)\simeq (0.5-1.1)\,10^{13}$~á¬$^{-2}$,
$N$(HN$^{13}$C$)\simeq (1.5-5.8)\,10^{12}$~á¬$^{-2}$,
$N$(HC$^{15}$N$)\simeq (0.8-1)\,10^{12}$~á¬$^{-2}$,
$N$(HNCO)$\simeq (0.2-2.6)\,10^{13}$~á¬$^{-2}$,
$N$(HCS$^+)\simeq (0.9-2)\,10^{12}$~á¬$^{-2}$.

We estimated the molecular abundances $X=N_{\rm MOL}/N($H$_2$) using molecular hydrogen column
densities taken from the literature. In 121.28+0.65, $N$(H$_2)\simeq 10^{23}$~cm$^{-2}$ \cite{Zin98, McCutcheon}. For 34.403+0.233 (G034.43 MM1), we adopted the value $N$(H$_2$)=$5.85\times 10^{22}$~cm$^{-2}$, calculated in \cite{Sanhueza} based on dust continuum observations at 1.2 mm \cite{Rathborne06}. Estimates of the hydrogen column density in 37.427+1.518 fall in the range (1.5--6.8)$\times 10^{23}$~cm$^{-2}$ \cite{McCutcheon,Beuther}.
Using the values of density and size of the region emitting at 1.2 mm from \cite{Faundez}
and correcting them for the distance to this source from Table 1, 
we found $N$(H$_2)\simeq 3.7\times 10^{23}$~cm$^{-2}$. 
For 77.462+1.759 we used $N$(H$_2)=2\times 10^{23}$~cm$^{-2}$ \cite{McCutcheon}. 
The hydrogen column density $N$(H$_2)=9.7\times 10^{23}$~cm$^{-2}$ for 99.982+4.17, 
determined from 2-mm continuum observations with a spatial resolution of 12$''$ \cite{Sugitani}, 
may be overestimated, since the authors used a gas-to-dust mass ratio that was 
a factor of 1.5 higher than the standard value, and possibly a slightly underestimated 
absorption coefficient for this wavelength. Using the column density of formaldehyde 
for this source (IC 1396 E1) \cite{HW}, the hydrogen column density was estimated 
in \cite{Walker} to be $N$(H$_2)=3.2\times 10^{23}$~cm$^{-2}$ for a 3$'$-sized region. 
We adopted $N$(H$_2)=5\times 10^{23}$~cm$^{-2}$.

The calculated molecular abundances are also presented in Table 5. They fall in the ranges:
$X$(C$_3$H$_2)\simeq (0.1-1.5)\,10^{-10}$,
$X$(CH$_3$C$_2$H$)\simeq (0.5-3.3)\,10^{-9}$,
$X$(C$_2$H$)\simeq (0.6-9.3)\,10^{-9}$,
$X$(NH$_2$D$)\simeq (0.5-20)\,10^{-9}$,
$X$(H$^{13}$CN$)\simeq (0.9-31)\,10^{-11}$,
$X$(H$^{13}$CO$^+)\simeq (0.9-11)\,10^{-11}$,
$X$(HN$^{13}$C$)\simeq (0.3-9.8)\,10^{-11}$,
$X$(HC$^{15}$N$)\simeq (0.2-1.7)\,10^{-11}$,
$X$(HNCO$)\simeq (0.4-43)\,10^{-11}$,
$X$(HCS$^+)\simeq (0.3-2.1)\,10^{-11}$.
The highest abundances were found for 34.403+0.233, and the lowest for 99.982+4.17.

\subsection{Sizes, velocity dispersions, and virial masses of the cores}
\label{sizes}

Table 6 presents the coordinate offsets of the centers of the regions emitting in various lines, the ratios of the axes of approximating ellipses (elongation factors), and the angular and linear sizes of the emitting regions determined from the map fitting (see Section 4.1). Apart from 34.403+0.233, the cores are close to spherically symmetric (the elongation factors of the regions emitting in the most of the lines are not larger than two). In these objects, the regions emitting in various lines that are probably optically thin have linear sizes $\sim 0.1-0.6$ pc. The smallest emitting regions were found in 99.982+4.17 ($\sim 0.1-0.2$ pc).

Column 7 of Table 6 gives the line widths averaged over the regions within the half-maximum contour. The line widths are significantly larger than the thermal widths. Line widths observed in a single object can be fairly different. The narrowest are the NH$_2$D(1$_{1,1}$--$1_{0,1}$) (1.3--1.6~km~s$^{-1}$), and the broadest the SiO(2--1) lines (3.1--7.5~km~s$^{-1}$), which trace shocks in the envelopes around young stars (see, e.g. cite{Harju98}). The widths of the other lines mainly fall in the range $\sim 2-3$~km~s$^{-1}$. The linewidths in 34.403+0.233 and 99.982+4.17 are $\sim 3-4$~km~s$^{-1}$ and $\sim 1.5-2.5$~km~s$^{-1}$, respectively.

The virial masses presented in column 8 of Table 6 were calculated from the sizes of the emitting regions
and the corresponding velocity dispersions via the expression
$M_{\rm vir}(M_{\odot})=105 \langle\Delta V\rangle^2 \cdot d$, where $\Delta V$ and $d$ are in km~s$^{-1}$ and pc (see, e.g. \cite{Pir03}). This expression is valid for spherically symmetric cores and in the absence of external pressure and magnetic fields. The virial masses of four cores calculated using molecular lines that are probably optically thin were found to be
 $\sim 100-150$~$M_{\odot}$ (121.28+0.65, $\sim 70$~$M_{\odot}$ for HN$^{13}$C), $\sim 130-180$~$M_{\odot}$ (37.427+1.518), $\sim 100-240$~$M_{\odot}$ (77.462+1.759) ¨ $\sim 40-90$~$M_{\odot}$ (99.982+4.17).
These values should be treated as upper limits. If the radial density profiles in the cores have the form $\sim r^{-\alpha}$, the expression for $M_{\rm vir}$ must be multiplied by the factor
$\frac{3}{5}\cdot\frac{5-2\alpha}{3-\alpha}$, which is $\la 1$ for $\alpha\ge 0$.

\subsection{Radial profiles of the velocity dispersions}
\label{sec:dvprof}

The line widths obtained as a result of the spectra fitting are significantly higher than the thermal line widths (for example, the NH$_2$D thermal widths are 0.16--0.23~km~s$^{-1}$ for T$_{\rm KIN}= 20-40$ K), and can therefore be used to approximately estimate the nonthermal velocity dispersion in the sources. This parameter is widely used in theoretical models devised to take into account non-thermal gas motions in dense cores; it is a parameter in the equation of state, and is applied in computations of stability and evolution of dense envelopes around young stars (see e.g. \cite{Pir09}).

To estimate the spatial variations of the velocity dispersions, we calculated dependences of the averaged
line widths on the impact parameters \cite{Caselli02,Pir03}. We used line widths exceeding $3\sigma$ for this averaging.
The impact parameters ($b$) were calculated as $\sqrt{A/\pi}$ where $A$ is the area of the region that includes all positions where intensity exceeds a given level. If the difference $b_i-b_{i-1}$ became larger than the map step (10$''$), we calculated the mean line width and its uncertainty for this region \cite{Pir03}.

The dependences of the averaged widths of four molecular lines whose images are spatially correlated
and fairly extended, so that the $\Delta V(b)$ dependences have at least four points, are shown in Fig. 6 for four objects of our sample. In the case of 37.427+1.518, there are not more than three points in the $\Delta V(b)$ dependences (apart from HCN(1--0)), and these dependences are not shown. The number of averaged line widths increases with $b$ from 1 in the center to $\sim10-15$ at the edge. The line widths virtually always decrease with increasing impact parameter. The clearest, close to linear, dependences were observed for 121.28+0.65. The differences between the line widths at the centers and at the edges of the
regions reach $\sim 1-1.5$~km~s$^{-1}$.

Line broadening toward the centers of the regions may be partly related to the line optical depths. Optical depths $\sim 1$ can broaden Gaussians by a factor of $\sim 1.2$ due to saturation (see, e.g. \cite{Phillips}). This effect can probably take place for the NH$_2$D lines, as well as for fairly strong lines, such as C$_2$H(1--0~3/2--1/2 $F=2-1$) and H$^{13}$CO$^+$(1--0). However, this explanation cannot be suitable for all the compared lines. The spectral indices $p$ of the $\Delta V\propto b^{-p}$ dependences for $b\ga 20''$ from the core centers (i.e. For $\ga 0.07$ pc for 99.982+4.17 and $\ga 0.15$ pc for 34.403+0.233), calculated using the c-C$_3$H$_2$(2$_{1,2}$--1$_{0,1}$), CH$_3$C$_2$H(5--4), H$^{13}$CN(1--0), H$^{13}$CO$^+$(1--0), HN$^{13}$C(1--0) and C$_2$H(1--0~3/2--1/2), lie in the ranges: $\sim 0.3-0.4$ for 121.28+0.65 ($\sim 0.7$ for HN$^{13}$C(1--0)), $\sim 0.2-0.3$ (34.403+0.233),  $\sim 0.3-0.4$ (77.462+1.759) and $\sim 0.4-0.5$ (99.982+4.17). The widths of the NH$_2$D(1$_{1,1}$--$1_{0,1}$) lines were not considered here, since the spatial distribution of their emission is often different from that of other molecules. The highest spectral indices were found in 99.982+4.17. It is possible that the inner source in this core, which is the weakest among our IRAS sources, increases the turbulence level at smaller distances from the center compared to the other cores, leading to a steeper decrease of the $\Delta V(b)$ dependence far from this source. 

One plausible explanation of the obtained trends is an increase in the dynamic activity in the core centers near IRAS sources, including differential rotation, turbulence, and systematic motions, such as inflows and outflows from young stellar objects.

\subsection{Systematic motions in the cores}
\label{vsys}

Figure 7 shows the profiles of the HCN(1--0) and HCO$^+$(1--0) lines toward positions near the peaks
of their emission. This figure also shows the (1--0) lines of isotopes of these molecules, which have lower optical depths, H$^{13}$CN and H$^{13}$CO$^+$. In all the objects except for 37.427+1.518, the optically thick HCN(1--0) and HCO$^+$(1--0) lines display self absorption, and the positions of absorption dips are close to the peaks of optically thin lines. The red and blue wings, separated by the dips, have different amplitudes. The blue wings are stronger in 121.28+0.65 and 99.982+4.17, while the red wings are stronger in 34.403+0.233 and 77.462+1.7. The asymmetry of the optically thick lines and symmetric, nearly Gaussian optically thin lines with peaks close to the absorption dips of the optically thick lines indicates the presence of systematic motions \cite{Evans} (collapse in the case of 121.28+0.65 and 99.982+4.17 and expansion in the case of 34.403+0.233 and 77.462+1.759). It is also possible that the absorption dips in the line profiles appear due to the presence of tenuous envelopes with smaller velocity dispersions around the cores (e.g., in 34.403+0.233 and 99.982+4.17). The optically thick lines sometimes have broad wings, suggesting the presence of high-velocity gas along the line of sight.

Based on the model presented in \cite{Myers}, we can crudely estimate the infall velocity $V_{\rm IN}$ from the velocities and intensities of the red and blue peaks of an optically thick line whose blue wing is stronger than its red wing. Using the HCO$^+$(1--0) and H$^{13}$CO$^+$(1--0) profiles, we found $V_{\rm IN}$ to be equal to $\sim 0.14$~km~s$^{-1}$ and $\sim 0.16$~km~s$^{-1}$ in 121.28+0.65 and 99.982+4.17, respectively. The inward mass flow rate (accretion rate) can be estimated from the equation
$dM/dt=4\pi\,r_{\rm in}^2\,m\,n_0\,V_{\rm IN}$ \cite{Myers, KW}, where $r_{\rm in}$ is the radius of the collapsing region, $m$ is the mean molecular mass (2.33 amu), and $n_0$ is the density of the surrounding gas. Taking the mean core radii $r_{\rm in}$ =  0.15 pc and $r_{\rm in}$ = 0.09 pc for 121.28+0.65 and 99.982+4.17,
respectively, yields approximate estimated accretion rates of $\sim 2\times 10^{-5}$~$M_{\odot}$~yr$^{-1}$ and $\sim 2\times 10^{-4}$~$M_{\odot}$~yr$^{-1}$  for 121.28+0.65 and $\sim 8\times 10^{-6}$~$M_{\odot}$~yr$^{-1}$ and $\sim 8\times 10^{-5}$~$M_{\odot}$~yr$^{-1}$ for 99.982+4.17 assuming plausible values $n_0=10^4$~cm$^{-3}$ and $10^5$~cm$^{-3}$, respectively. These estimates lie in the range of accretion rates determined for a number of regions where stars with masses exceeding a solar mass are forming (see, e.g. \cite{KW,Fuller}). In the cores studied here, the line broadening due to collapse is $\sim 0.3$~km~s$^{-1}$, which is less than the observed difference between the line widths toward the centers and the peripheries of the cores (Fig. 6), and thus cannot be the only reason for the observed $\Delta V(b)$ trends.

\section{Discussion}

The observations show the rich molecular compositions of the observed regions (see Table
5), typical for ``hot cores'' that are associated with massive star formation regions or regions of cluster star formation. The 37.427+1.518 core has a slightly poorer composition (in particular, the HC$^{15}$N and NH$_2$D lines detected in other objects were not found in this core). 

The kinetic temperatures derived by us are, in general, close to the temperatures obtained in other studies \cite{Molinari,Zin98,Alakoz,Sanchez,Cyganowski}, given the uncertainties in our estimates. Note that the kinetic temperatures determined from
the ammonia observations are slightly below our estimates. The kinetic temperatures toward the positions
close to IRAS point sources are close to the dust temperatures that we calculated adopting the spectral
index for the frequency dependence of the dust absorption coefficient $\beta=1-2$. The kinetic temperature
in 34.403+0.233 (20$''$,0$''$) is slightly above the dust temperature (32 K for $\beta=1$); given the spatial
resolution of our observations is different from that of IRAS, this can be explained by temperature gradients
in this core. 

A comparison of the virial masses from Table 6 with masses estimated 
in other studies shows both agreements and differences. 
Our virial masses for 121.28+0.65 are close to estimates based on observations 
of ammonia \cite{Zin97} and HCN(3--2) \cite{Wu10}. Similar estimates obtained 
using optically thick lines, CS(2--1) and HCN(1--0) \cite{Zin98,Wu10}, 
are a factor of $\sim 2-3$ higher. Our mass for 99.982+4.17 agrees with 
the mass determined in \cite{Sugitani}, although the latter may be overestimated 
(see Section 5.2). 
Our virial mass for
37.427+1.518 is appreciably lower than the mass derived from the dust observations \cite{Beuther} scaled to the distance presented in Table 1. Our virial mass estimates for 77.462+1.759 are close to the estimates of 
\cite{Wu10}.
The differences between the masses obtained using different molecular lines or dust are probably related
to the differences in the sizes of the emitting regions, which, in turn, probably result from differences in the
methods used to determine the sizes, the conditions for exciting different lines, and the chemical variations
in the cores.

The molecular abundances in the core centers vary from source to source by a factor of a few (see Table 5),
being the highest in 34.403+0.233 and the lowest in 99.982+4.17. This may partly be related to the fact that, in the latter case, we used an overestimated $N$(H$_2$) value to calculate the abundances. The core in 99.982+4.17 is associated with the least luminousIRAS object in our sample, is the most compact core
in our sample, and has the lowest virial mass.

An increased level of non-thermal motion near the core centers (including 37.427+1.518, not shown in Fig. 6) shows the impact of the inner sources on the surrounding gas at distances $b\la 30-50''$ ($\la 0.15-0.25$ pc). The spectral indices for the $\Delta V(b)$ dependences calculated for different lines are mostly similar within a particular object. This may indicate that these lines trace the same regions inside the cores. The $\Delta V(b)$ dependences in 99.982+4.17 have the highest indices, where the region of enhanced
dynamic activity is smaller compared to the other cores. The $\Delta(b)$ dependences for the SO, SiO,
and NH$_2$D lines sometimes differ from the other dependences. For example, the widths of the NH2D
lines in 121.28+0.65 and 99.982+4.17 do not change significantly with distance from their centers, while
these widths decrease with increasing impact parameter in 34.403+0.233 and 77.462+1.759.

The core structures and kinematics are obviously different, as is indicated by the differences in the shapes of the optically thick HCN(1--0) and HCO$^+$(1--0) lines. Two cores demonstrate inward motions (collapse), the gas in two other objects is expanding, and one core (37.427+1.518) showed no clear evidence for systematic radial motions. Further studies of the sample objects should be aimed at detailed comparative analysis of their densities and kinematics, which requires computations of molecular excitation using inhomogeneous source models, taking into account possible gradients of temperature, density, and turbulent velocity dispersion, as well as systematic velocity fields. The results of these studies will be presented in subsequent publications.

\section{Conclusions}

We have used the 20-m radio telescope of the Onsala Space Observatory (Sweden) to obtain spectral
observations of five regions of massive star formation at frequencies $\sim 85-89$~GHz in the
c-C$_3$H$_2$(2$_{1,2}$--1$_{0,1}$), CH$_3$C$_2$H(5--4), SO(2$_2$--$1_1$), H$^{13}$CN(1--0), H$^{13}$CO$^+$(1--0), SiO(2--1), HN$^{13}$C(1--0), C$_2$H(1--0~3/2--1/2), HCN(1--0), HCO$^+$(1--0), NH$_2$D(1$_{1,1}$--$1_{0,1}$) and some other lines.
These regions are associated with methanol maser sources observed previously in the HCN(1--0) line with the CrAO 22-m telescope, IRAS point sources, near-IR and mid-IR sources, and radio and submillimeter
sources. The maps in different molecular lines show that these objects have dense cores containing young
stellar objects. We have obtained the following results.

1. We have determined the main physical parameters of the dense cores, including the sizes of the emitting regions ($\sim 0.1-0.6$ pc), kinetic temperatures ($\sim 20-40$ K), and virial masses ($\sim 40-500 M_{\odot}$). The line widths ($\ga 2$~km~s$^{-1}$) are significantly larger than the thermal widths. Four cores are nearly spherically-symmetric.

2. Column densities and abundances of different molecules were calculated in the LTE approximation.
The core in 34.403+0.233 has the highest molecular abundances, while the core in 99.982+4.17, which
is associated with the weakest IRAS source among the objects in our sample, displays reduced molecular
abundances.

3. The widths of molecular lines ($\Delta V$) that are probably optically thin decrease with increasing impact
parameter ($b$), indicating the influence of the inner sources on the dynamical activity of the gas in the cores. The $\Delta V(b)$ dependences for distances $\ga 0.1$ pc from the core centers are close to power laws ($\propto b^{-p}$), where $p$ varies from $\sim 0.2$ to $\sim 0.5$, depending on the object. The highest $p$ values were found for the $\Delta V(b)$ dependences in 99.982+4.17.

4. The shapes of the asymmetric profiles of the optically thick HCN(1--0) and HCO$^+$(1--0) lines indicate
collapse in 121.28+0.65 and 99.982+4.17 and expansion in 34.403+403 and 77.462+1.759. We used two plausible density values of the gas around the regions displaying systematic motions to estimate the inward velocities and accretion rates in 121.28+0.65 and 99.982+4.17. It is likely that stars with masses exceeding the solar mass are forming in these cores.

\begin{acknowledgments}

The work was performed in the framework of the government program 0035-2014-0030 (theme 16.30,
``Spectral radio astronomy studies at millimeter and submillimeter waves''). This work was supported by the Russian Foundation for Basic Research (grants 14-02-90441-Ukr\_a, 15-02-06098-a, 
16-02-00761, 16-32-00873-mol\_a).

\end{acknowledgments}

\vspace{3mm}
\begin{flushright}
\it{Translated into English by S.~Kalenskii}
\end{flushright}

\newpage

\begin{table}[p]
\setcaptionmargin{0mm}
\onelinecaptionsfalse
\centering
\caption{Source list}
\vskip 2mm
\footnotesize
\begin{tabular}{l|c|c|r|l}
\noalign{\hrule}\noalign{\smallskip}
Source       &$\alpha$(2000)  &$\delta$(2000)     & $D$ (kpc) & Association with other objects\\
\noalign{\hrule}\noalign{\smallskip}
121.28+0.65     & 00:36:42.2   & $+$63:28:30.0  & 0.93 \cite{Rygl}  &  IRAS 00338+6312, RNO1B, L1287 \\
34.403+0.233   & 18:53:17.4   & $+$01:24:55.0  & 1.56 \cite{Kurayama} &  IRAS 18507+0121, G034.43 MM1\\
37.427+1.518    & 18:54:14.3   & $+$04:41:39.0  & 1.88 \cite{Wu14} &  IRAS 18517+0437, G037.43 \\
77.462+1.759    & 20:20:39.3   & $+$39:37:52.0  & 1.4 \cite{Lumsden}  &  IRAS 20188+3928  \\
99.982+4.17      & 21:40:42.3   & $+$58:16:09.7  & 0.75 \cite{Molinari} &  IRAS 21391+5802, IC1396, L1121 \\
\noalign{\hrule}\noalign{\smallskip}
\end{tabular}

\label{list}
\end{table}
\normalsize

\newpage

\begin{table}[p]
\setcaptionmargin{0mm}
\onelinecaptionsfalse

\centering
\caption{List of observed molecular lines}
\vskip 2mm
\scriptsize
\begin{tabular}{l|c|r|r}
\noalign{\hrule}\noalign{\smallskip}
Molecule  &  Transition  & Frequency (MHz) & $E_u/k$ (K) \\
\noalign{\smallskip}\hline\noalign{\smallskip}

c-C$_3$H$_2$ 	&	 2$_{1,2}$--1$_{0,1}$ 	&	 85338.906 	&	6.45 	\\
 HCS$^+$     	&	 2--1           	&	 85347.869 	&	6.14 	\\
CH$_3$C$_2$H 	&	 5$_3$--4$_3$     	&	 85442.600 	&	77.34 	\\
CH$_3$C$_2$H 	&	 5$_2$--4$_2$     	&	 85450.765 	&	41.21 	\\
CH$_3$C$_2$H 	&	 5$_1$--4$_1$     	&	 85455.665 	&	19.53 	\\
CH$_3$C$_2$H 	&	 5$_0$--4$_0$     	&	 85457.299 	&	12.30 	\\
C$_4$H      	&	 9--8~19/2--17/2   	&	 85634.000 	&	20.55 	\\
NH$_2$D     	&	 1$_{1,1}$--1$_{0,1}$ $F$=0--1 	&	 85924.747 	&	20.68 	\\
NH$_2$D     	&	 1$_{1,1}$--1$_{0,1}$ $F$=2--1 	&	 85925.684 	&	20.68 	\\
NH$_2$D     	&	 1$_{1,1}$--1$_{0,1}$ $F$=2--2 	&	 85926.263 	&	20.68 	\\
NH$_2$D     	&	 1$_{1,1}$--1$_{0,1}$ $F$=1--2 	&	 85926.858 	&	20.68 	\\
NH$_2$D     	&	 1$_{1,1}$--1$_{0,1}$ $F$=1--0 	&	 85927.721 	&	20.68 	\\
HC$^{15}$N  	&	 1--0           	&	 86054.967 	&	4.13 	\\
SO          	&	 2$_2$--1$_1$     	&	 86093.983 	&	19.31 	\\
C$_2$S      	&	 7$_6$--6$_5$       	&	 86181.413 	&	23.35 	\\
H$^{13}$CN  	&	 1--0  $F$=1--1  	&	 86338.737 	&	4.14 	\\
H$^{13}$CN  	&	 1--0  $F$=2--1  	&	 86340.176 	&	4.14 	\\
H$^{13}$CN  	&	 1--0  $F$=0--1  	&	 86342.255 	&	4.14 	\\
HCO         	&	 1$_{0,1}$--0$_{0,0}$ 3/2--1/2 $F=2-1$ 	&	 86670.82 	&	4.18 	\\
HCO         	&	 1$_{0,1}$--0$_{0,0}$ 3/2--1/2 $F=1-0$ 	&	 86708.35 	&	4.16 	\\
H$^{13}$CO$^+$ 	&	1--0         	&	 86754.288 	&	4.16 	\\
HCO         	&	 1$_{0,1}$--0$_{0,0}$ 1/2--1/2 $F=1-1$ 	&	 86777.43 	&	4.18 	\\
SiO         	&	 2--1           	&	 86846.995 	&	6.25 	\\
HN$^{13}$C  	&	 1--0 $F$=2--1   	&	 87090.859 	&	4.18 	\\
C$_2$H      	&	 1--0 3/2--1/2 $F$=1--1 	&	 87284.156 	&	4.19 	\\
C$_2$H      	&	 1--0 3/2--1/2 $F$=2--1 	&	 87316.925 	&	4.19 	\\
C$_2$H      	&	 1--0 3/2--1/2 $F$=1--0 	&	 87328.624 	&	4.19 	\\
C$_2$H      	&	 1--0 1/2--1/2 $F$=1--1 	&	 87402.004 	&	4.20 	\\
C$_2$H      	&	 1--0 1/2--1/2 $F$=0--1 	&	 87407.165 	&	4.20 	\\
C$_2$H      	&	 1--0 1/2--1/2 $F$=1--0 	&	 87446.512 	&	4.20 	\\
HNCO        	&	 4$_{0,4}$--3$_{0,3}$ 	&	 87925.238 	&	10.55 	\\
HCN         	&	 1--0  $F$=1--1  	&	 88630.416 	&	4.25 	\\
HCN         	&	 1--0  $F$=2--1  	&	 88631.847 	&	4.25 	\\
HCN         	&	 1--0  $F$=0--1  	&	 88633.936 	&	4.25 	\\
H$^{15}$NC  	&	 1--0           	&	 88865.692  	&	4.26 	\\
HCO$^+$     	&	 1--0           	&	 89188.526 	&	4.28 	\\

\noalign{\smallskip}\hline\noalign{\smallskip}
\end{tabular}
\label{lines}
\end{table}

\newpage


\begin{table}[p]
\setcaptionmargin{0mm}
\onelinecaptionsfalse

\centering
\caption{Parameters of observed lines}
\vskip 2mm
\tiny
\begin{tabular}{l|l|l|l|ll|l|l|l|ll|l|l|l|l}
\noalign{\hrule}\noalign{\smallskip}
       & \multicolumn{4}{c} {121.28+0.65 (40$''$,20$''$)} &
       & \multicolumn{4}{c} {34.403+0.233 (20$''$,0$''$)} & 
       & \multicolumn{4}{c} {37.427+1.518 (0$''$,0$''$)} \\
\noalign{\smallskip}\cline{2-5}\cline{7-10}\cline{12-15}\noalign{\smallskip}
Line & $I$ & $T_{\rm MB}$ & $V_{\rm LSR}$ & $\Delta V$ &
      & $I$ & $T_{\rm MB}$ & $V_{\rm LSR}$ & $\Delta V$ &
      & $I$ & $T_{\rm MB}$ & $V_{\rm LSR}$ & $\Delta V$ \\
      & (K km s$^{-1}$) & (K) & (km s$^{-1}$) & (km s$^{-1}$) &
      & (K km s$^{-1}$) & (K) & (km s$^{-1}$) & (km s$^{-1}$) &
      & (K km s$^{-1}$) & (K) & (km s$^{-1}$) & (km s$^{-1}$) \\
\noalign{\smallskip}\hline\noalign{\smallskip}

c-C$_3$H$_2$(2$_{1,2}$--1$_{0,1}$)
                             & 1.68(0.05) & 0.64(0.02) & --17.05(0.04) & 2.5(0.1) &
                             & 1.8(0.1)   & 0.44(0.03) & 57.4(0.1)     & 4.71(0.3) &
                             & 0.8(0.2)   & 0.29(0.05) & 44.5(0.2)     & 2.5(0.5) \\
HCS$^+$(2--1)                & 0.37(0.05) & 0.15(0.02) & --16.8(0.2)   & 2.6(0.4) &
                             & 0.4(0.1)   & 0.15(0.03) & 56.9(0.3)     & 2.9(0.7) &
                             & 0.7(0.2)   & 0.16(0.05) & 44.2(0.6)     & 4.2(1.5) \\
CH$_3$C$_2$H(5$_0$--$4_0$)      & 2.5(0.1)   & 0.51(0.02) & --17.08(0.03) & 2.1(0.1) &
                             & 3.2(0.2) & 0.41(0.03) & 57.6(0.1)     & 2.9(0.2)      &
                             & 1.7(0.2) & 0.31(0.05) & 44.0(0.2)     & 2.8(0.4) \\
C$_4$H(9--8~19/2--17/2)  & 0.12(0.04)& 0.12(0.04) & --16.8(0.1)   & 0.8(0.3) &
                             && $<0.2$  &&&
                             && $<0.3$ \\
NH$_2$D(1$_{1,1}$--$1_{0,1}$~$F$=2--2)
                             & 1.9(0.1) & 0.41(0.02) & --17.3(0.1) & 1.7(0.1) &
                             & 2.9(0.2) & 0.42(0.05) & 58.3(0.1)   & 2.1(0.2) &
                             && $<0.3$ \\
HC$^{15}$N(1--0)             & 0.5(0.05) & 0.22(0.02) & --17.3(0.1)   & 2.1(0.2) &
                             & 0.5(0.1)  & 0.25(0.04) & 57.5(0.2)     & 2.1(0.4) &
                             && $<0.35$ \\
SO(2$_2$--$1_1$)                 & 0.8(0.1) & 0.32(0.02) & --17.2(0.1)   & 2.3(0.2) &
                             & 0.7(0.1) & 0.20(0.03) & 57.8(0.2)     & 3.4(0.6)     &
                             & 1.4(0.2) & 0.50(0.05) & 43.8(0.1)     & 2.4(0.3) \\
H$^{13}$CN(1--0~$F$=2--1)     & 3.9(0.1) & 0.64(0.02) & --17.49(0.04) & 3.0(0.1)    &
                             & 9.2(0.2) & 0.94(0.03) & 57.6(0.1)     & 4.4(0.1)     &
                             & 2.7(0.2) & 0.55(0.05) & 44.0(0.1)     & 2.9(0.2) \\
H$^{13}$CO$^+$(1--0)         & 4.5(0.1) & 1.86(0.02) & --17.72(0.01) & 2.27(0.04)   &
                             & 5.7(0.2) & 1.32(0.03) & 57.57(0.04)   & 4.0(0.1)     &
                             & 3.2(0.2) & 1.17(0.05) & 43.9(0.1)     & 2.6(0.1) \\
SiO(2--1)                    & 1.8(0.1) & 0.53(0.02) & --17.36(0.05) & 3.0(0.1)     &
                             & 7.8(0.2) & 0.73(0.02) & 57.9(0.1)     & 8.8(0.3)     &
                             & 1.0(0.2) & 0.23(0.04) & 44.0(0.3)     & 3.7(0.8) \\
HN$^{13}$C(1--0)             & 1.8(0.1) & 0.67(0.02) & --17.59(0.04) & 2.5(0.1)     &
                             & 3.1(0.1) & 0.82(0.03) & 57.8(0.1)     & 3.6(0.2)     &
                             & 0.8(0.1) & 0.33(0.06) & 44.0(0.2)     & 2.3(0.5) \\
C$_2$H(1--0~3/2--1/2~$F$=2--1) & 4.38(0.05) & 1.64(0.02) & --17.70(0.02) & 2.53(0.04)& 
                              & 4.8(0.1)   & 1.06(0.02) & 57.9(0.1)     & 4.3(0.1)   &
                              & 6.2(0.2)   & 1.88(0.05) & 43.92(0.04)   & 3.0(0.1) \\
HNCO(4$_{0,4}$--$3_{0,3}$) & 0.8(0.3) & 0.3(0.1)   & --16.5(0.5) & 3.8(1.3)          &
                     & 2.5(0.3) & 0.39(0.05) & 56.8(0.3)   & 5.8(0.8)                &
                     & 1.1(0.2) & 0.26(0.05) & 44.1(0.4)   & 4.7(1.1) \\
HCN(1--0~$F$=2--1)   & 38.7(0.7) &&&&
                     & 31.6(0.5) &&&&
                     & 43.0(0.2) & 6.19(0.04) & 43.84(0.01) & 3.69(0.02) \\
HCO$^+$(1--0)        & 34.5(0.5) &&&&
                     & 12.4(0.3) &&&&
                     & 35.1(0.3) & 9.3(0.1)  & 43.86(0.01)  & 3.46(0.03) \\

\noalign{\smallskip}\hline\noalign{\smallskip}
\end{tabular}
\label{parln}

Integrated intensities of the CH$_3$C$_2$H(5--4),
NH$_2$D(1$_{1,1}$-1$_{0,1}$), H$^{13}$CN(1--0) and HCN(1--0)
lines were calculated over all spectral components.

\end{table}

\newpage

\small

\addtocounter{table}{-1}

\begin{table}[p!]
\setcaptionmargin{0mm}
\onelinecaptionsfalse

\centering
\caption{Continued}
\vskip 2mm
\scriptsize
\begin{tabular}{l|l|l|l|ll|l|l|l|l}
\noalign{\hrule}\noalign{\smallskip}
       & \multicolumn{4}{c} {77.462+1.759 (--20$''$,0$''$)} &
       & \multicolumn{4}{c} {99.982+4.17 (0$''$,0$''$)} \\
\noalign{\smallskip}\cline{2-5}\cline{7-10}\noalign{\smallskip}
‹¨­¨ï  & $I$ & $T_{\rm MB}$ & $V_{\rm LSR}$ & $\Delta V$ &
       & $I$ & $T_{\rm MB}$ & $V_{\rm LSR}$ & $\Delta V$ \\
       & (K ª¬ á$^{-1}$) & (K) & (ª¬ á$^{-1}$) & (ª¬ á$^{-1}$) &
       & (K ª¬ á$^{-1}$) & (K) & (ª¬ á$^{-1}$) & (ª¬ á$^{-1}$) \\
\noalign{\smallskip}\hline\noalign{\smallskip}

c-C$_3$H$_2$(2$_{1,2}$--$1_{0,1}$)
                             & 2.3(0.1) & 0.96(0.03) & 2.03(0.03)  & 2.2(0.1) &
                             & 1.4(0.1) & 0.47(0.02) & 1.01(0.05)  & 2.5(0.1) \\
HCS$^+$(2--1)
                             & 0.3(0.1)   & 0.20(0.03) & 2.3(0.1)  & 1.6(0.3) &
                             & 0.24(0.05) & 0.14(0.03) & 0.9(0.1)  & 1.4(0.3) \\
CH$_3$C$_2$H(5$_0$-$4_0$)
                             & 3.4(0.1) & 0.71(0.03) & 1.78(0.03)    & 2.0(0.1) &
                             & 0.8(0.1) & 0.17(0.03) & 0.8(0.1)      & 1.5(0.2) \\
NH$_2$D(1$_{1,1}$--$1_{0,1}$~$F$=2--2)
                                & 3.5(0.1) & 0.80(0.03) & 1.5(0.04)   & 1.7(0.1) &
                                & 0.7(0.1) & 0.14(0.03) & 0.1(0.2)    & 2.0(0.3)\\
HC$^{15}$N(1--0)
                             & 0.4(0.1) & 0.16(0.03) & 2.0(0.2)      & 2.6(0.5) &
                             & 0.5(0.1) & 0.23(0.02) & 0.7(0.1)      & 2.0(0.2) \\
SO(2$_2$--$1_1$)
                             & 0.4(0.1) & 0.23(0.03) & 1.9(0.1)      & 1.9(0.3) &
                             & 1.0(0.1) & 0.48(0.02) & 0.9(0.1)      & 1.9(0.1) \\
CCS(7$_6$--$6_5$)
                             & 0.2(0.1) & 0.10(0.03) & 1.7(0.3)      & 2.4(0.7) &
                             & 0.4(0.1) & 0.08(0.02) & 0.5(0.4)      & 4.8(1.0) \\
H$^{13}$CN(1--0~$F$=2--1)
                             & 2.8(0.1) & 0.60(0.03) & 1.78(0.04)    & 2.3(0.1) &
                             & 2.2(0.1) & 0.60(0.02) & 0.64(0.03)    & 2.0(0.1) \\
HCO(1$_{0,1}$--$0_{0,0}$~3/2--1/2~$F$=2--1)
                                     & 0.2(0.1)   & 0.13(0.03) & 1.3(0.2) & 1.5(0.4) &
                                     & 0.31(0.01) & 0.10(0.02) & 1.7(0.3) & 3.0(0.6) \\
H$^{13}$CO$^+$(1--0)
                             & 4.8(0.1) & 1.93(0.03) & 1.44(0.02)    & 2.21(0.04) &
                             & 3.9(0.1) & 2.05(0.03) & 0.48(0.01)    & 1.8(0.03) \\
SiO(2--1)
                             & 1.6(0.2) & 0.28(0.02) & 1.7(0.2)      & 5.1(0.4) &
                             & & $<0.15$ \\
HN$^{13}$C(1--0)
                             & 2.2(0.1) & 1.03(0.03) & 1.45(0.02)    & 2.0(0.1) &
                             & 0.8(0.1) & 0.42(0.03) & 0.40(0.05)    & 1.7(0.1) \\
C$_2$H(1--0~3/2--1/2~$F$=2--1)
                              & 6.2(0.1) & 2.25(0.04) & 1.37(0.02)    & 2.37(0.04) &
                              & 4.0(0.1) & 1.47(0.02) & 0.48(0.02)    & 2.47(0.05) \\
HNCO(4$_{0,4}$--$3_{0,3}$)
                     & 0.6(0.1) & 0.26(0.05) & 2.3(0.2)  & 1.8(0.4) &
                     & 0.2(0.1) & 0.26(0.06) & 0.9(0.1)  & 0.9(0.2) \\
HCN(1--0)            & 32.6(0.2) &&&&
                     & 30.7(0.2) \\
H$^{15}$NC(1--0)
                     & 0.7(0.1) & 0.22(0.03) & 1.7(0.2) & 2.9(0.5)   &
                     & 0.3(0.1) & 0.16(0.05) & --0.1(0.2) & 1.4(0.5) \\
HCO$^+$(1--0)        & 33.2(0.2) &&&&
                     & 17.5(0.2) \\

\noalign{\smallskip}\hline\noalign{\smallskip}
\end{tabular}

\end{table}


\newpage

\begin{table}[p]
\setcaptionmargin{0mm}
\onelinecaptionsfalse
\centering
\caption{Kinetic temperatures}
\vskip 2mm
\small
\begin{tabular}{l|r|r|r}
\noalign{\hrule}\noalign{\smallskip}
Source     & $\Delta\alpha ('')$  & $\Delta\delta ('')$  & $T_{\rm KIN}$ (K)  \\
 \noalign{\hrule}\noalign{\smallskip}
121.28+0.65   & 40    &  20  & 26(4) \\
                       & 60    &  40  & 22(6) \\
                       & 20    &  60   & 32(8) \\
34.403+0.233      &   0  &  40  & 42(8)   \\
                         & 20   & 0  &  37(7) \\
                         &  20   & 40  & 29(7) \\
77.462+1.759     & --20 & 0  & 29(3) \\
                           & 20   &  20 & 22(6) \\
                           & 20   & --20 & 36(5) \\
                           & 0     & --20  & 24(3) \\
99.982+4.17       & --20    & 0  & 33(12) \\
\noalign{\hrule}\noalign{\smallskip}
\end{tabular}
\normalsize
\label{tkin}
\end{table}

\newpage


\begin{table}[p]
\setcaptionmargin{0mm}
\onelinecaptionsfalse

\centering
\caption
{Molecular column densities and abundances}
\vskip 2mm
\scriptsize
\begin{tabular}{l|l|ll|l|ll|l|ll|l|ll|l|l}
\noalign{\hrule}\noalign{\smallskip}
Molecule &\multicolumn{2}{c}{121.28(40$''$,20$''$)} &&\multicolumn{2}{c}{34.403(20$''$,0$''$)}&
&\multicolumn{2}{c}{37.427(0$''$,0$''$)} &&\multicolumn{2}{c}{77.462(--20$''$,0$''$)}
&&\multicolumn{2}{c}{99.982(0$''$,0$''$)} \\
\noalign{\smallskip}\cline{2-3}\cline{5-6}\cline{8-9}\cline{11-12}\cline{14-15}\noalign{\smallskip}
              & $N$, cm$^{-2}$    & $X$   && $N$, cm$^{-2}$ & $X$ && $N$, cm$^{-2}$ & $X$ &
              & $N$, cm$^{-2}$     & $X$  && $N$, cm$^{-2}$ & $X$ \\
              & ($\times 10^{12}$) & ($\times 10^{-11}$)  && ($\times 10^{12}$) & ($\times 10^{-11}$) &
               & ($\times 10^{12}$) & ($\times 10^{-11}$)  && ($\times 10^{12}$) & ($\times 10^{-11}$) &
& ($\times 10^{12}$) & ($\times 10^{-11}$) \\
\noalign{\smallskip}\hline\noalign{\smallskip}
c-C$_3$H$_2$  & 8.3  & 8.3  && 8.9  & 15.1 && 3.9 & 1.1 && 11.3 & 5.7 && 6.9 & 1.4 \\
CH$_3$C$_2$H  & 123  & 123  && 193  & 327  && 119 & 201 && 182  & 308 && 28  & 47  \\
C$_2$H        & 447  & 447  && 550  & 932  && 550 & 149 && 550  & 275 && 302 & 60.4 \\
C$_4$H        & 13.8 & 13.8 \\
NH$_2$D       & 347  & 347  && 1170 & 1983 &&     &     && 739  & 370 && 259 & 51.8 \\
H$^{13}$CN    & 7.7  & 7.7  && 18.1 & 30.7 && 5.3 & 1.4 && 5.5  & 2.8 && 4.3 & 0.9 \\
H$^{13}$CO$^+$& 5.2  & 5.2  && 6.6  & 11.2 && 3.7 & 1.0 && 5.5  & 2.8 && 4.5 & 0.9 \\
HN$^{13}$C    & 3.4  & 3.4  && 5.8  &  9.8 && 1.5 & 0.4 && 4.1  & 2.1 && 1.5 & 0.3 \\
H$^{15}$NC    &      &      &&      &      &&     &     && 1.6  & 0.8 && 0.7 & 0.14 \\
HC$^{15}$N    & 1.0  & 1.0  && 1.0  &  1.7 &&     &     && 0.8  & 0.4 && 1.0 & 0.2 \\
HNCO          & 8.2  & 8.2  && 25.7 & 43.6 && 11.3& 3.1 && 6.2  & 3.1 && 2.1 & 0.4 \\
HCO           &      &      &&      &      &&     &     && 1.2  & 0.6 && 1.9 & 0.4 \\
HCS$^+$       & 2.1  &  2.1 && 0.8  & 1.4  && 0.9 & 0.2 && 1.7  & 0.9 && 1.3 & 0.3 \\
CCS           &      &     & &      &      &&     &     && 1.4  & 0.7 && 2.7 & 0.5 \\

\noalign{\smallskip}\hline\noalign{\smallskip}
\end{tabular}

\label{colden}
\end{table}


\newpage

\begin{table}[p]
\setcaptionmargin{0mm}
\onelinecaptionsfalse

\centering
\caption{Physical parameters of the cores}
\vskip 2mm
\tiny
\begin{tabular}{l|l|l|l|l|l|l|l}
\noalign{\hrule}\noalign{\smallskip}
Line  & $\Delta\alpha$ & $\Delta\delta$ & Axis & $\Delta\Theta$ & $d$
       & $\langle\Delta V\rangle$ & $M_{\rm vir}$ \\
       & ($'',''$)      & ($'',''$)    &  ratio    & ($''$)         & (pc)
       & (km s$^{-1}$)  & ($M_\odot$) \\
\noalign{\smallskip}\hline\noalign{\smallskip}
{\bf 121.28+0.65} \\
c-C$_3$H$_2$  & 40.5(2.1) & 23.0(1.7) & 1.5(0.3) & 58.9(9.7) & 0.27(0.05) & 1.9(0.1) & 101(20) \\
CH$_3$C$_2$H  & 27.0(3.7) & 29.6(3.0) & 1.3(0.3) & 79.1(8.4) & 0.36(0.04) & 2.0(0.1) & 150(21)  \\
NH$_2$D       &           &           &          &           & & 1.3(0.1) \\
SO            & 25.2(3.6) & 34.9(3.1) & 1.5(0.5) & 50.6(8.5) & 0.23(0.04) & 2.4(0.2) & 138(33) \\
H$^{13}$CN    & 35.4(1.7) & 33.8(1.8) & 1.1(0.3) & 36.6(5.0) & 0.17(0.02) & 2.8(0.1) & 136(21) \\
H$^{13}$CO$^+$& 36.0(1.5) & 30.3(1.3) & 1.5(0.2) & 60.3(3.5) & 0.27(0.02) & 2.0(0.1) &  114(13) \\
SiO           &           &           &          &           & & 3.1(0.2) \\
HN$^{13}$C    & 37.6(1.8) & 32.3(1.8) & 1.0(0.3) & 37.4(5.2) & 0.17(0.02)& 2.0(0.2) & 71(17) \\
C$_2$H        & 31.2(1.2) & 27.6(1.1) & 1.2(0.1) & 55.7(2.8) & 0.25(0.01) & 2.1(0.1) & 116(13) \\
HCN           & 33.6(0.9) & 33.9(0.9) & 1.4(0.1) & 50.7(2.3)  &  0.23(0.01) \\
HCO$^+$       & 35.2(1.0) & 38.9(1.1) & 1.1(0.1) & 65.7(2.5)  &  0.30(0.01) \\
\noalign{\smallskip}\hline\noalign{\smallskip}
{\bf 34.403+0.233} \\
c-C$_3$H$_2$  &           &           &          &           & & 2.9(0.2) \\
CH$_3$C$_2$H  &           &           &          &           & & 2.9(0.1) \\
NH$_2$D       &           &           &          &           & & 1.6(0.1) \\
SO            &           &           &          &           & & 3.7(0.3) \\
H$^{13}$CN    & 15.1(2.2) & 13.9(2.1) & 4.0(8.2) & 22.9(23.5) & 0.17(0.18) & 3.9(0.1) & 280(280) \\
H$^{13}$CO$^+$&           &           &          &           & & 3.1(0.1) \\
SiO           &           &           &          &           & & 7.5(0.4) \\
HN$^{13}$C    & 19.2(0.9) & 11.2(2.4) & 3.5(1.4) & 62.6(12.5) & 0.5(0.1) & 2.9(0.1) & 420(90) \\
C$_2$H        & 19.1(1.2) & 5.5(2.1)  & 3.7(2.6) & 36.2(12.7) & 0.3(0.1) & 4.1(0.1) & 480(170) \\
\noalign{\smallskip}\hline\noalign{\smallskip}
{\bf 37.427+1.518} \\
c-C$_3$H$_2$  &           &           &          &           & & 2.0(0.2) \\
CH$_3$C$_2$H  &           &           &          &           & & 1.7(0.6) \\
SO            &           &           &          &           & & 2.9(0.5) \\
H$^{13}$CN    & --2.2(3.4)& 1.3(3.6)  & 1.4(1.9) & 29.7(21.3)& 0.3(0.2) & 2.5(0.2) & 180(130) \\
H$^{13}$CO$^+$& 4.0(2.0)  & 3.0(1.9)  & 1.9(2.7) & 24.7(17.4) &  0.2(0.2)& 2.3(0.1) & 130(90) \\
C$_2$H        &           &           &          &           & & 2.8(0.1) \\
HCN           & --1.7(1.0)& 6.0(0.9)  & 1.2(0.2) & 50.5(4.7) &  0.46(0.04) & 3.31(0.09) &  530(57) \\
\noalign{\smallskip}\hline\noalign{\smallskip}
{\bf 77.462+1.759} \\
c-C$_3$H$_2$  & --16.8(3.0) & --0.2(2.9) & 1.4(0.5) & 93.1(15.9) & 0.6(0.1) & 1.9(0.1) & 240(48) \\
CH$_3$C$_2$H  & --8.1(2.1)  & --2.1(2.4) & 1.7(0.6) & 41.1(7.3)  & 0.28(0.05) & 1.8(0.1) & 95(20) \\
NH$_2$D       & --16.9(3.0) & 10.7(3.6)  & 4.0(1.8) & 58.3(13.2) & 0.40(0.09) & 1.6(0.1) & 106(28) \\
SO            & & & & & & 2.1(0.2) \\
H$^{13}$CN    & --14.1(2.3) & 4.1(3.0) & 1.9(0.8) & 45.0(8.9) & 0.31(0.06) & 2.2(0.1) &  155(34) \\
H$^{13}$CO$^+$& --11.7(1.0) & 1.8(1.1) & 1.7(0.2) & 53.4(3.4) & 0.36(0.02) & 2.0(0.1) & 152(18) \\
SiO           & --20.5(2.1) & 30.4(2.2) & 1.3(0.3) & 49.5(6.3) & 0.34(0.04) & 5.5(0.4) & 1070(206) \\
HN$^{13}$C    & --12.8(1.8) & 0.2(2.0) & 1.4(0.3) & 56.8(6.0) & 0.39(0.04) & 1.9(0.1)  & 146(22)\\
C$_2$H        & --8.9(1.2)  & 0.9(1.4) & 1.4(0.2) & 46.6(3.9) & 0.32(0.03) & 2.5(0.1) & 208(24) \\
HCN           & --9.0(1.9) & 11.4(1.9) & 1.1(0.2) & 44.3(5.2) & 0.30(0.04) \\
HCO$^+$       & --10.7(0.9) & 9.0(1.1) & 1.9(0.2) & 51.0(3.0) & 0.46(0.04) & 0.35(0.02) \\
\noalign{\smallskip}\hline\noalign{\smallskip}
{\bf 99.982+4.17} \\
c-C$_3$H$_2$  &  &  &  & & & 1.6(0.1) \\
CH$_3$C$_2$H  &  &  &  & & & 2.4(0.4) \\
NH$_2$D       &  &  &  & & & 1.4(0.2) \\
SO            & 11.1(5.2) & --2.6(9.3) & 2.5(0.9) & 108.4(20.7) & 0.39(0.08) & 1.5(0.1) & 93(22) \\
H$^{13}$CN    & --12.5(3.0) & 6.9(3.3) & 2.3(1.2) & 45.5(11.5) & 0.17(0.04) & 2.0(0.1) & 70(19) \\
H$^{13}$CO$^+$& --4.4(1.5) & 7.6(1.7) & 2.2(0.6) & 42.7(6.0) & 0.16(0.02) & 1.7(0.1) & 47(9) \\
SiO           & &  & &  & & 4.0(0.4) \\
HN$^{13}$C    & --5.7(2.7) & 11.6(4.1) & 3.0(2.6) & 36.7(16.1) & 0.13(0.06) & 1.7(0.1) & 41(18) \\
C$_2$H        & --6.4(2.5)  & 3.1(2.4) & 2.4(1.1) & 41.2(9.9) & 0.15(0.04) & 2.0(0.1) & 63(16) \\
HCN           & --8.9(1.8) & 11.3(1.6) & 1.3(0.2) & 62.7(5.6) & 0.23(0.02) \\
HCO$^+$       & --4.7(1.5) & 12.0(1.5) & 1.2(0.2) & 46.8(4.4) &  0.17(0.02) \\

\noalign{\smallskip}\hline\noalign{\smallskip}
\end{tabular}
\label{phys}
\end{table}

\newpage

\begin{figure}[t!]
\setcaptionmargin{5mm}
\onelinecaptionsfalse

\centering
\resizebox{\hsize}{!}{
\includegraphics[width=1.3\textwidth]{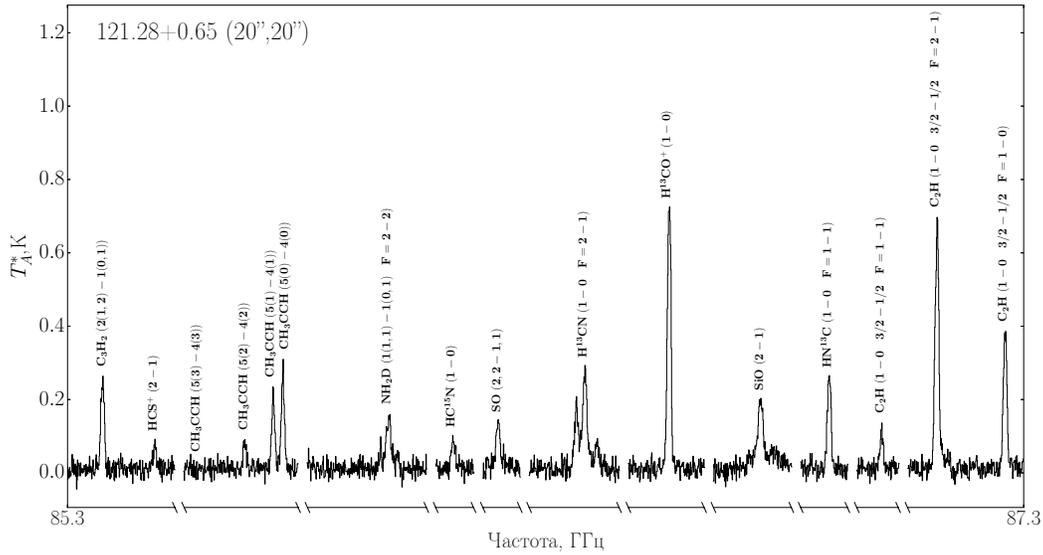}}
\captionstyle{flushleft}
\caption{
Spectrum toward the peak of the molecular emission in 121.28+0.65, observed using a Fourier spectrum analyser with a 2.5 GHz bandwidth, a 76 kHz resolution, and a central frequency of 86.6 GHz. A part of the spectrum analyser band, divided into segments with molecular lines, is shown in the figure.
}
\label{121}
\end{figure}

\newpage

\begin{figure}[t!]
\setcaptionmargin{5mm}
\onelinecaptionsfalse
\captionstyle{flushleft}

\centering
\begin{minipage}[b]{0.31\textwidth}
    \includegraphics[width=\textwidth]{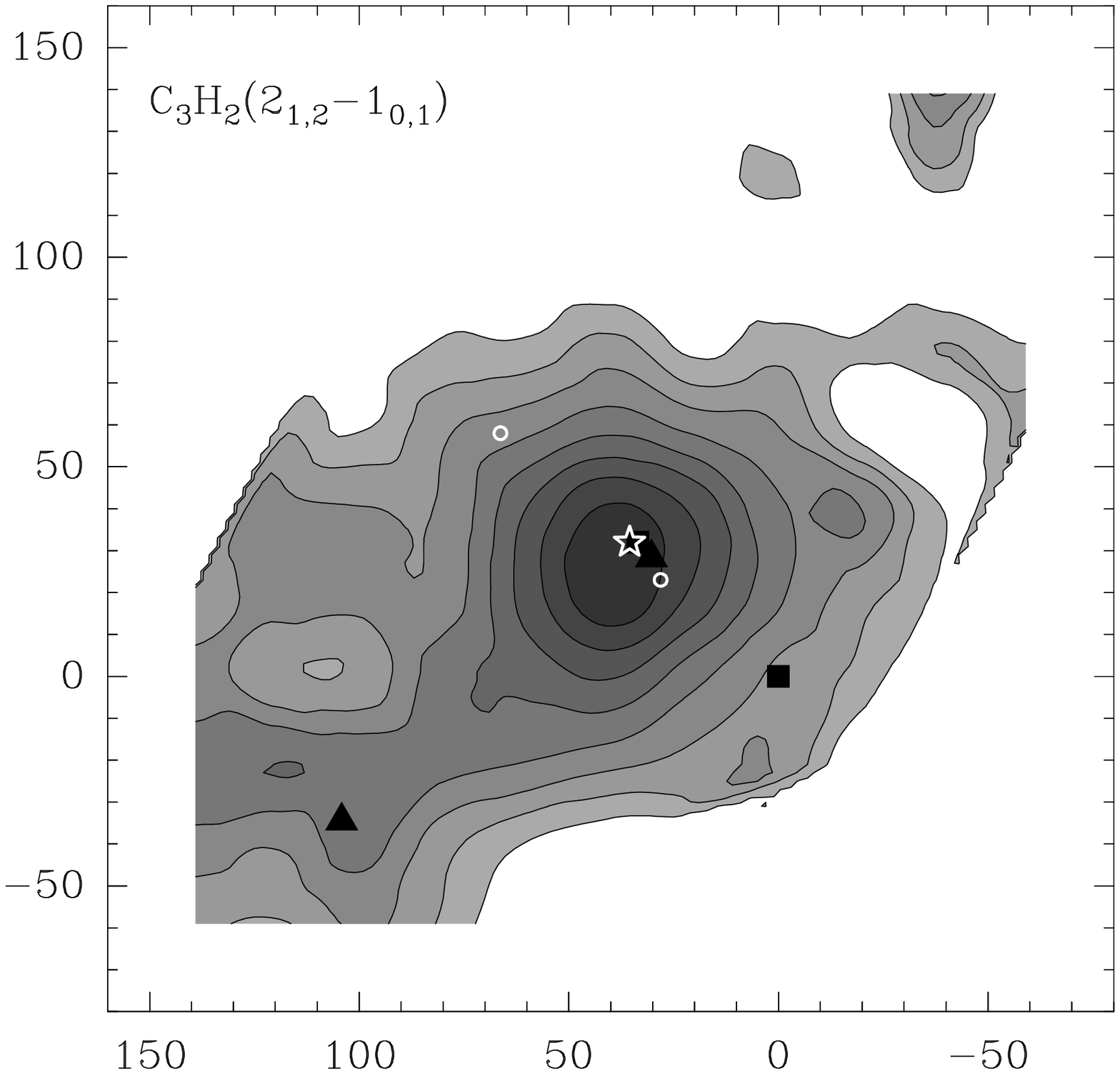}
\end{minipage}
\begin{minipage}[b]{0.31\textwidth}
    \includegraphics[width=\textwidth]{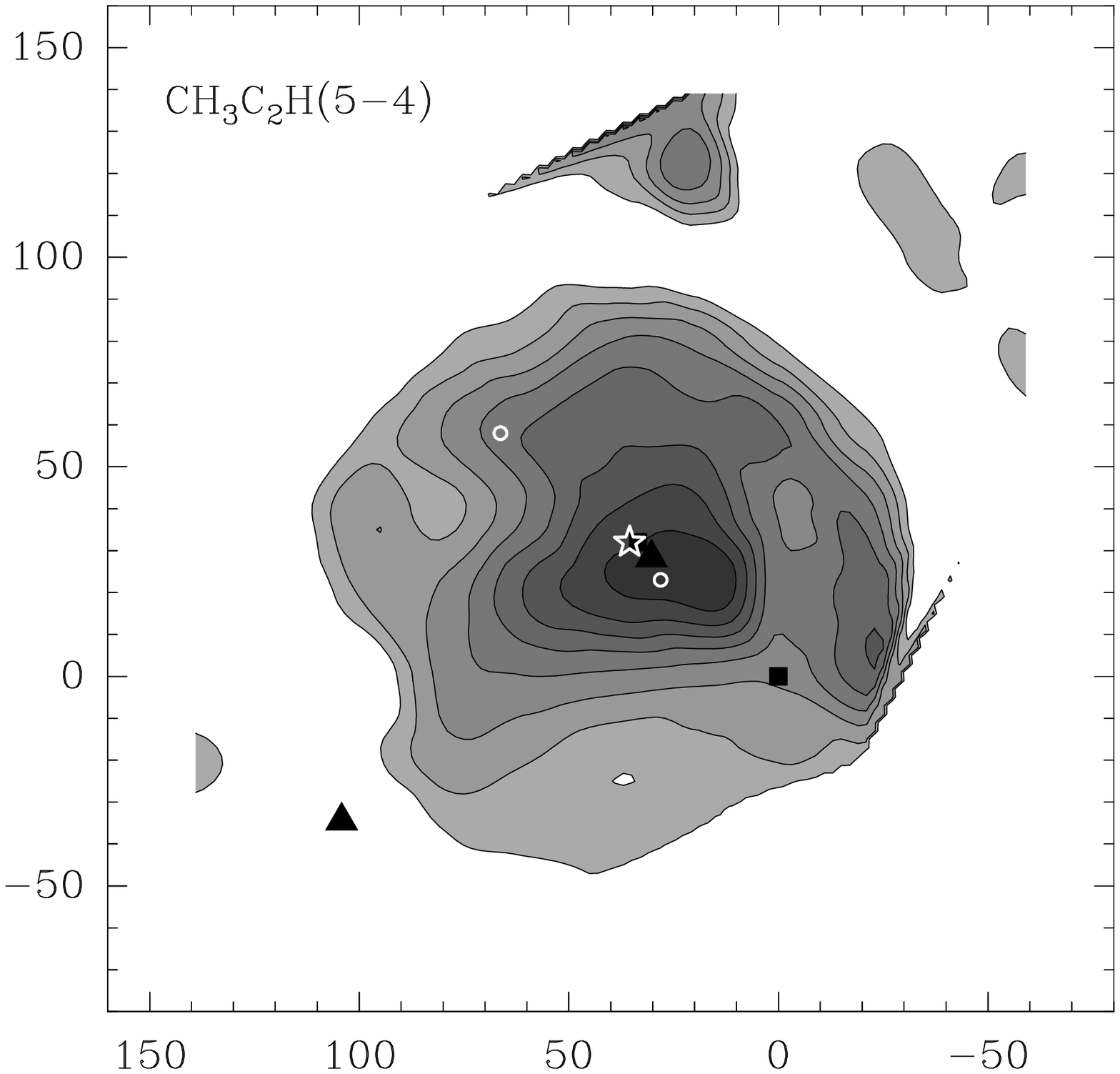}
\end{minipage}
\begin{minipage}[b]{0.31\textwidth}
    \includegraphics[width=\textwidth]{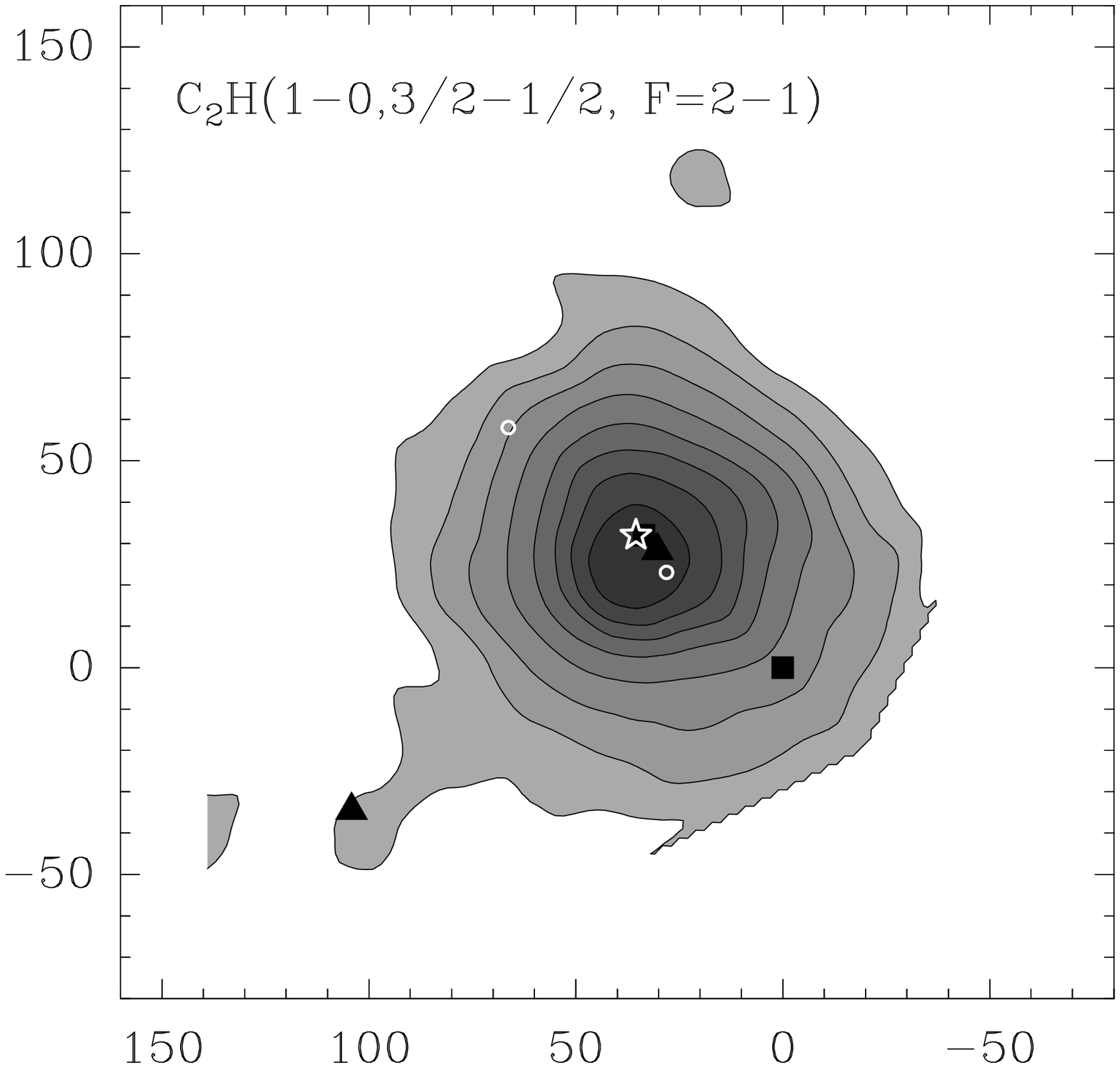}
\end{minipage}

\vskip 1mm

\begin{minipage}[b]{0.31\textwidth}
    \includegraphics[width=\textwidth]{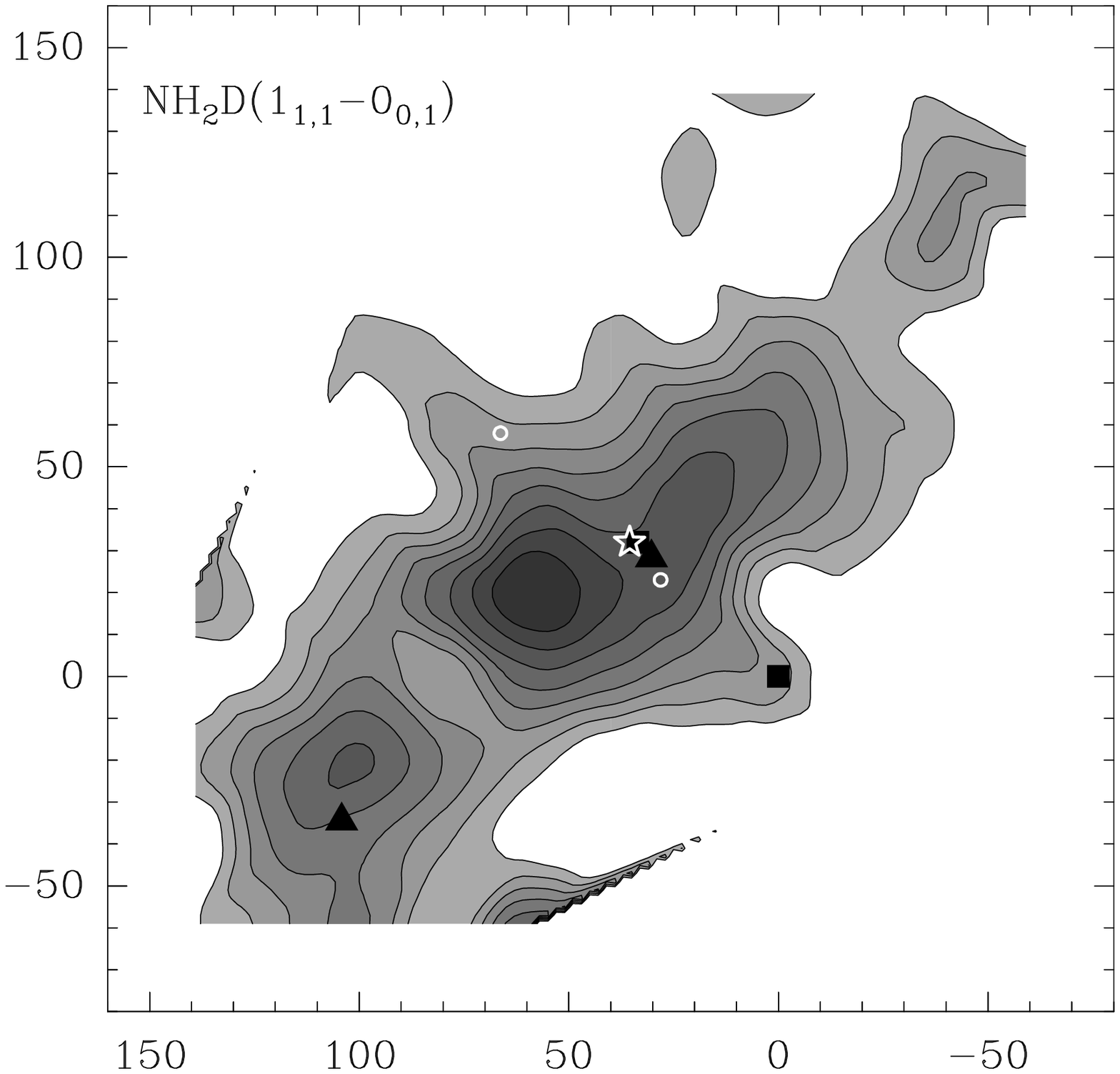}
\end{minipage}
\begin{minipage}[b]{0.31\textwidth}
    \includegraphics[width=\textwidth]{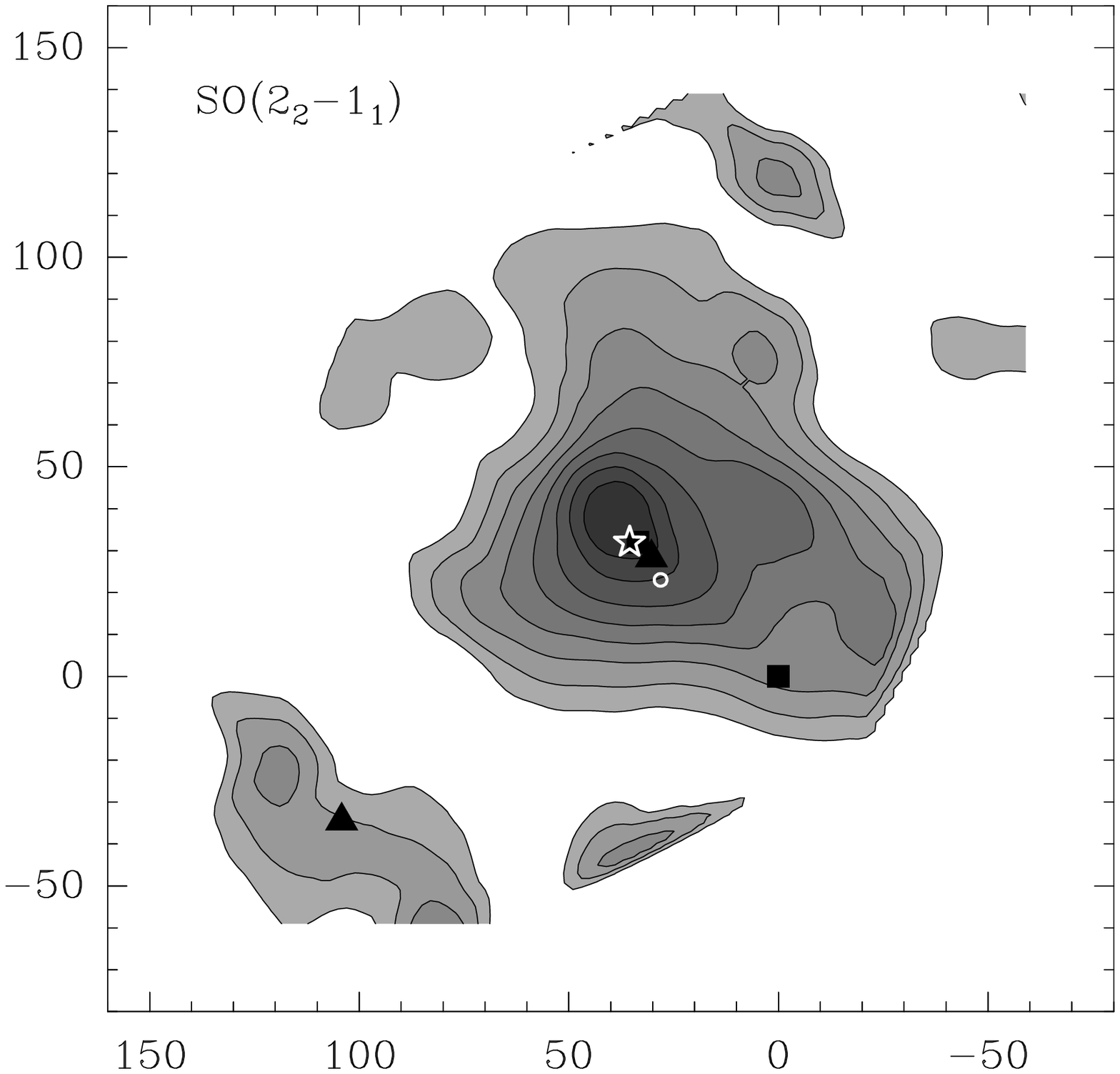}
\end{minipage}
\begin{minipage}[b]{0.31\textwidth}
    \includegraphics[width=\textwidth]{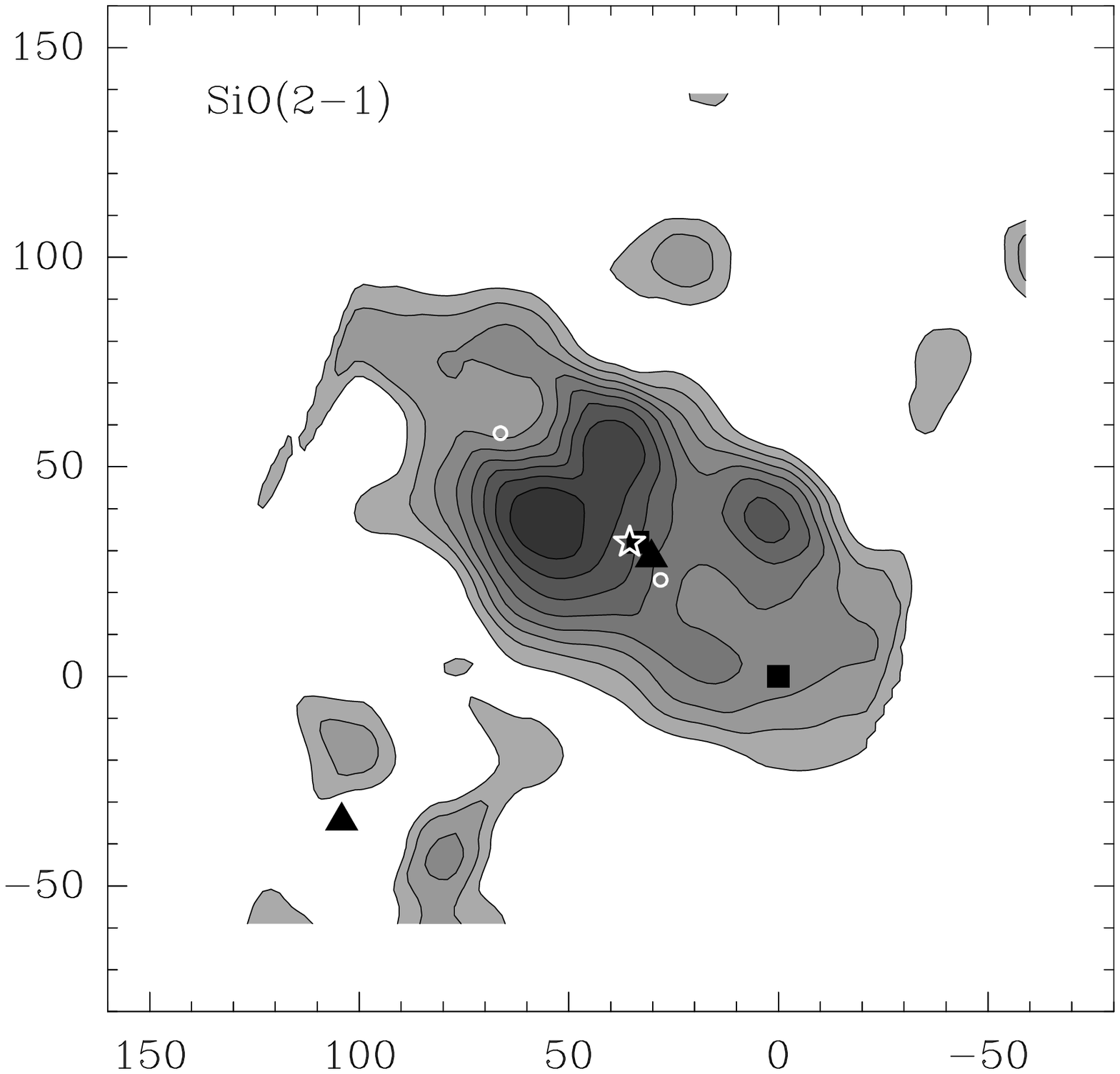}
\end{minipage}

\vskip 1mm

\begin{minipage}[b]{0.31\textwidth}
    \includegraphics[width=\textwidth]{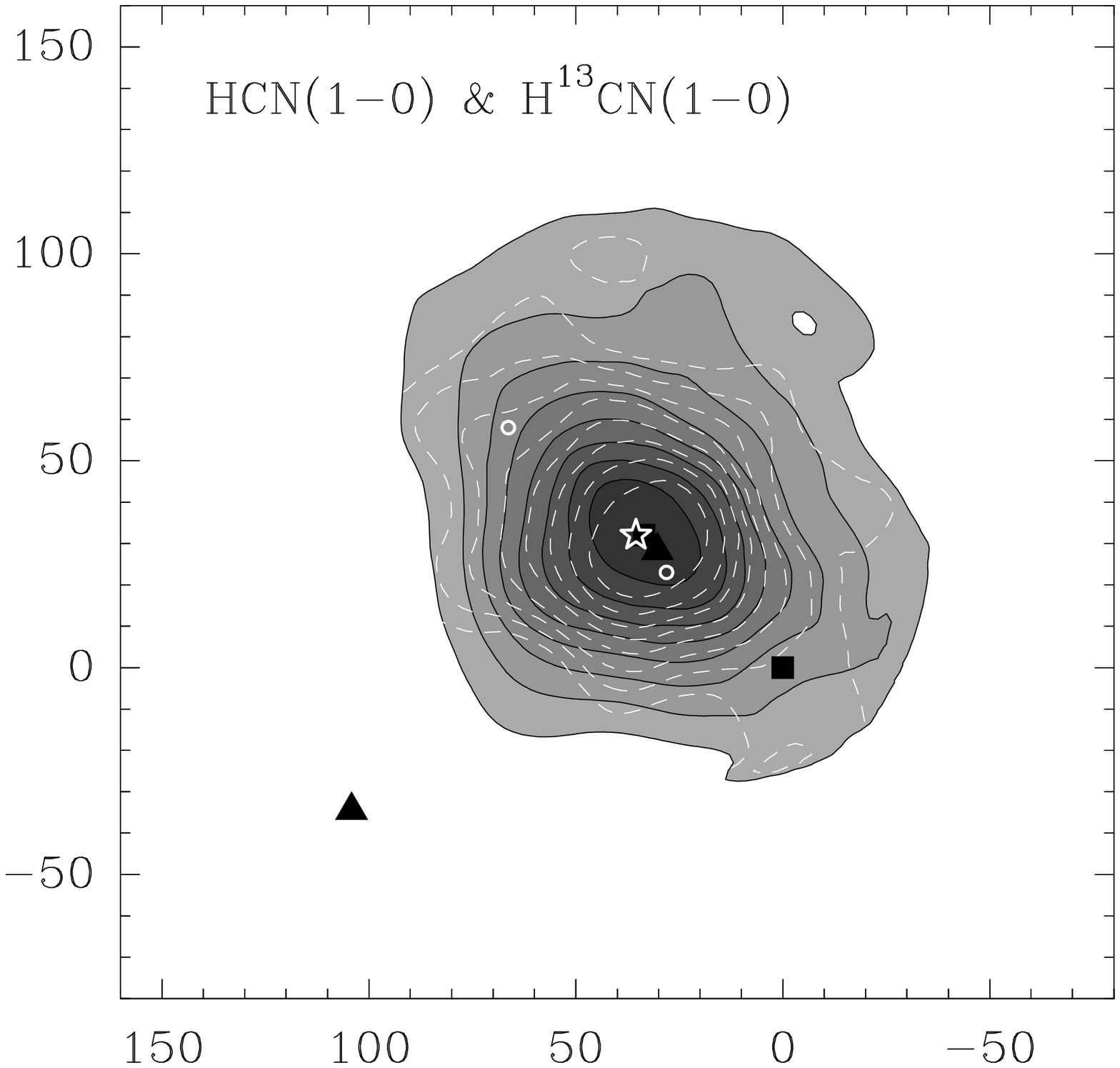}
\end{minipage}
\begin{minipage}[b]{0.31\textwidth}
    \includegraphics[width=\textwidth]{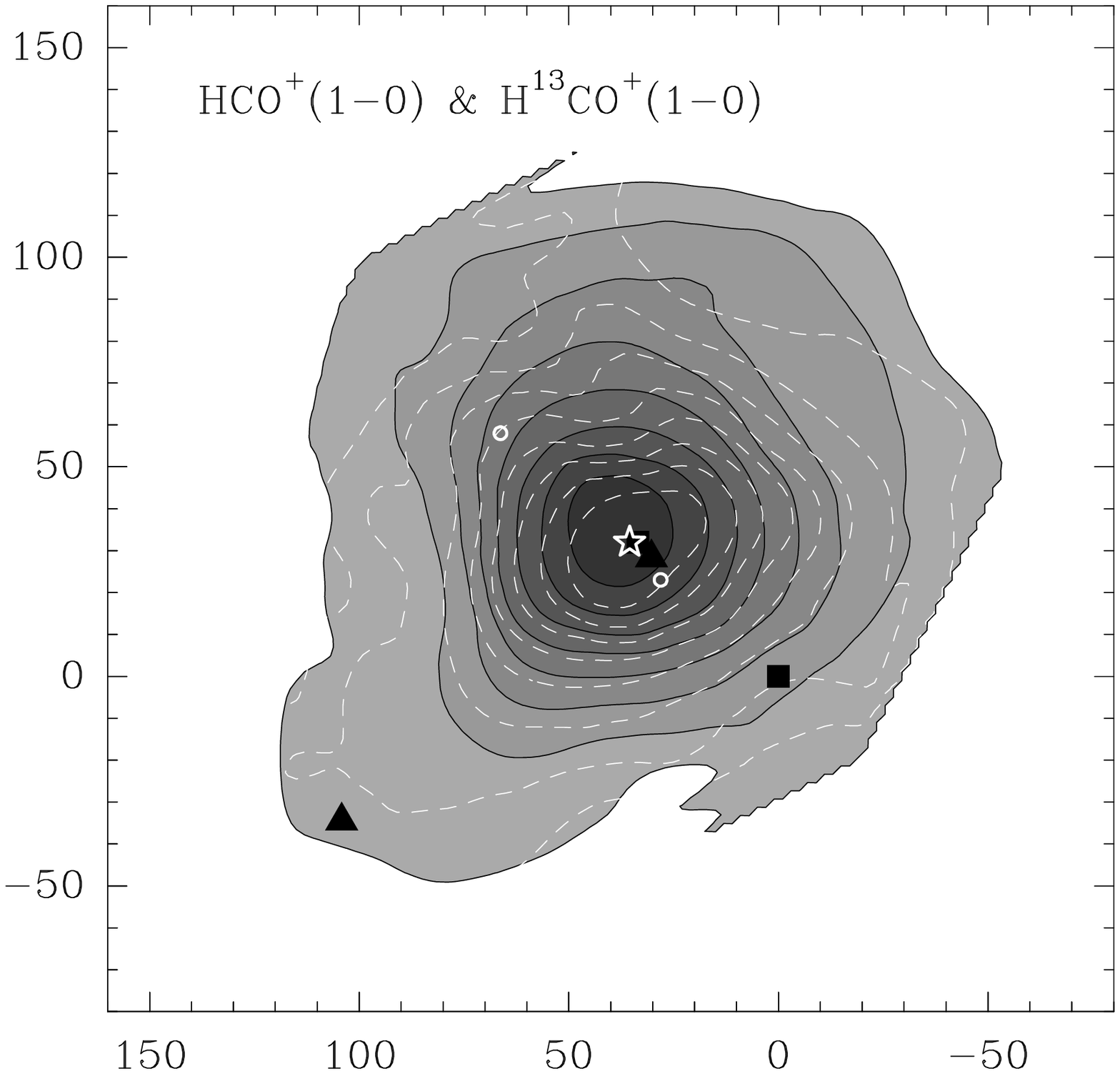}
\end{minipage}
\begin{minipage}[b]{0.31\textwidth}
    \includegraphics[width=\textwidth]{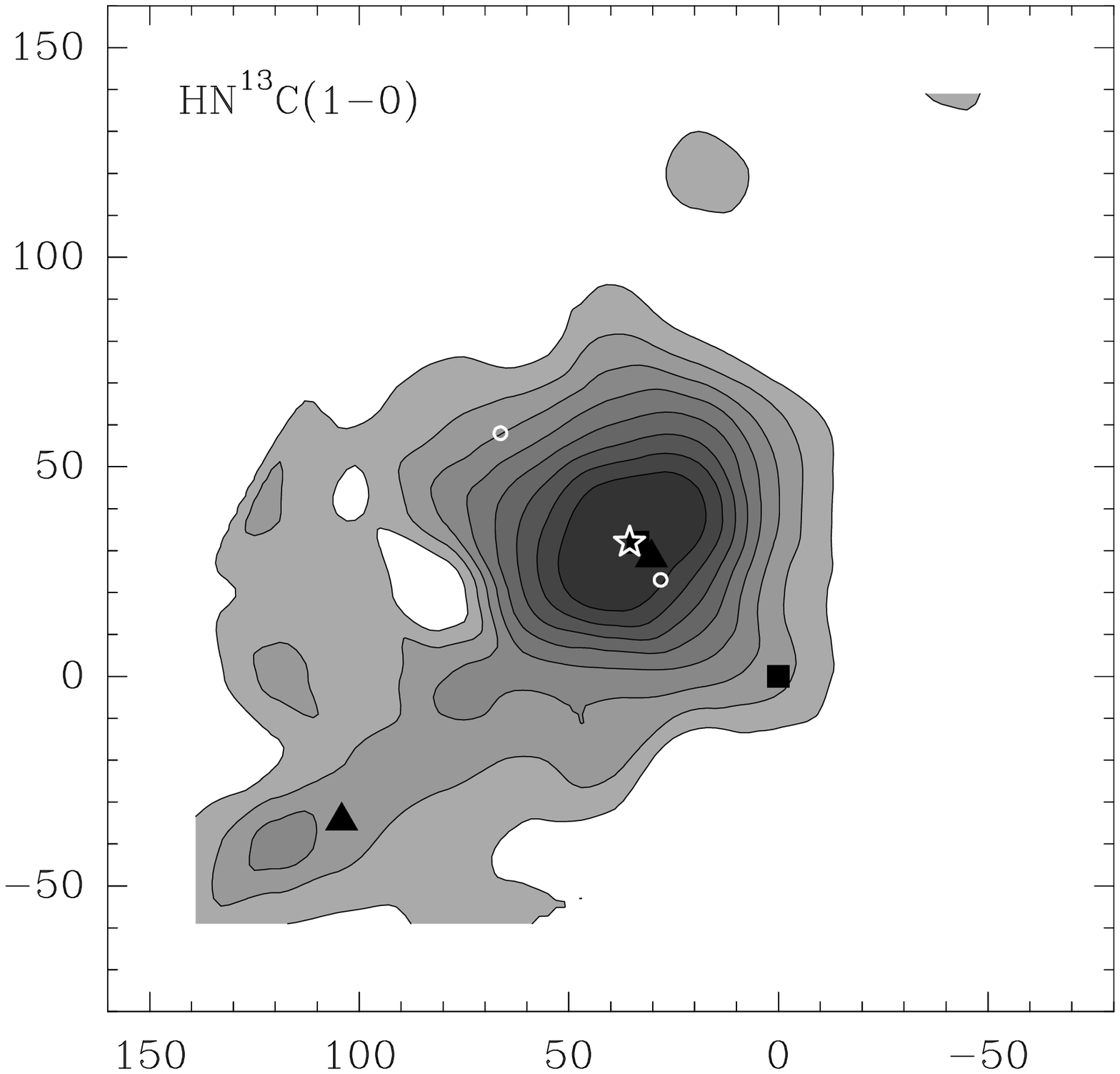}
\end{minipage}

\caption{\small
Maps of molecular lines observed in 121.28+0.65. The axes plot the offsets $\Delta\alpha('')$ and $\Delta\delta('')$ relative to the coordinates from Table 1. The contours of the integrated intensity correspond to 20--90\% of the peak intensities (Table 3). The dashed contours in the HCN and HCO$^+$ (1--0) images show the H$^{13}$CN and H$^{13}$CO$^+$ (1--0) maps, respectively. An open white star denotes the IRAS source, filled squares denote methanol and water-vapor masers \cite{Gan,Pestalozzi,Valdettaro}, filled triangles denote submillimeter sources \cite{Sandell01},
and open white circles denote MSX IR sources.
}
\label{fig:121}
\end{figure}

\newpage

\begin{figure}[t!]
\setcaptionmargin{5mm}
\onelinecaptionsfalse
\captionstyle{flushleft}

\begin{minipage}[b]{0.31\textwidth}
    \includegraphics[width=\textwidth]{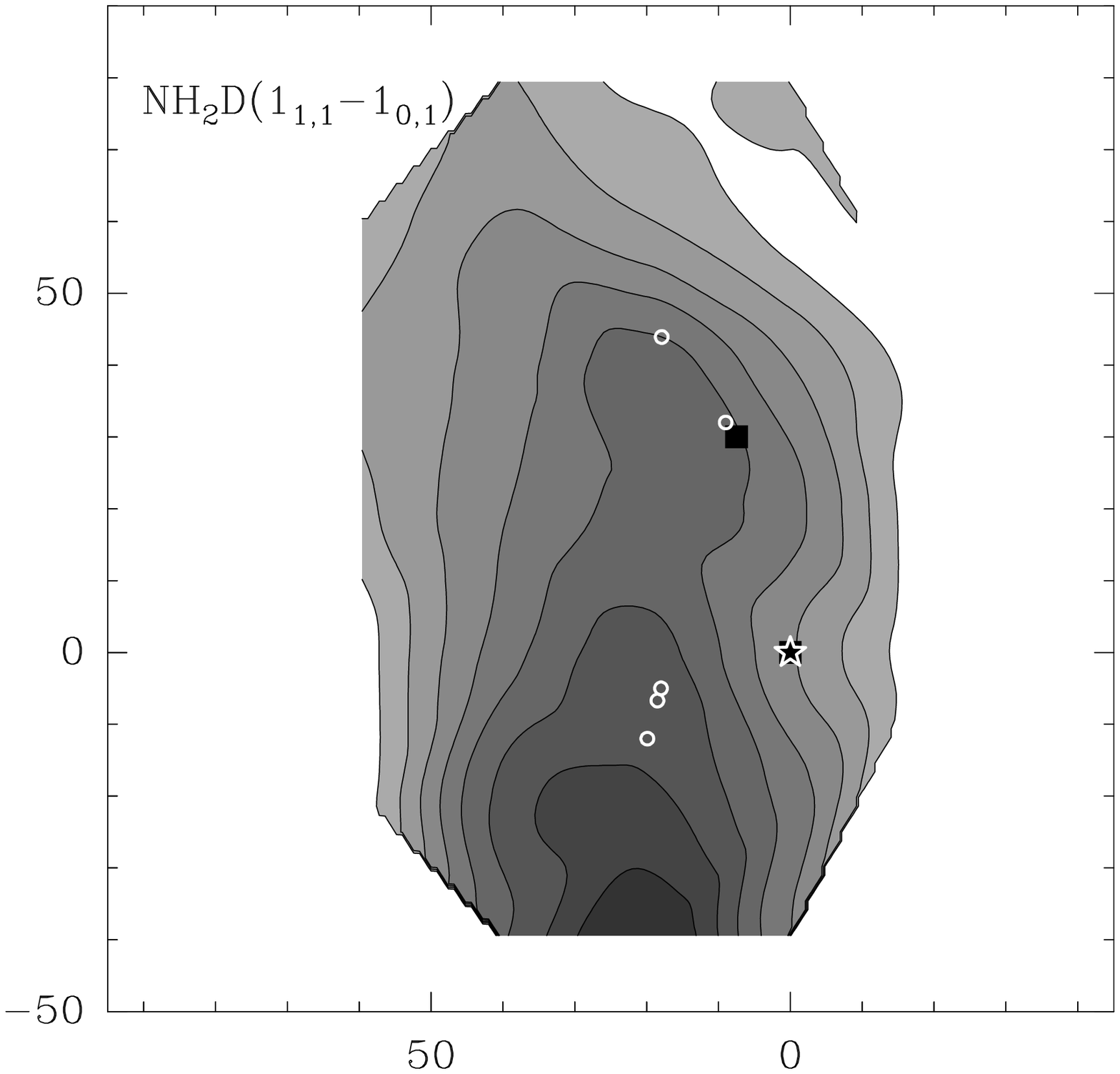}
\end{minipage}
\begin{minipage}[b]{0.31\textwidth}
    \includegraphics[width=\textwidth]{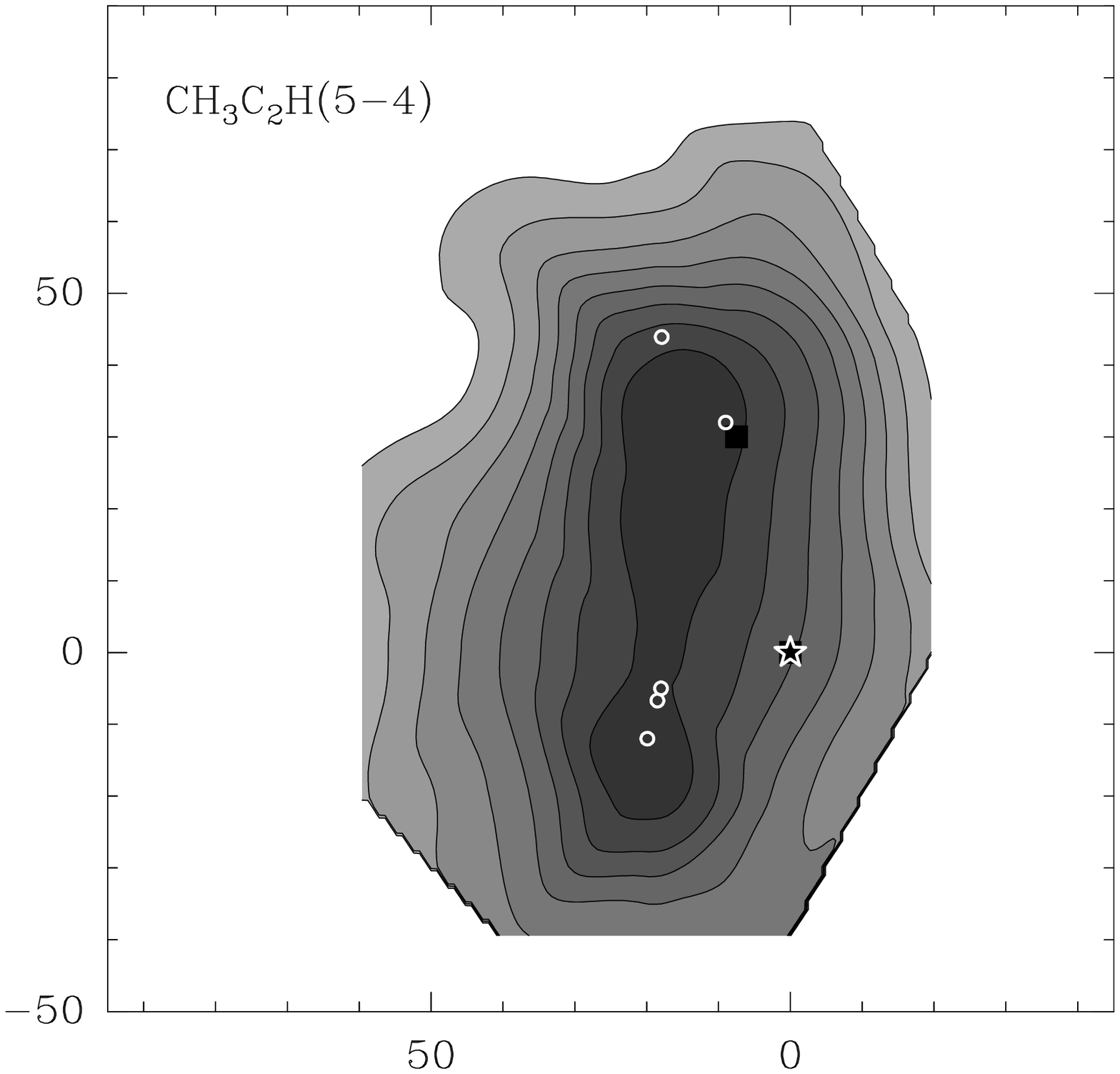}
\end{minipage}
\begin{minipage}[b]{0.31\textwidth}
    \includegraphics[width=\textwidth]{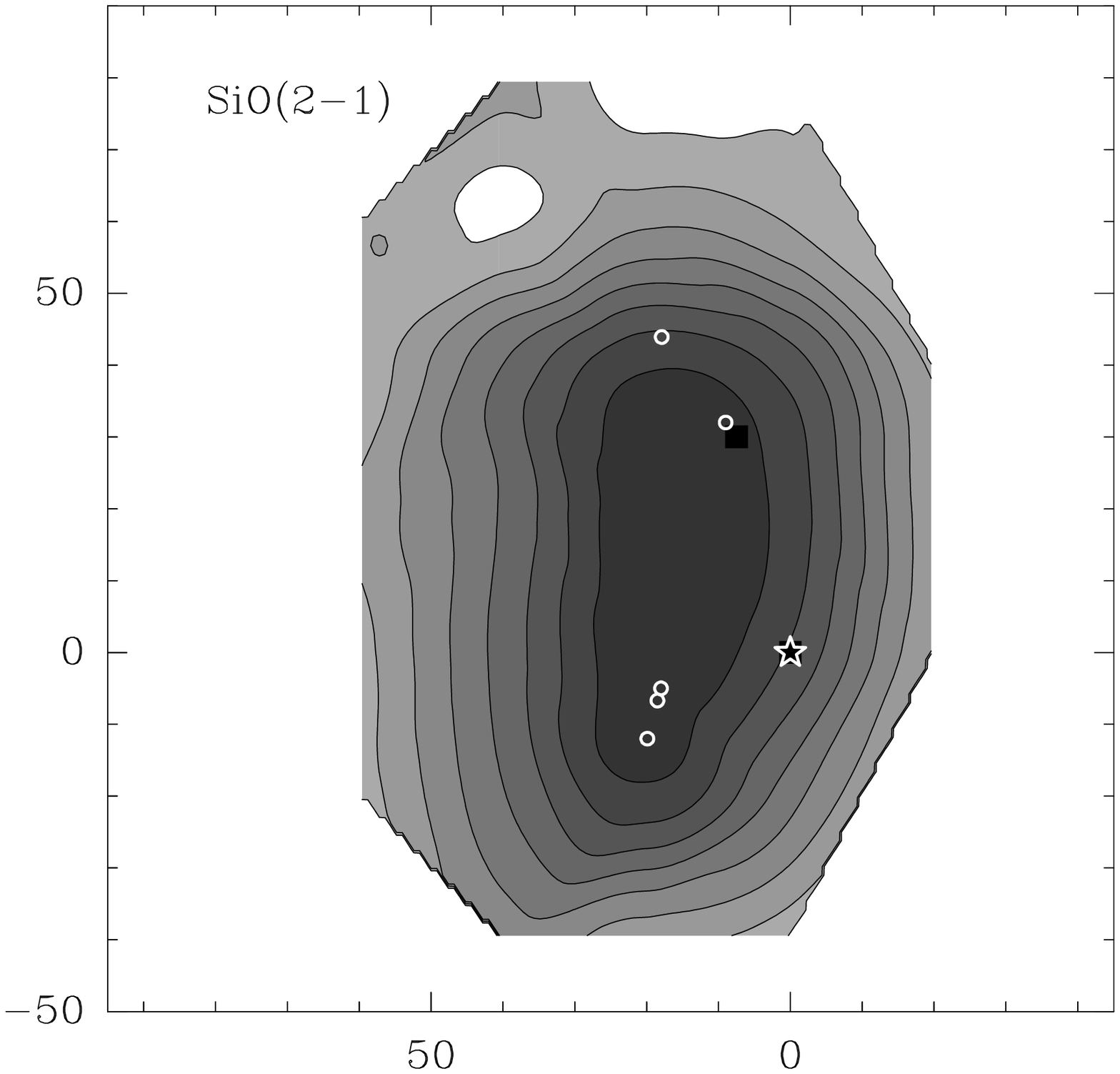}
\end{minipage}

\vskip 1mm

\begin{minipage}[b]{0.31\textwidth}
    \includegraphics[width=\textwidth]{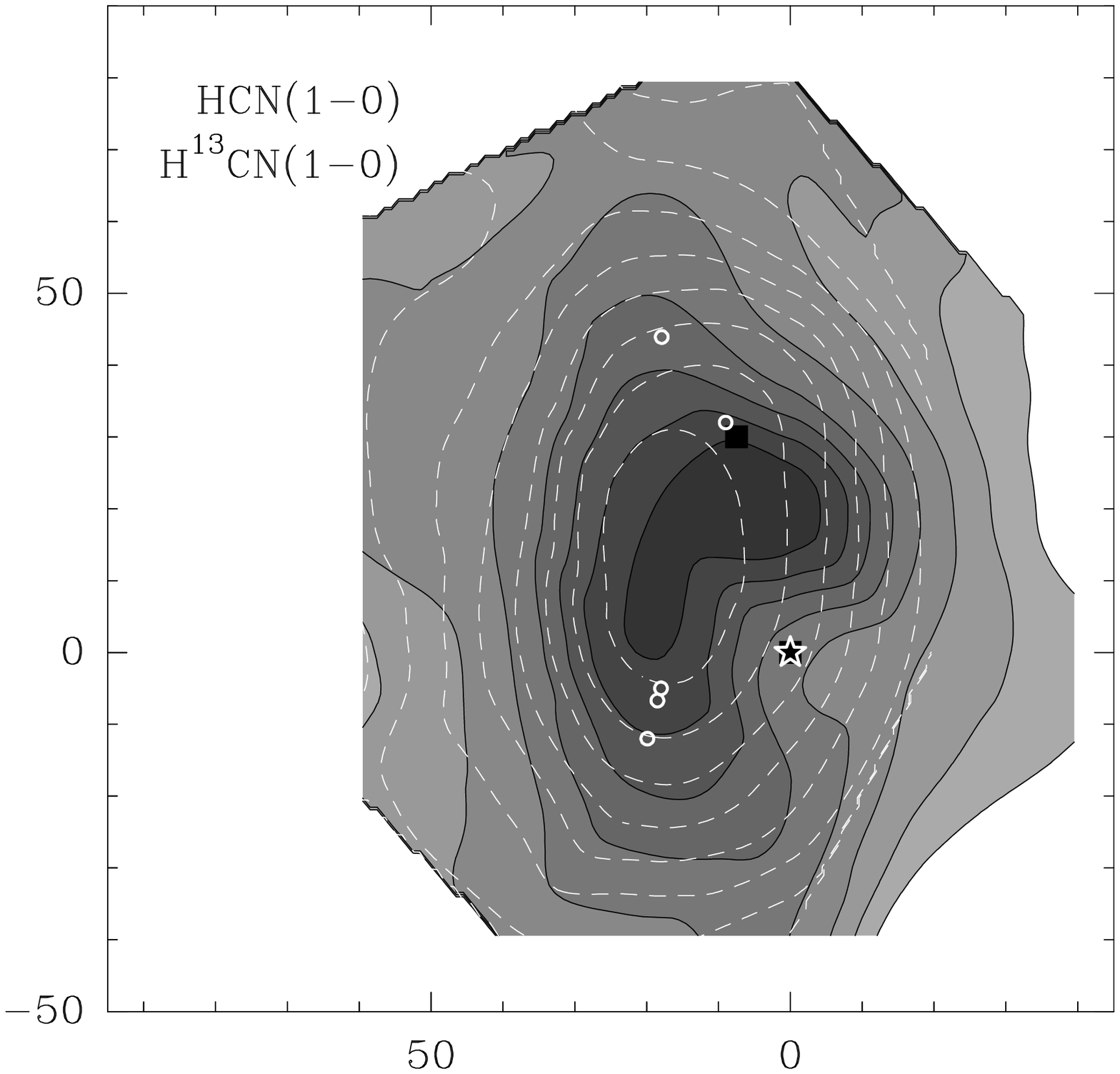}
\end{minipage}
\begin{minipage}[b]{0.31\textwidth}
    \includegraphics[width=\textwidth]{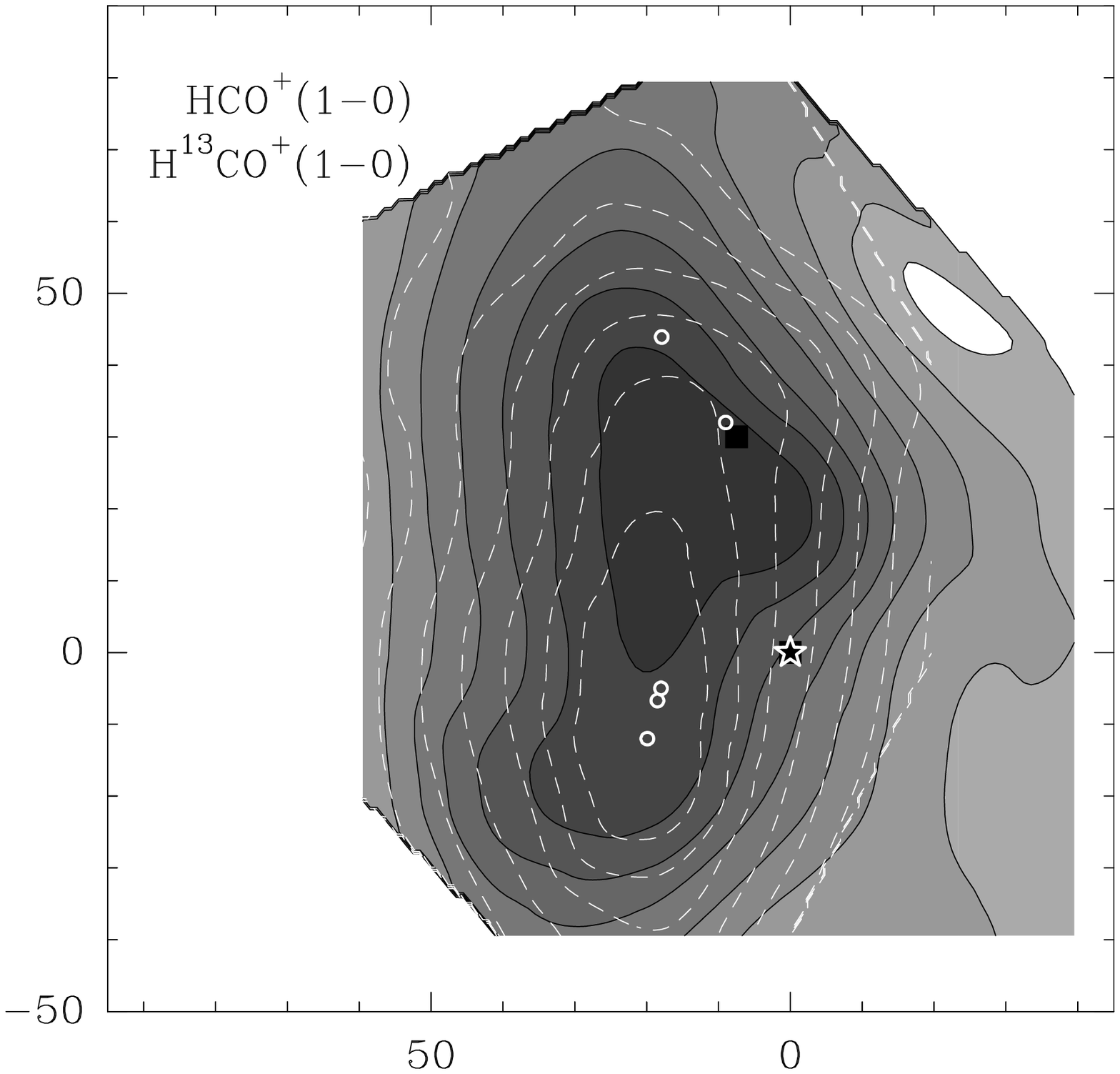}
\end{minipage}
\begin{minipage}[b]{0.31\textwidth}
    \includegraphics[width=\textwidth]{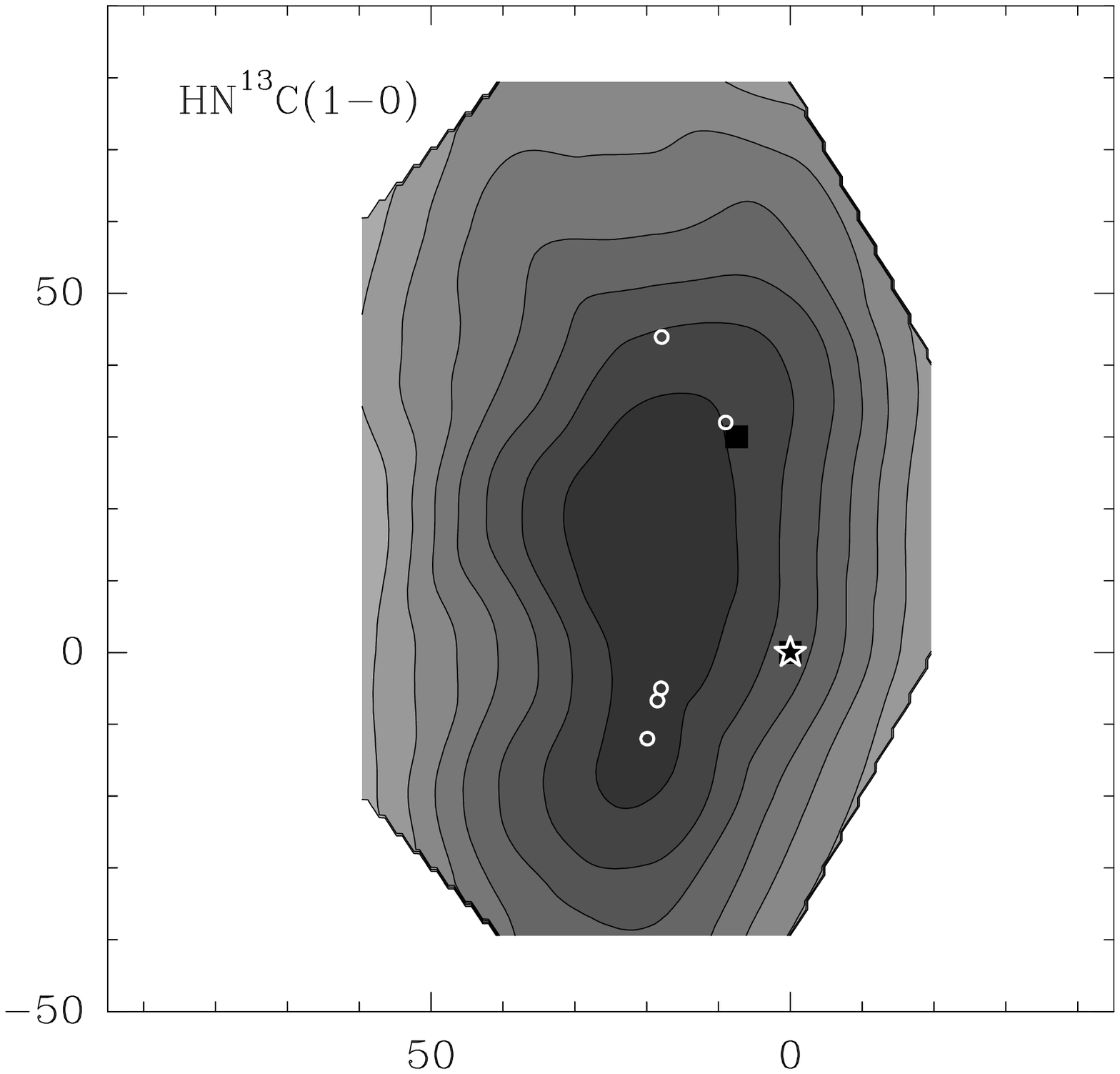}
\end{minipage}

\vskip 1mm

\begin{minipage}[b]{0.31\textwidth}
    \includegraphics[width=\textwidth]{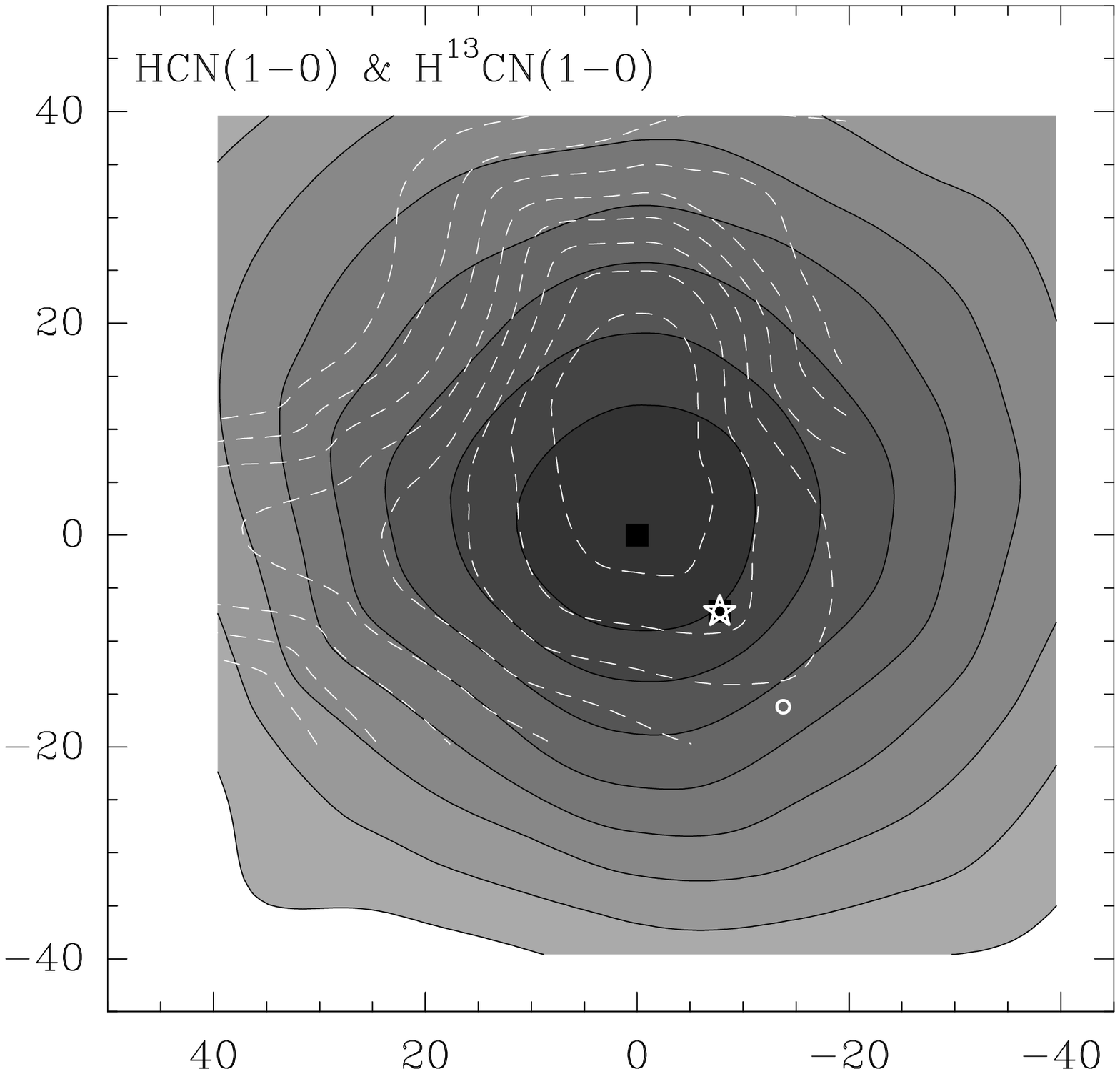}
\end{minipage}
\begin{minipage}[b]{0.31\textwidth}
    \includegraphics[width=\textwidth]{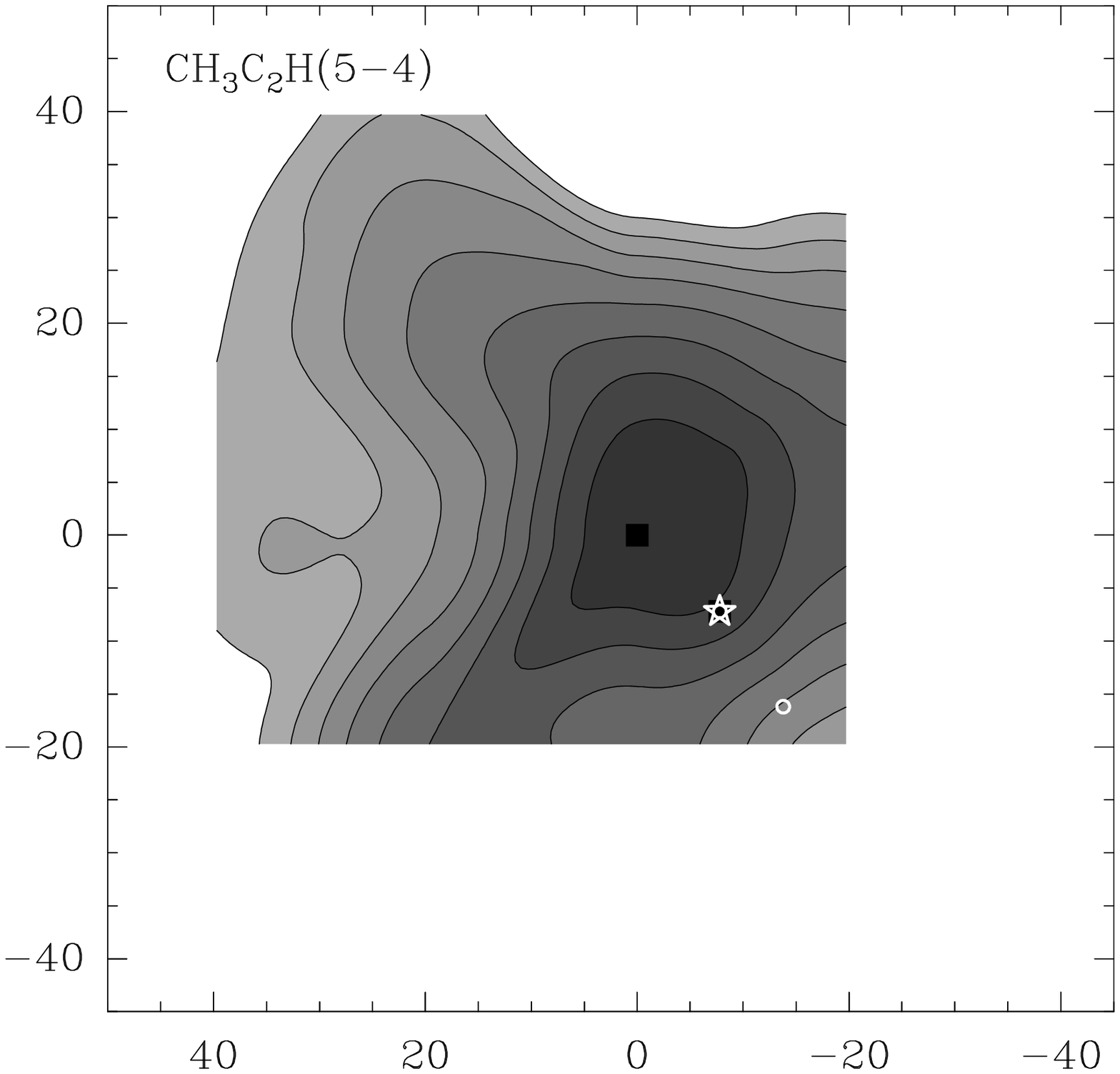}
\end{minipage}
\begin{minipage}[b]{0.31\textwidth}
    \includegraphics[width=\textwidth]{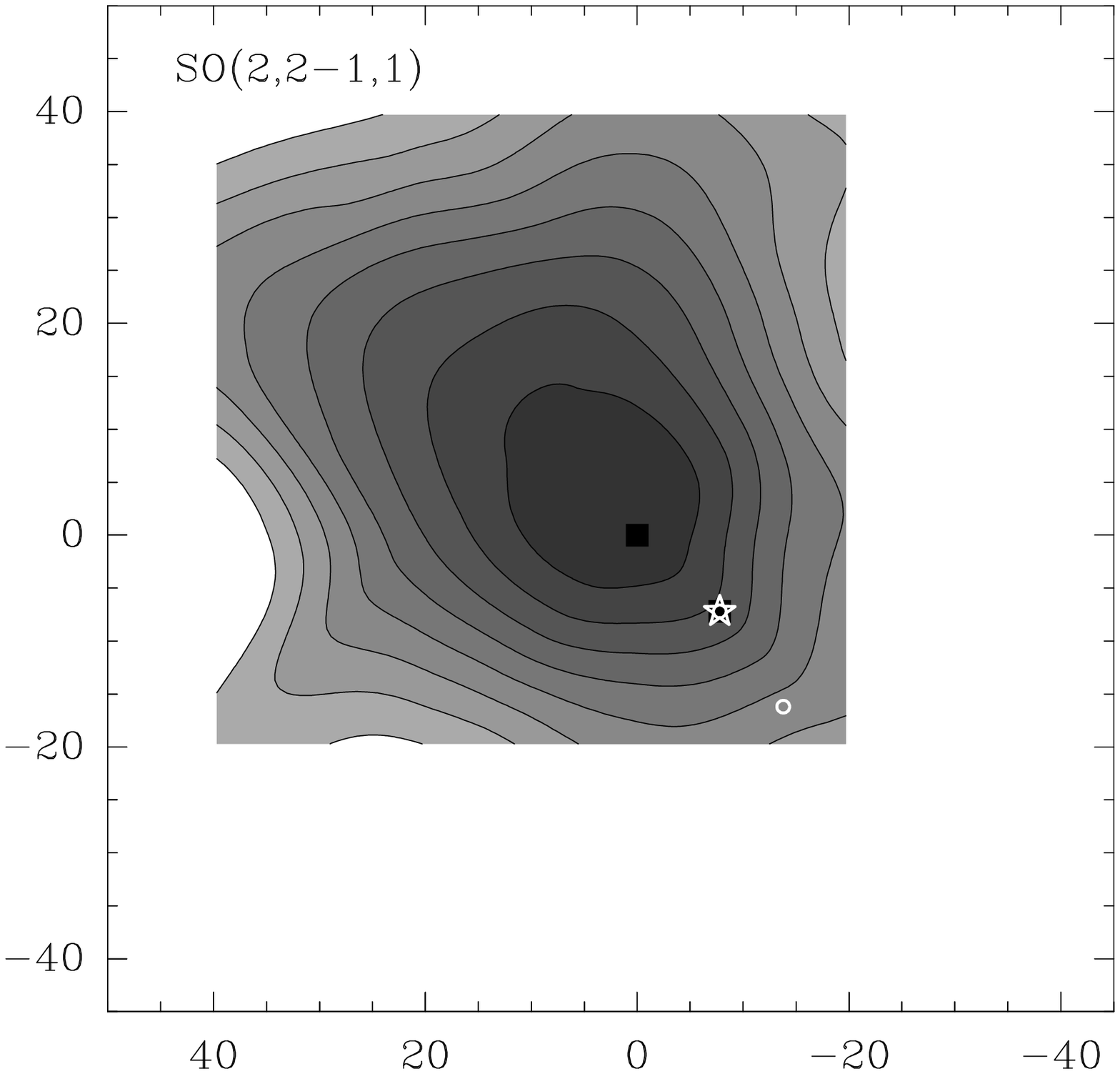}
\end{minipage}

\caption{\small
Maps of molecular lines observed in 34.403+0.233 (upper and central panels) and in 37.427+1.518 (lower panel). The notation is the same as in Fig. 2. The open white circles in the 34.403+0.233 maps denote IR MSX and Spitzer sources \cite{Shepherd}. The methanol-maser positions were taken from \cite{Xu03,Szymczak,Chen11,Blaszkiewicz,Fontani,Wu14}. The 37.427+1.518 center corresponds to the position of water vapor and methanol masers \cite{Wu14}. The IRAS position in 37.427+1.518 coincides with the positions of methanol \cite{Blaszkiewicz} and H$_2$O \cite{Valdettaro} masers.
}
\label{34-37}
\end{figure}

\newpage

\begin{figure}[t!]
\setcaptionmargin{5mm}
\onelinecaptionsfalse
\captionstyle{flushleft}
\captionstyle{flushleft}

\begin{minipage}[b]{0.3\textwidth}
    \includegraphics[width=\textwidth]{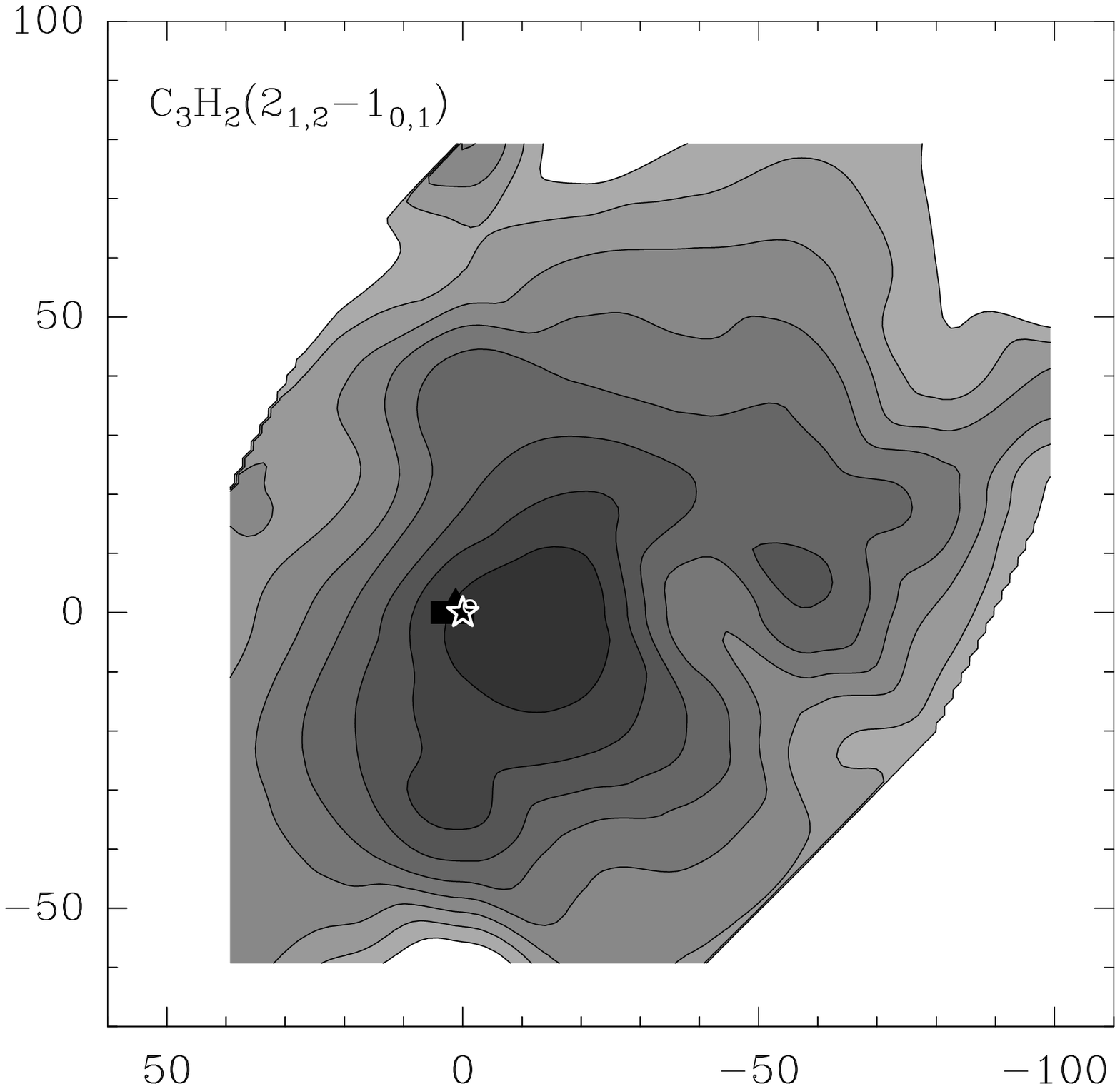}
\end{minipage}
\begin{minipage}[b]{0.3\textwidth}
    \includegraphics[width=\textwidth]{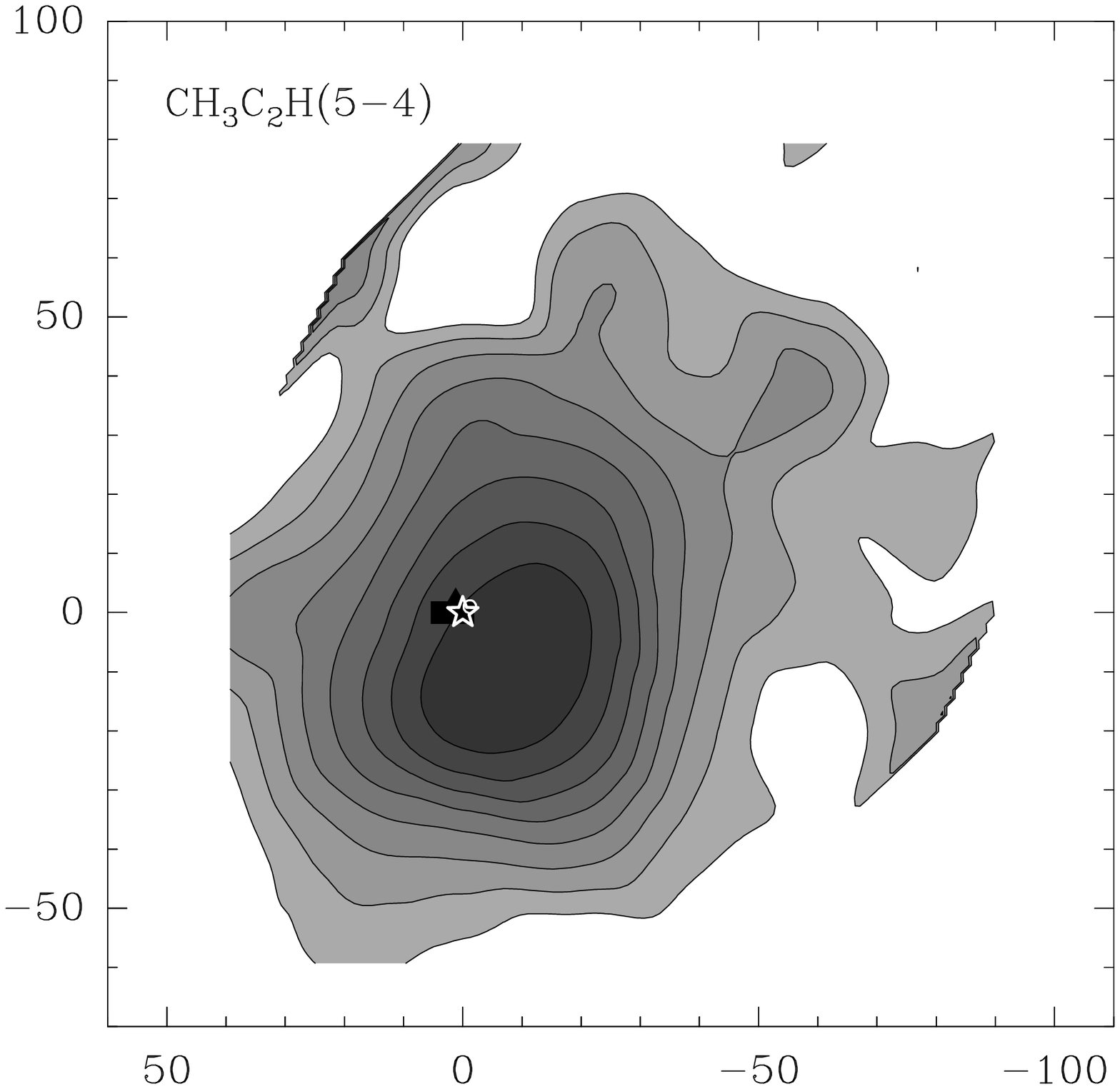}
\end{minipage}
\begin{minipage}[b]{0.3\textwidth}
    \includegraphics[width=\textwidth]{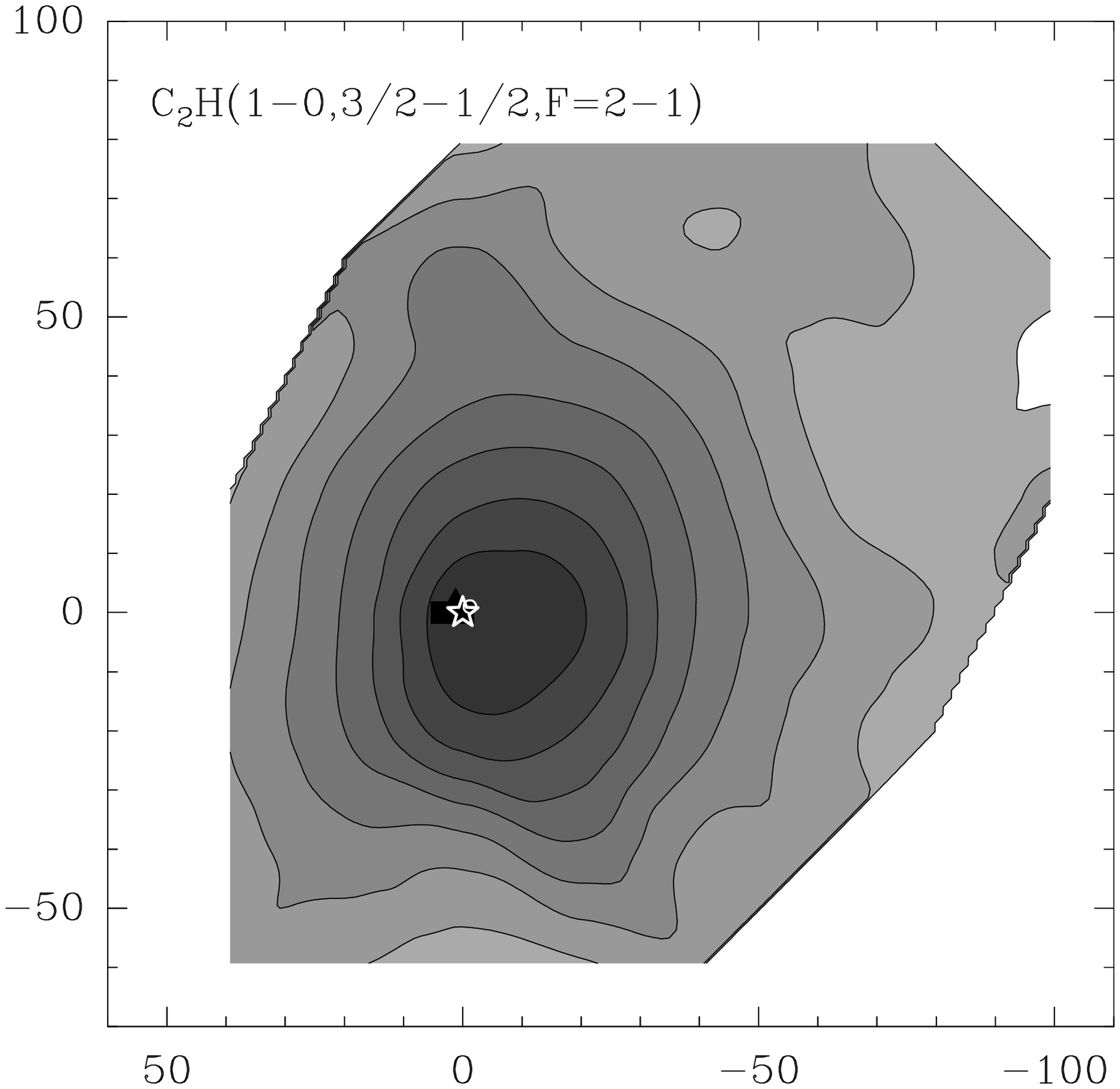}
\end{minipage}

\vskip 1mm

\begin{minipage}[b]{0.3\textwidth}
    \includegraphics[width=\textwidth]{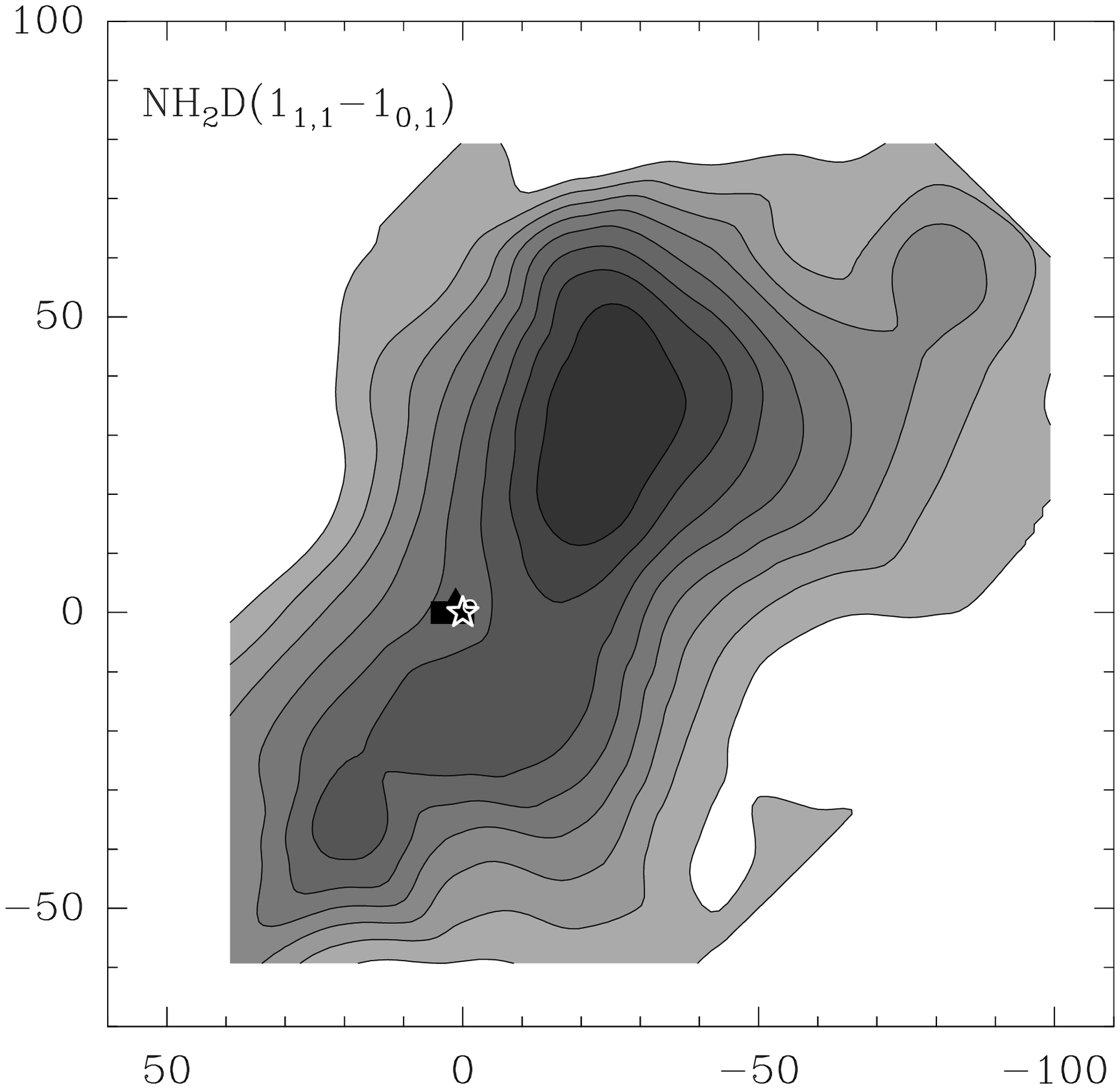}
\end{minipage}
\begin{minipage}[b]{0.3\textwidth}
    \includegraphics[width=\textwidth]{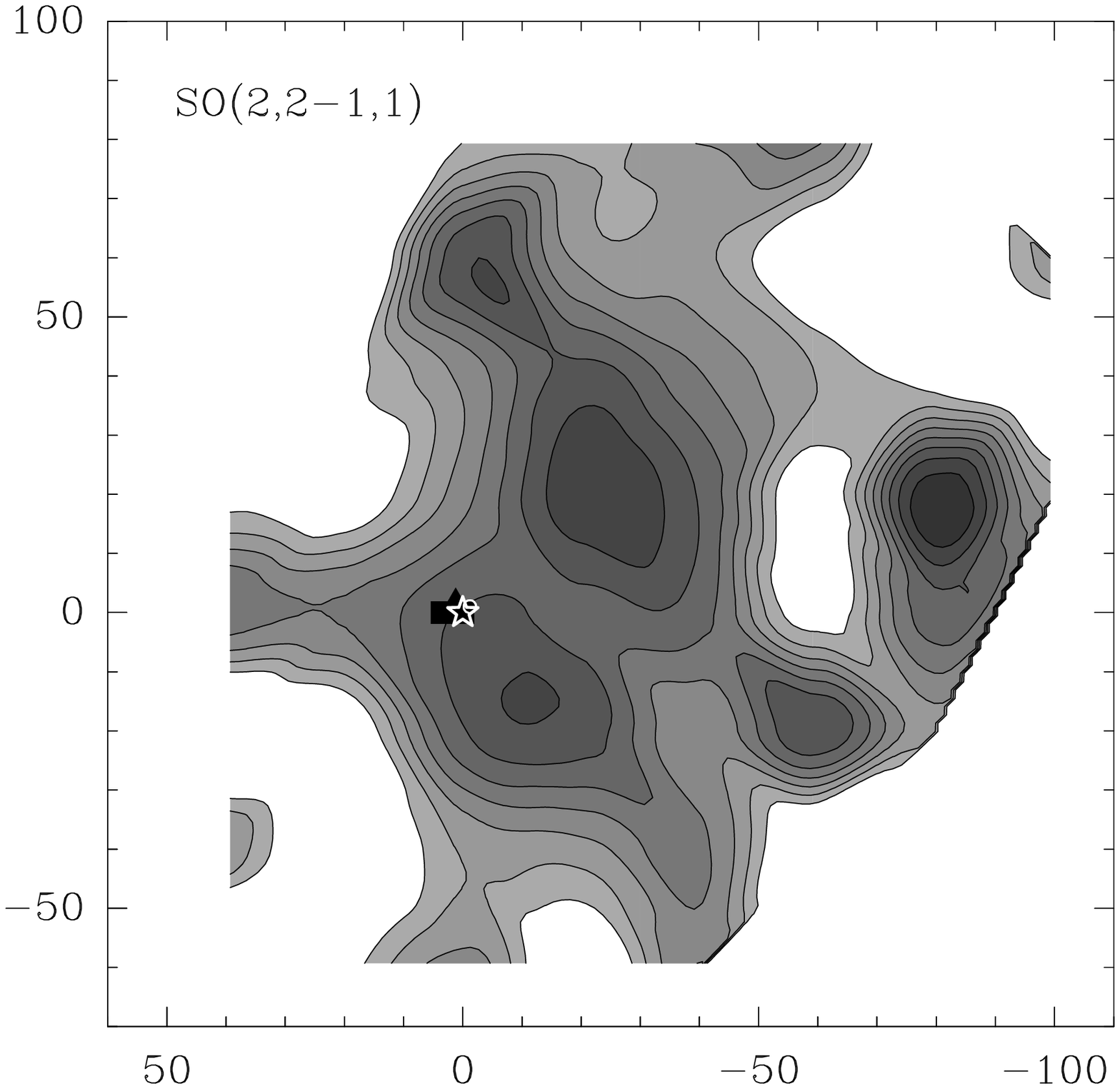}
\end{minipage}
\begin{minipage}[b]{0.3\textwidth}
    \includegraphics[width=\textwidth]{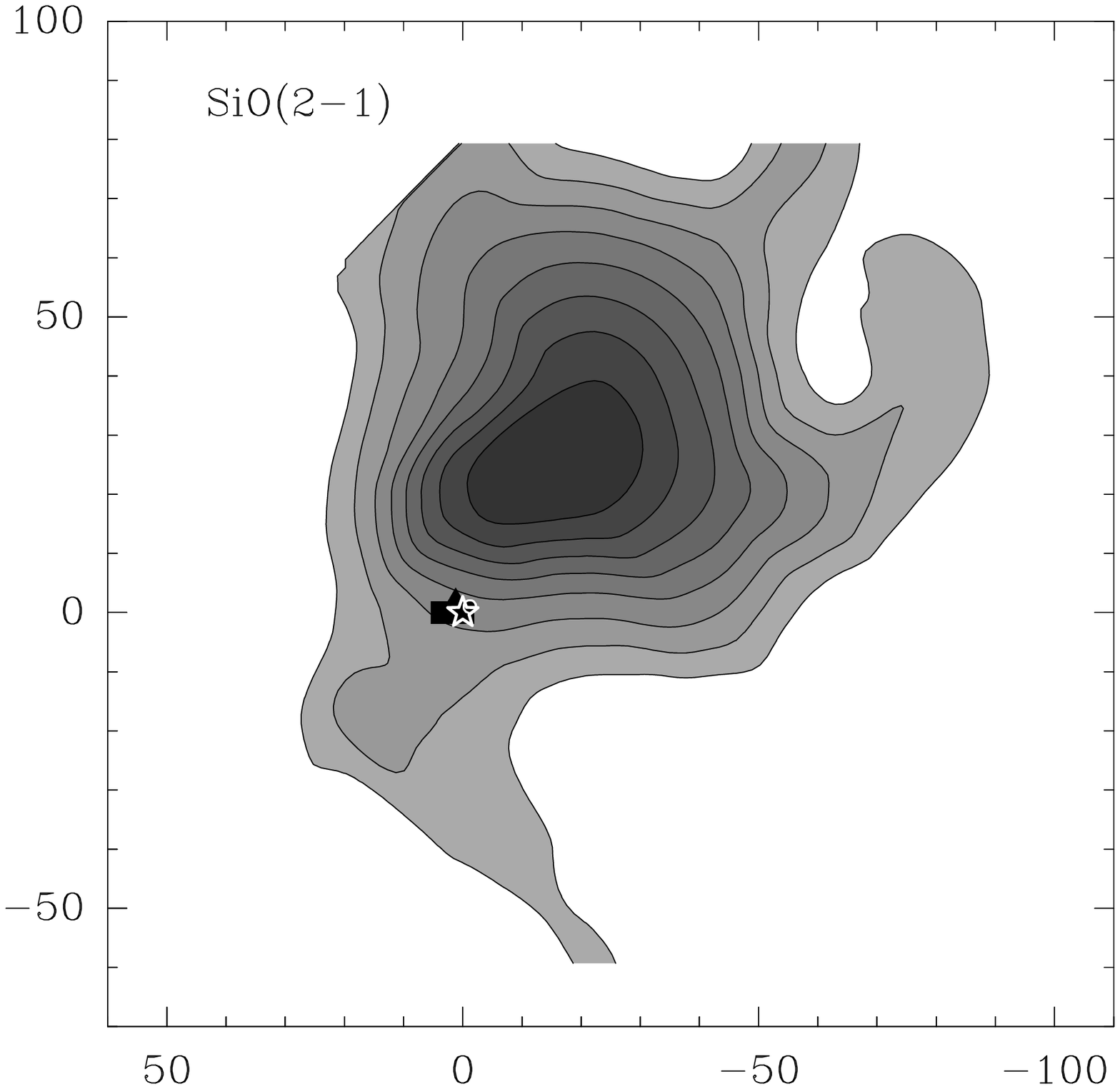}
\end{minipage}

\vskip 1mm

\begin{minipage}[b]{0.3\textwidth}
    \includegraphics[width=\textwidth]{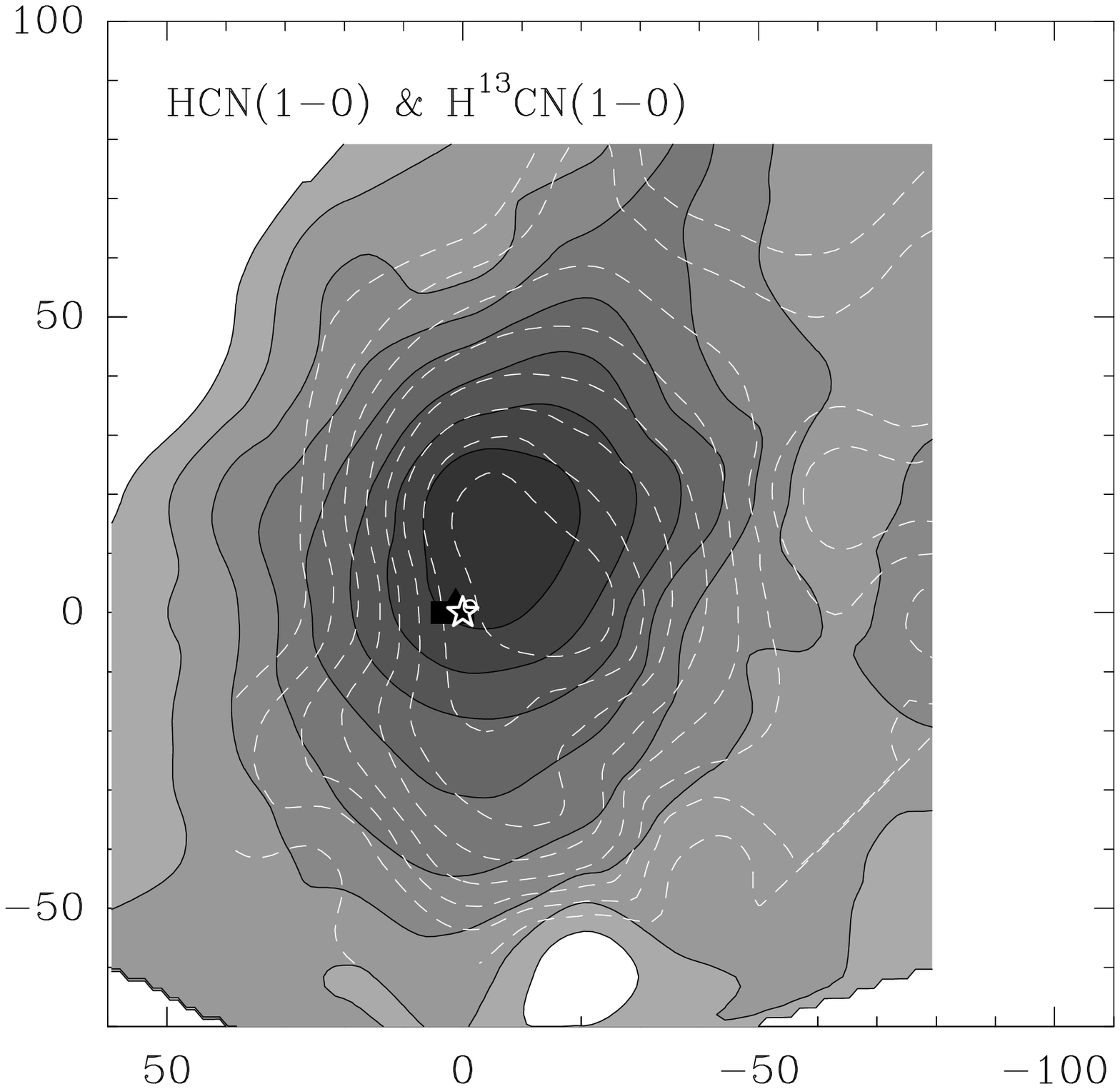}
\end{minipage}
\begin{minipage}[b]{0.3\textwidth}
    \includegraphics[width=\textwidth]{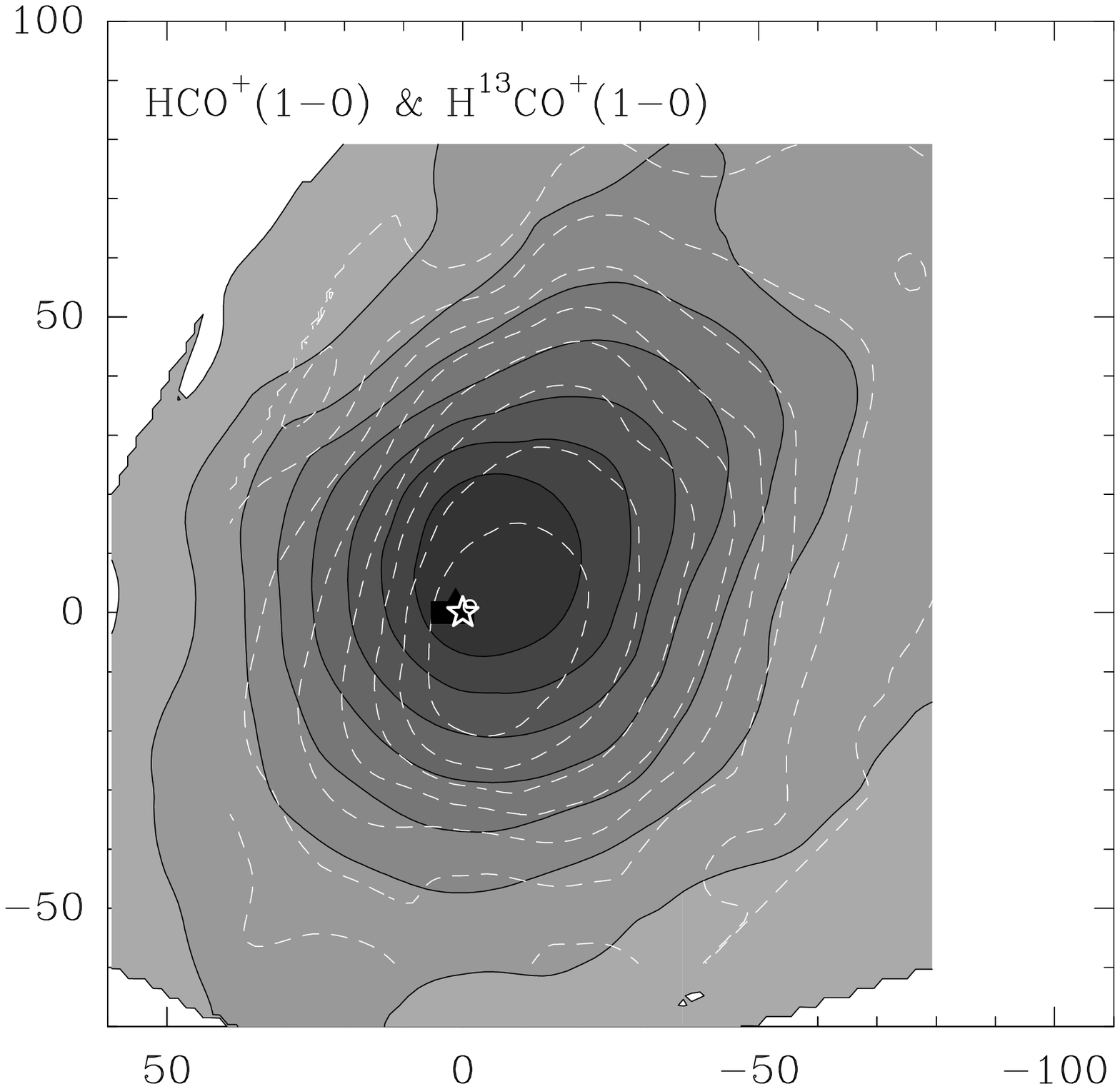}
\end{minipage}
\begin{minipage}[b]{0.3\textwidth}
    \includegraphics[width=\textwidth]{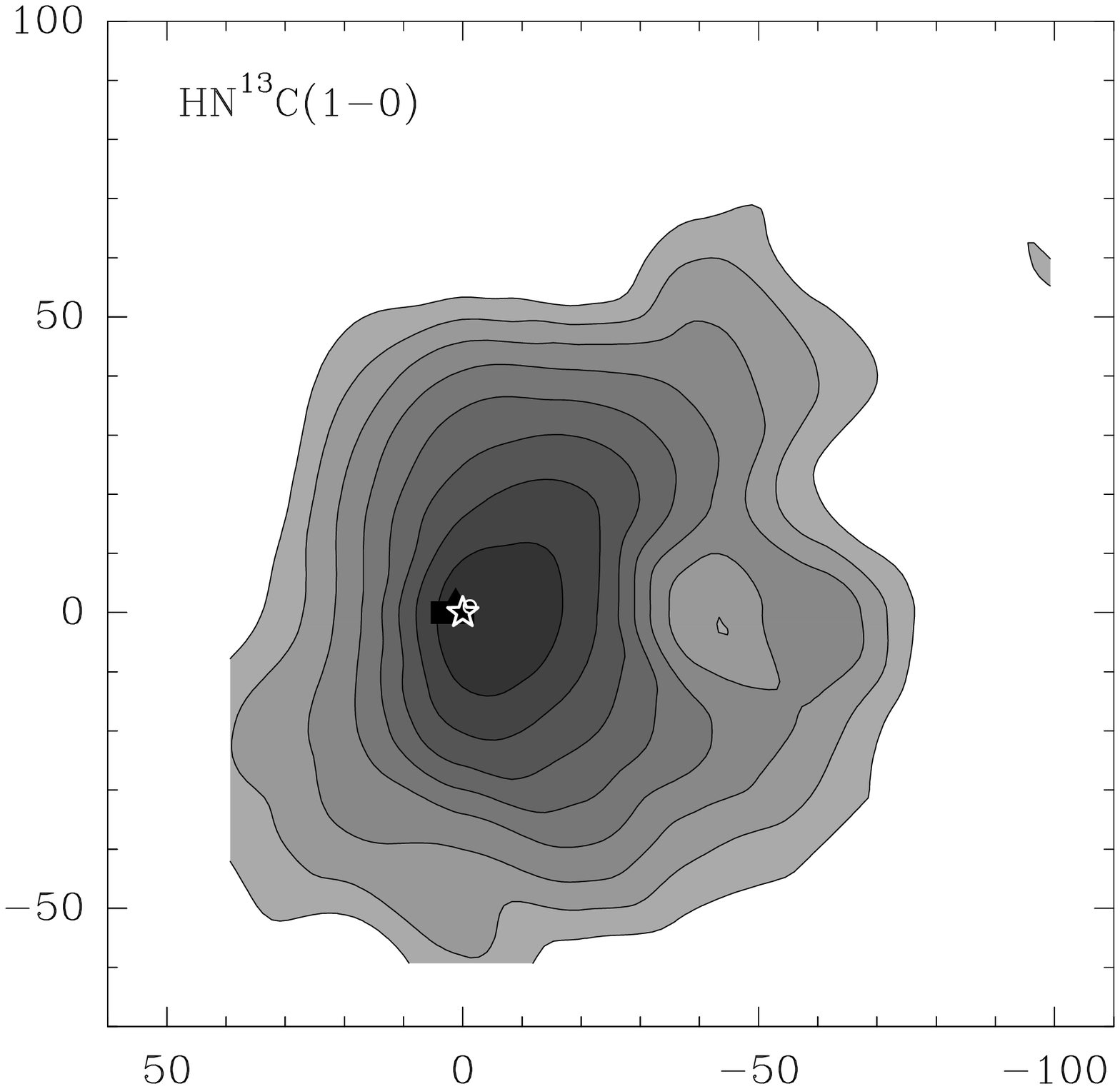}
\end{minipage}

\caption{\small
Maps of molecular lines observed in 77.462+1.759. The notation is the same as in Fig. 2. Two methanol masers \cite{Gan,Fontani}, a water-vapor maser \cite{Palla}, a 6 cm VLA radio source \cite{Urquhart}, an IRAS source, and an IR MSX source are located near the center of the core.
}
\label{77}
\end{figure}

\newpage

\begin{figure}[t!]
\setcaptionmargin{5mm}
\onelinecaptionsfalse
\captionstyle{flushleft}

\begin{minipage}[b]{0.3\textwidth}
    \includegraphics[width=\textwidth]{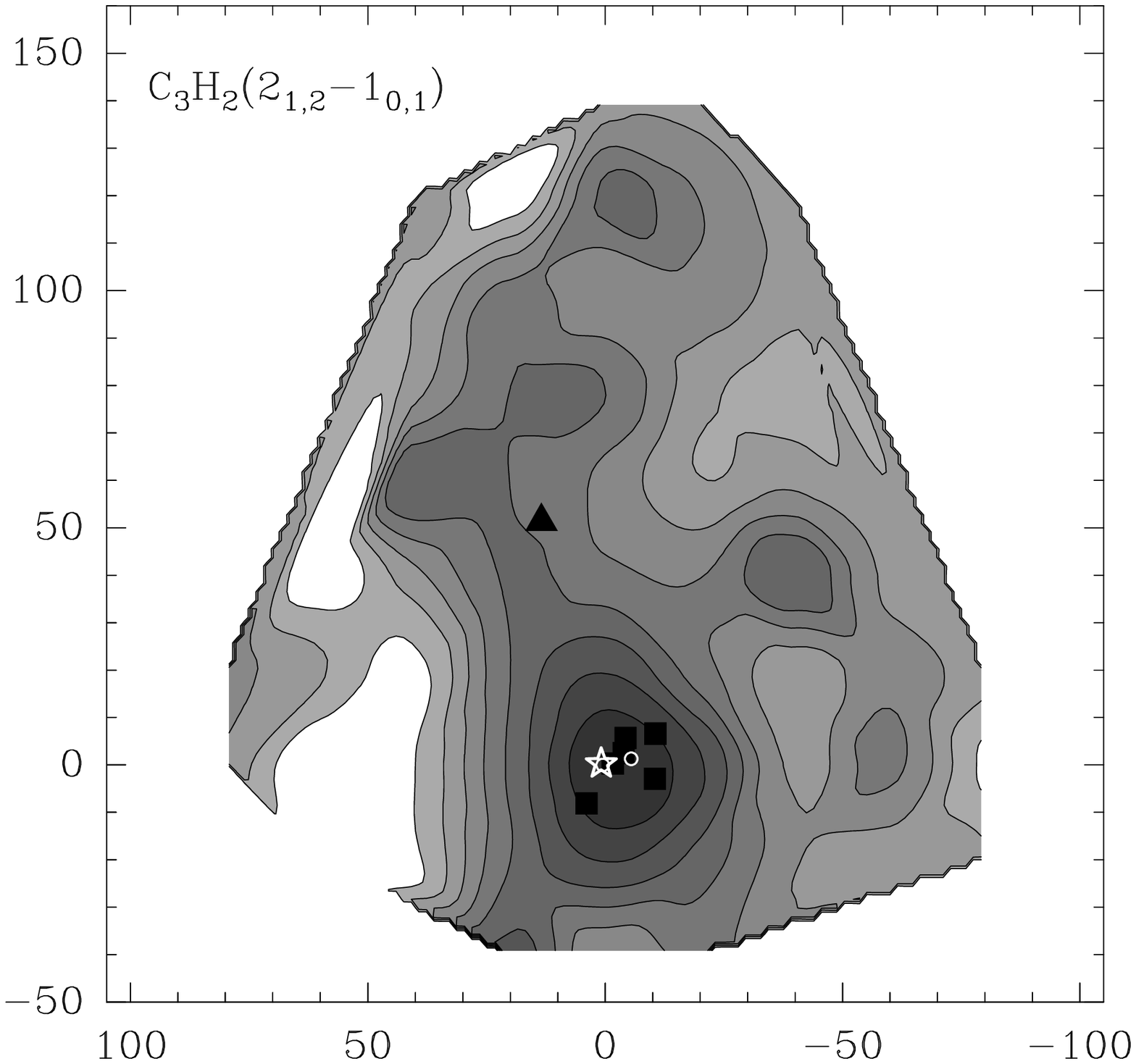}
\end{minipage}
\begin{minipage}[b]{0.3\textwidth}
    \includegraphics[width=\textwidth]{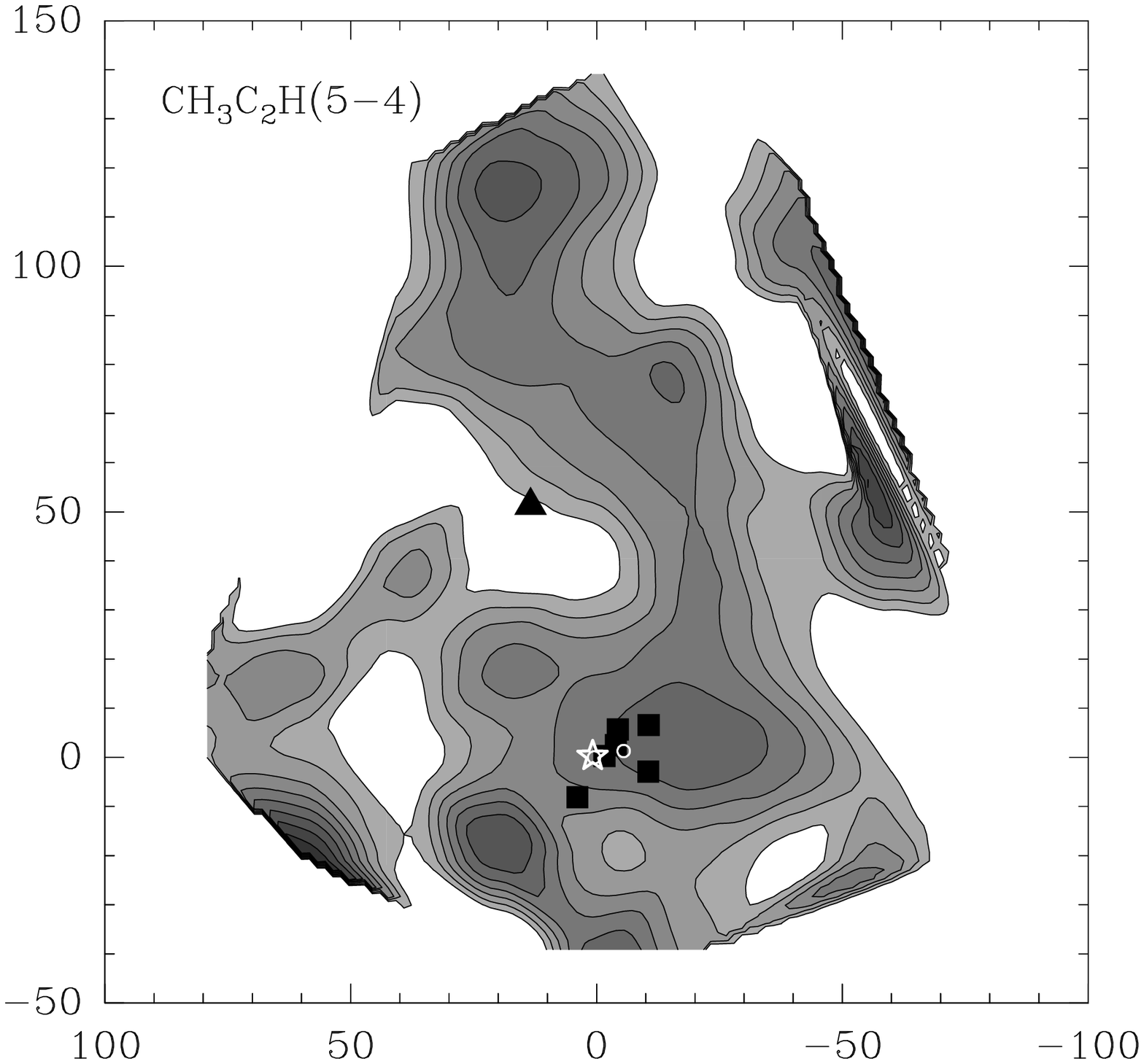}
\end{minipage}
\begin{minipage}[b]{0.3\textwidth}
    \includegraphics[width=\textwidth]{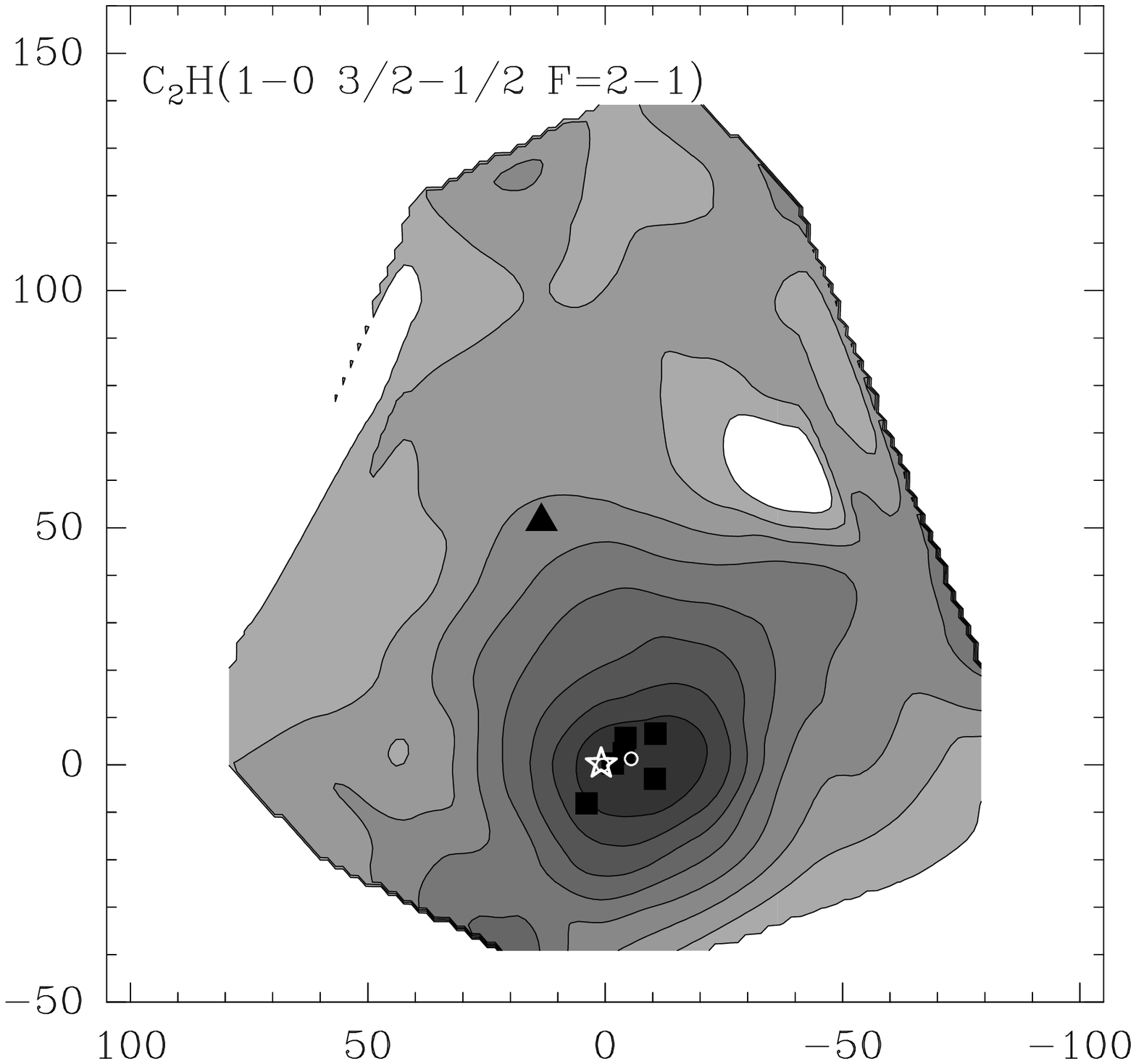}
\end{minipage}

\vskip 1mm

\begin{minipage}[b]{0.3\textwidth}
    \includegraphics[width=\textwidth]{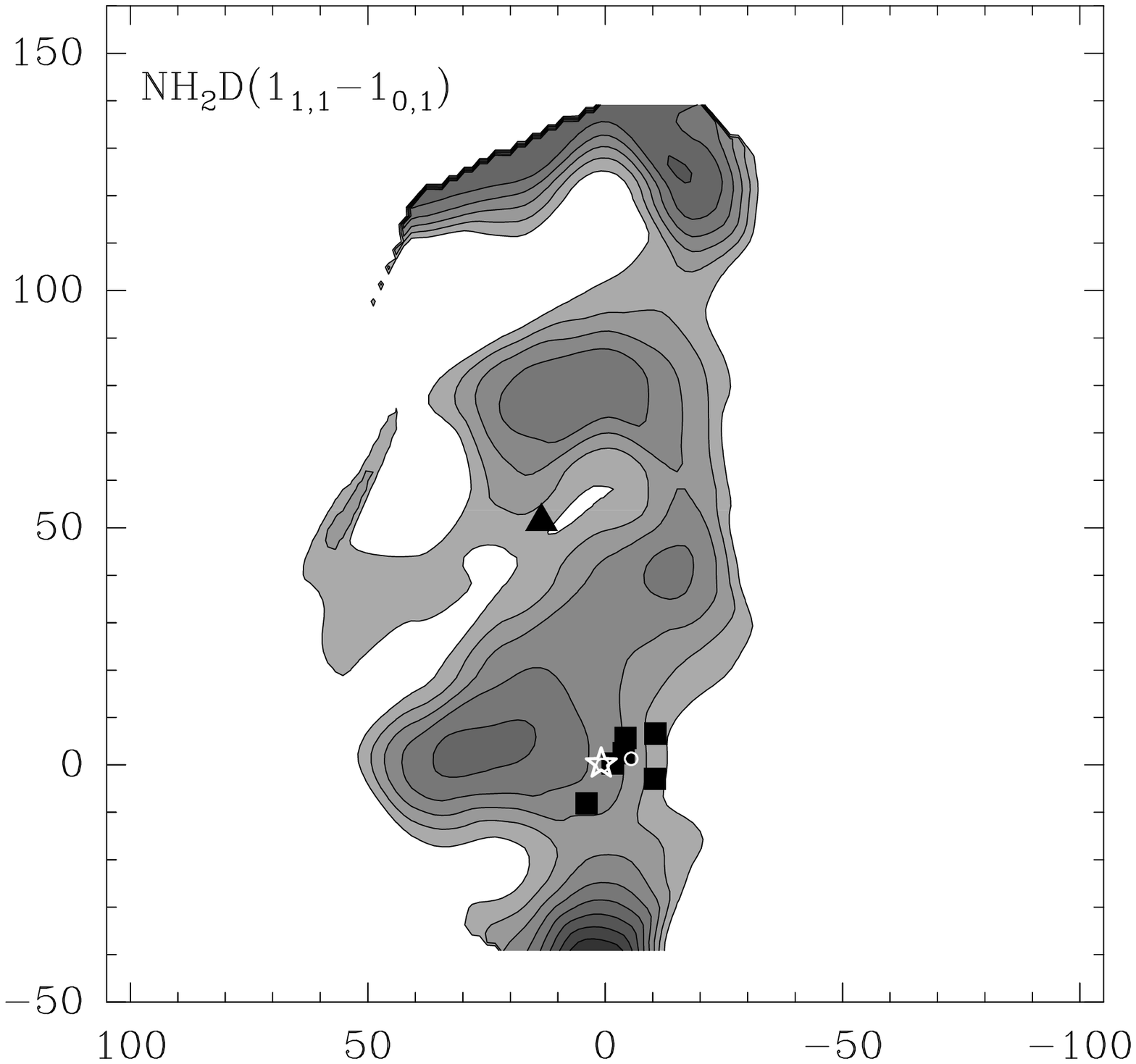}
\end{minipage}
\begin{minipage}[b]{0.3\textwidth}
    \includegraphics[width=\textwidth]{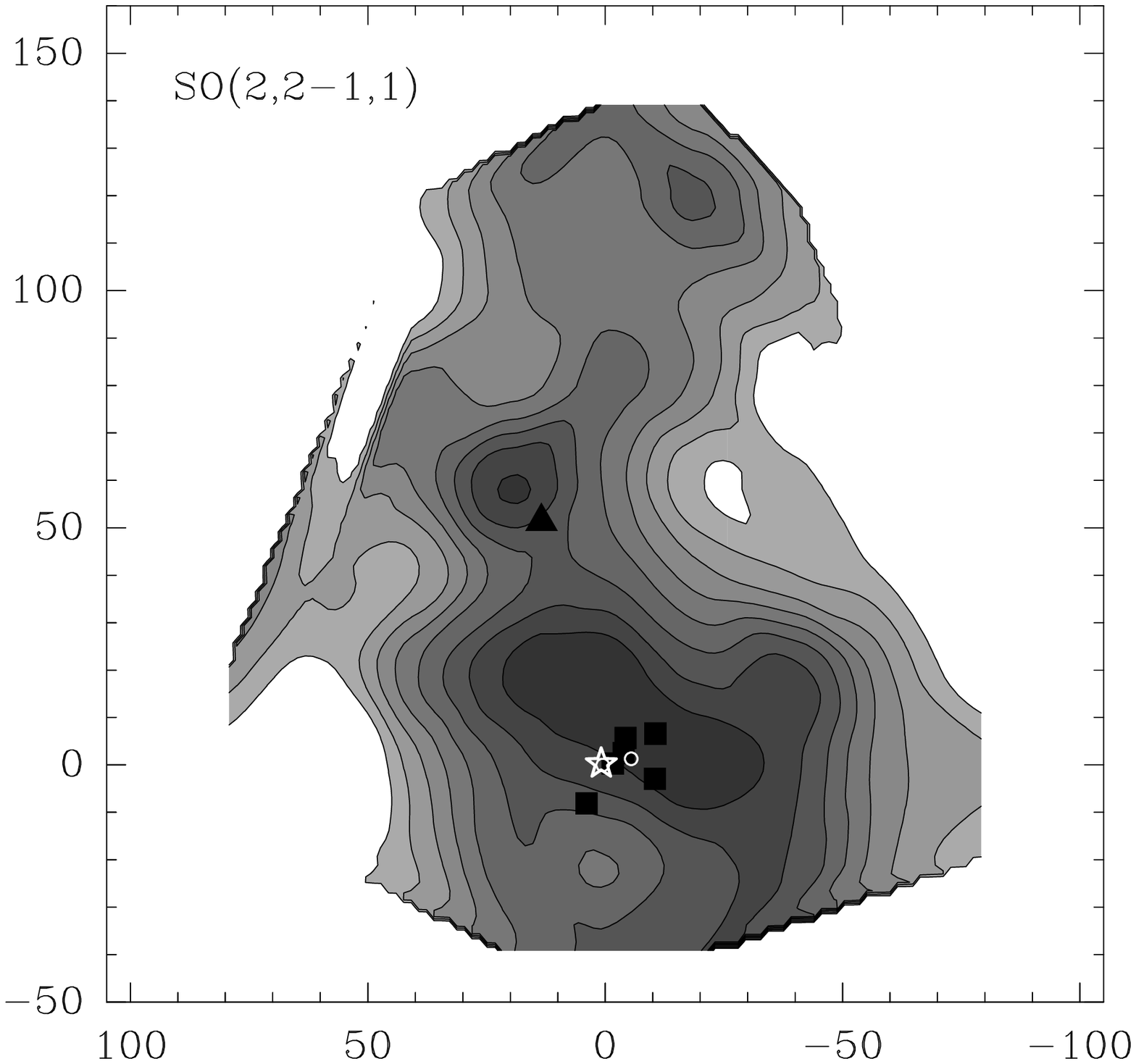}
\end{minipage}
\begin{minipage}[b]{0.3\textwidth}
    \includegraphics[width=\textwidth]{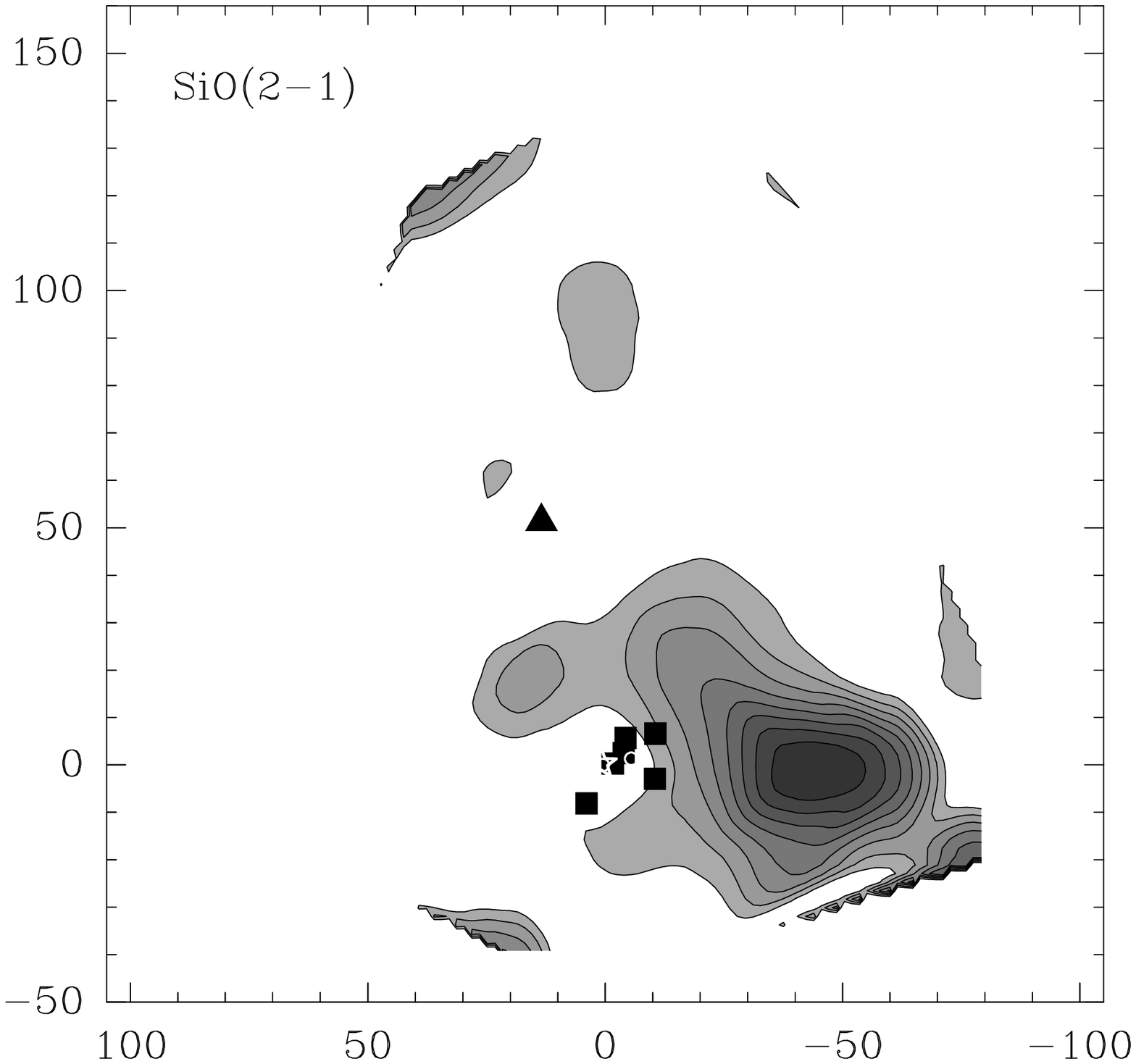}
\end{minipage}

\vskip 1mm

\begin{minipage}[b]{0.3\textwidth}
    \includegraphics[width=\textwidth]{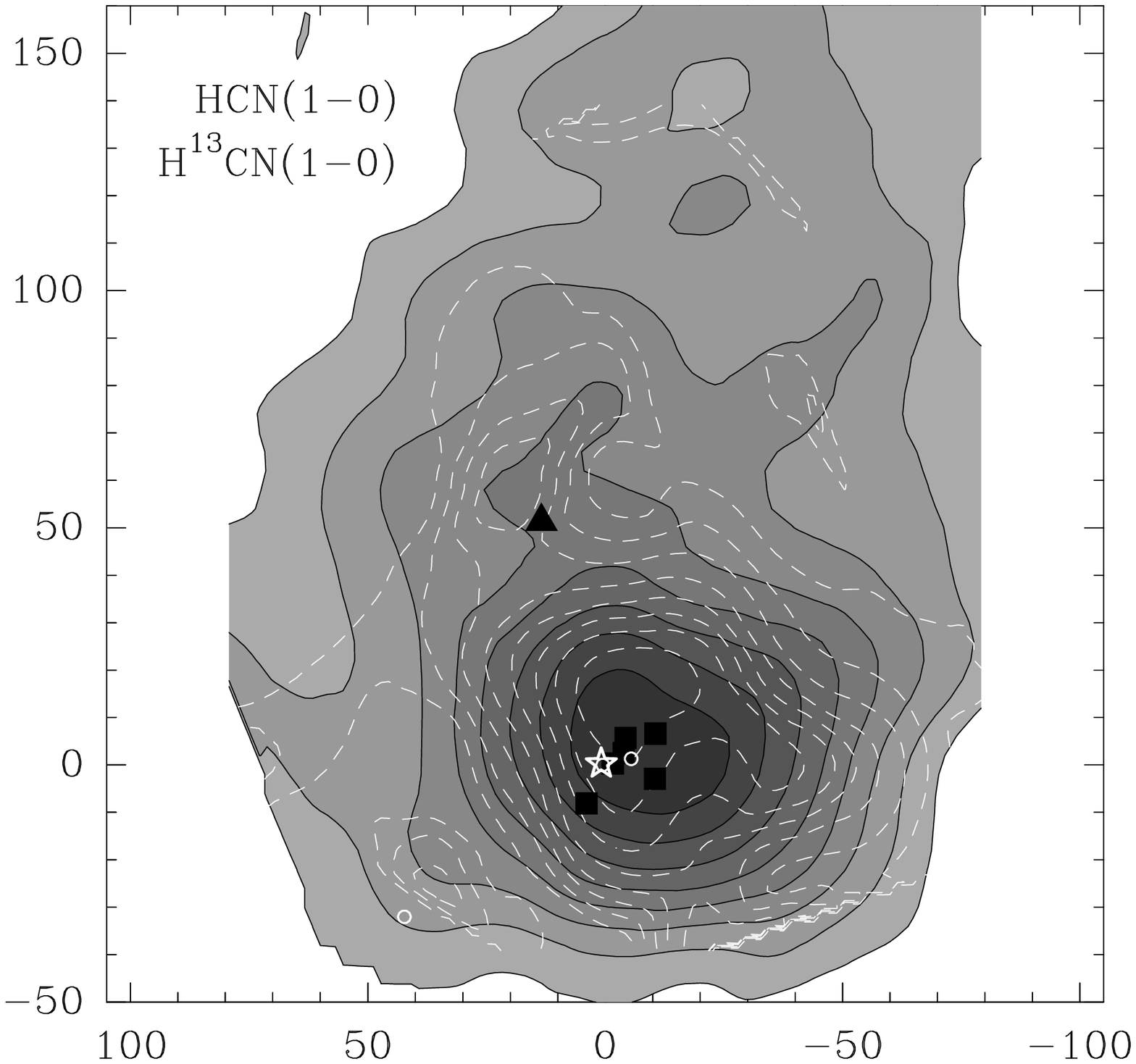}
\end{minipage}
\begin{minipage}[b]{0.3\textwidth}
    \includegraphics[width=\textwidth]{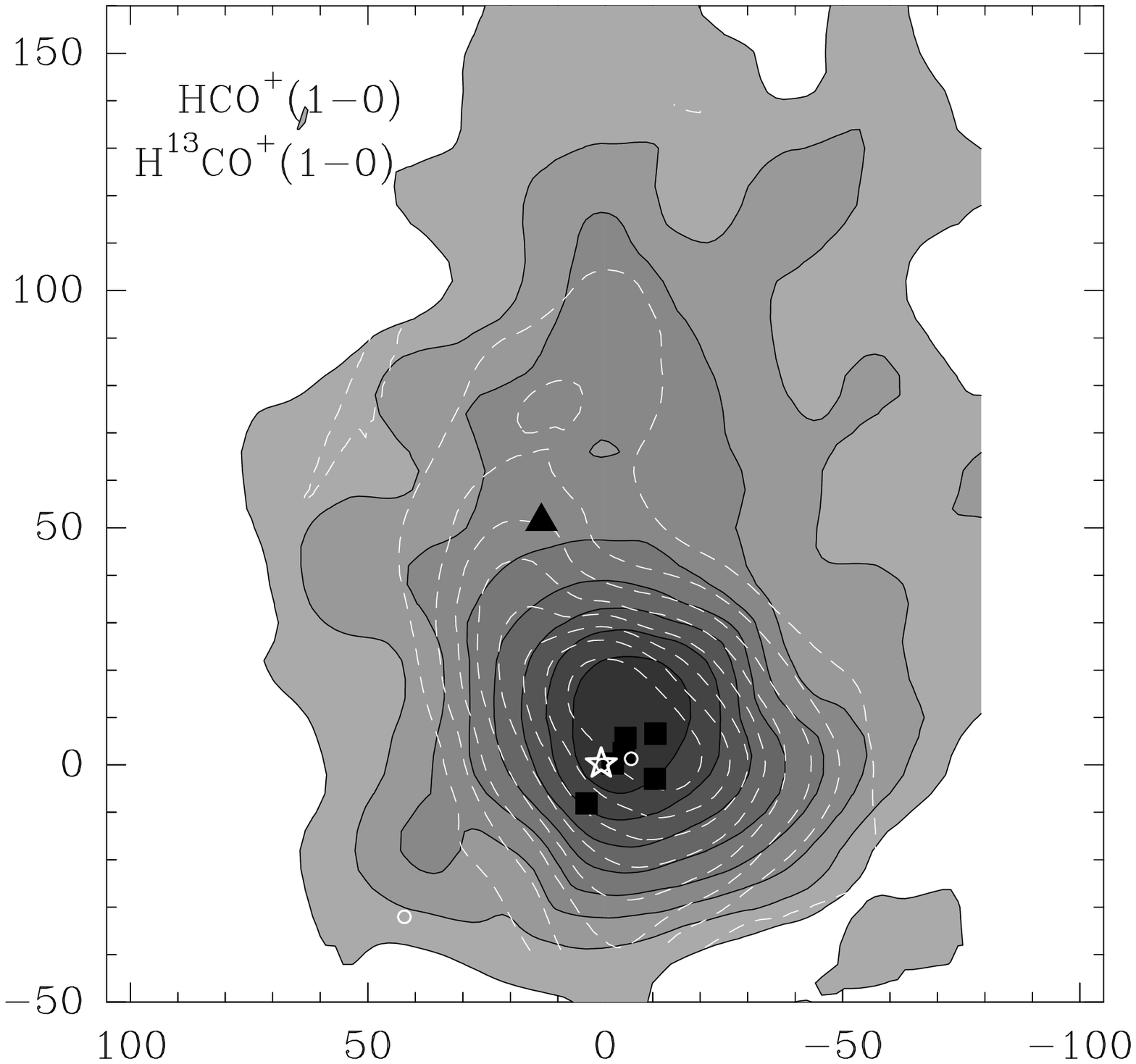}
\end{minipage}
\begin{minipage}[b]{0.3\textwidth}
    \includegraphics[width=\textwidth]{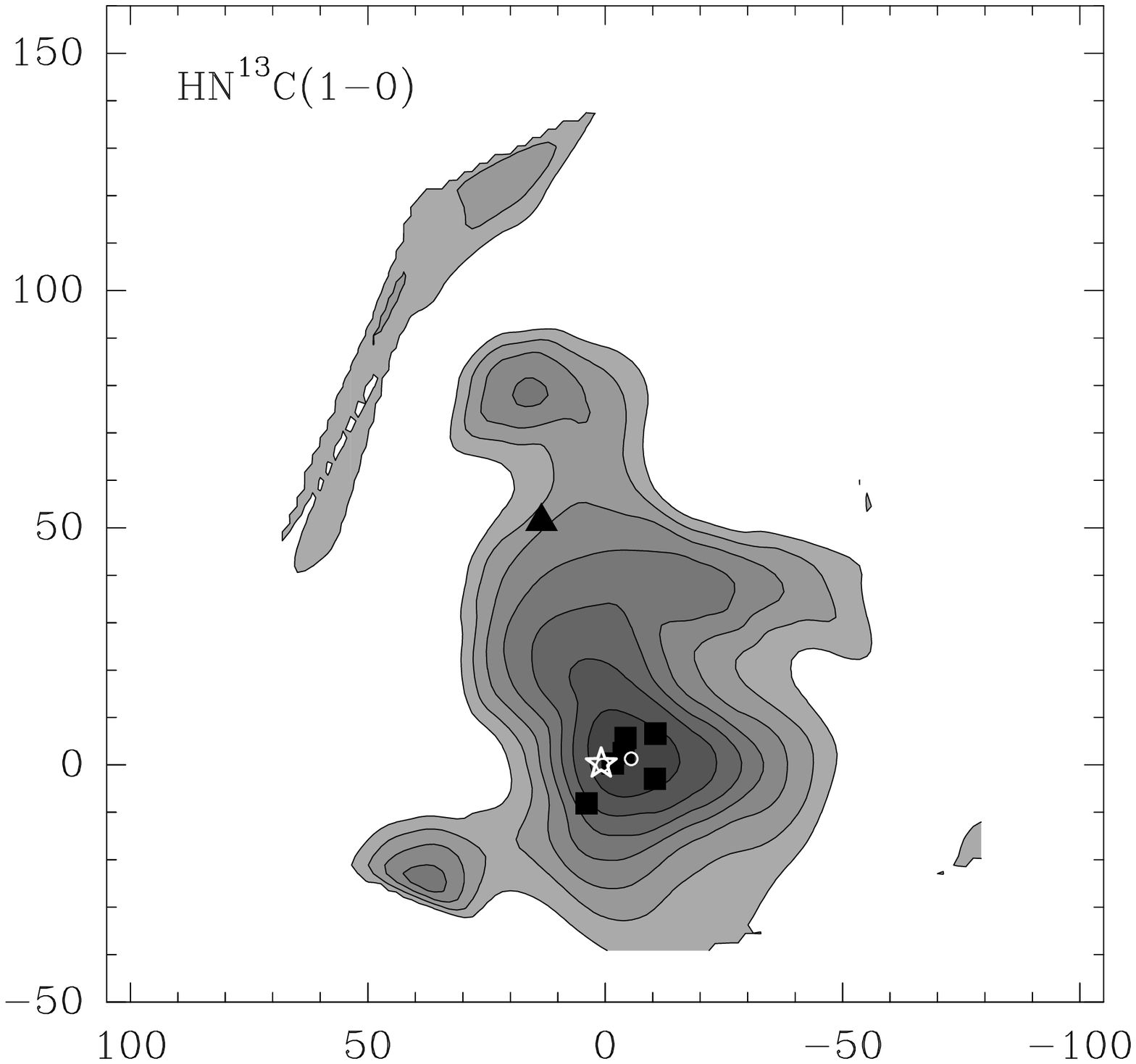}
\end{minipage}

\caption{\small
Maps of molecular lines observed in 99.982+4.17. The notation is the same as in Fig. 2. The positions of methanol masers were taken from \cite{Gan,Kurtz}, and those of water-vapor masers from \cite{Palla, Slysh}. Near-IR and mid-IR sources from \cite{Pollanen,Chauhan,Choudhury} and the submillimeter source from \cite{DiFrancesco} are shown.
}
\label{99}
\end{figure}

\newpage

\begin{figure}[t!]
\setcaptionmargin{5mm}
\onelinecaptionsfalse
\captionstyle{flushleft}

\begin{minipage}[b]{0.24\textwidth}
    \includegraphics[width=\textwidth]{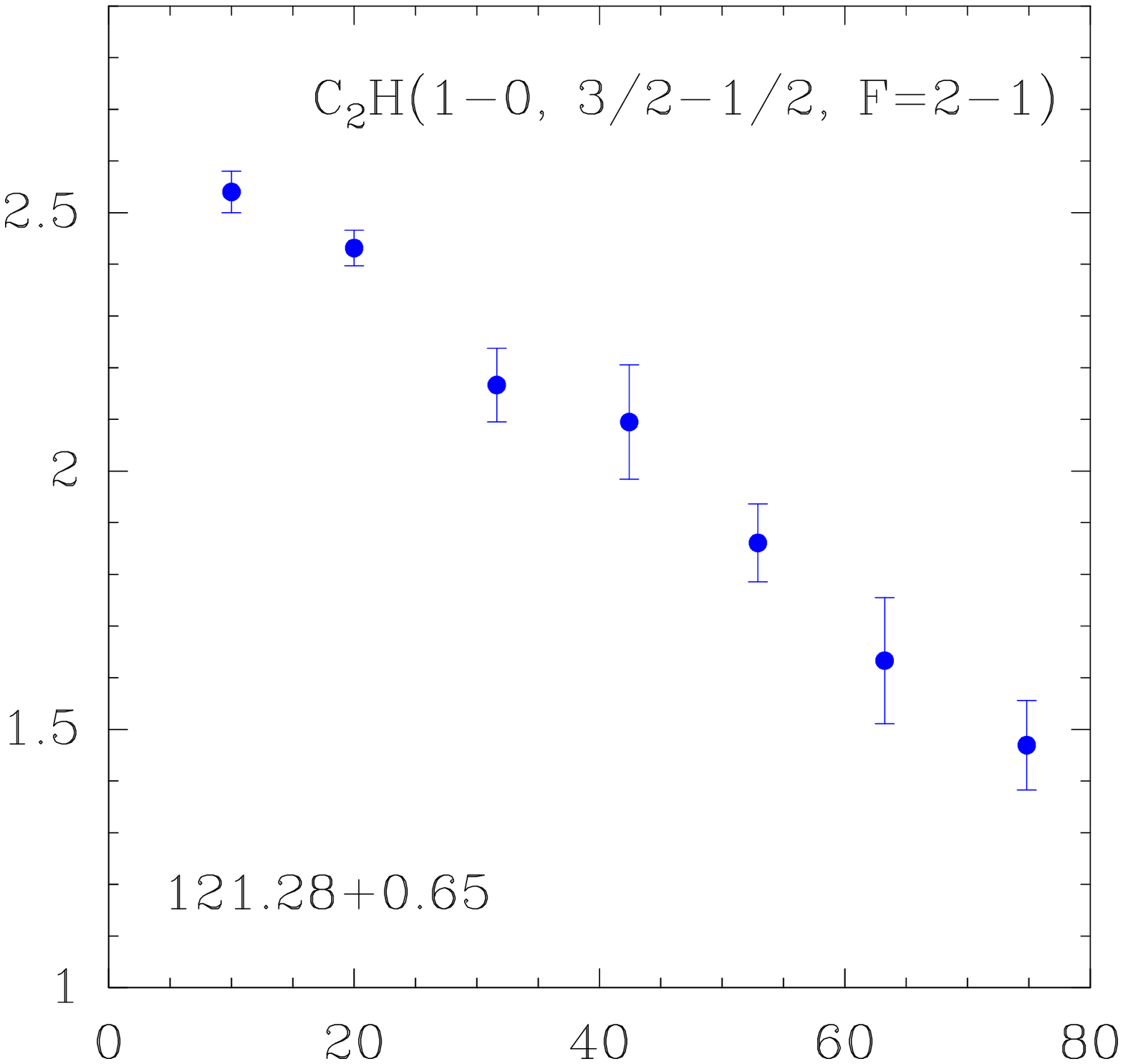}
\end{minipage}
\begin{minipage}[b]{0.24\textwidth}
    \includegraphics[width=\textwidth]{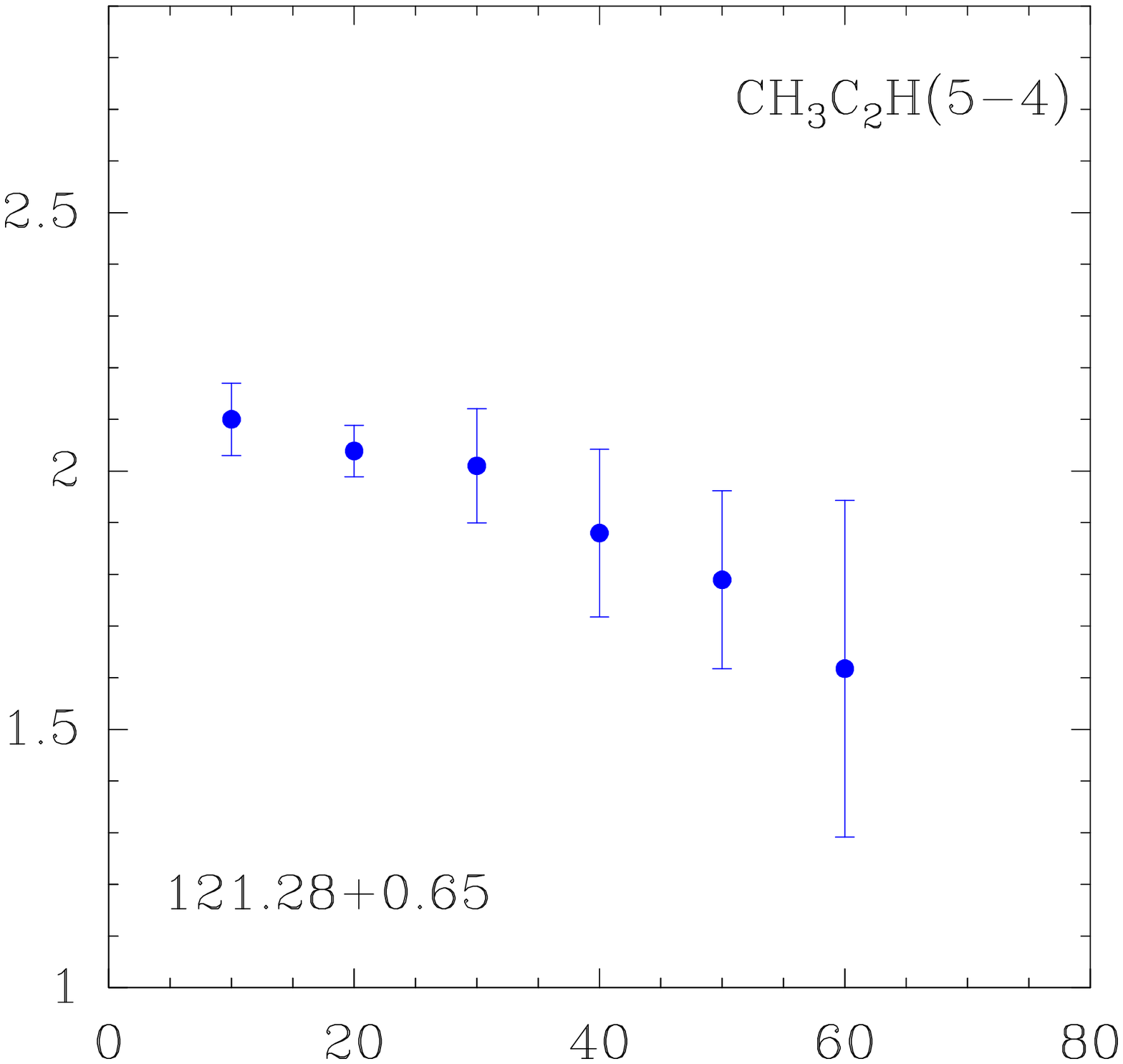}
\end{minipage}
\begin{minipage}[b]{0.24\textwidth}
    \includegraphics[width=\textwidth]{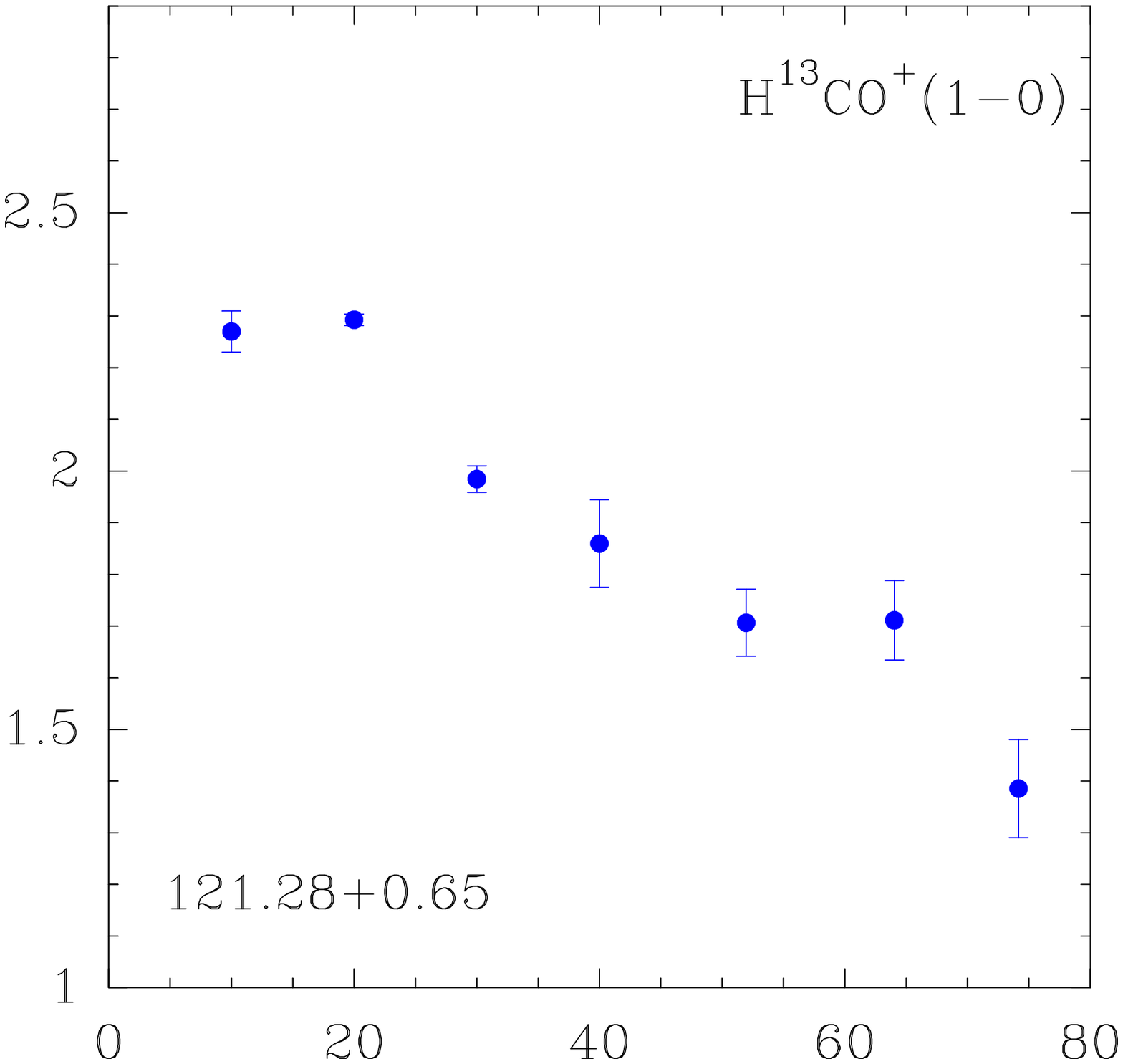}
\end{minipage}
\begin{minipage}[b]{0.24\textwidth}
    \includegraphics[width=\textwidth]{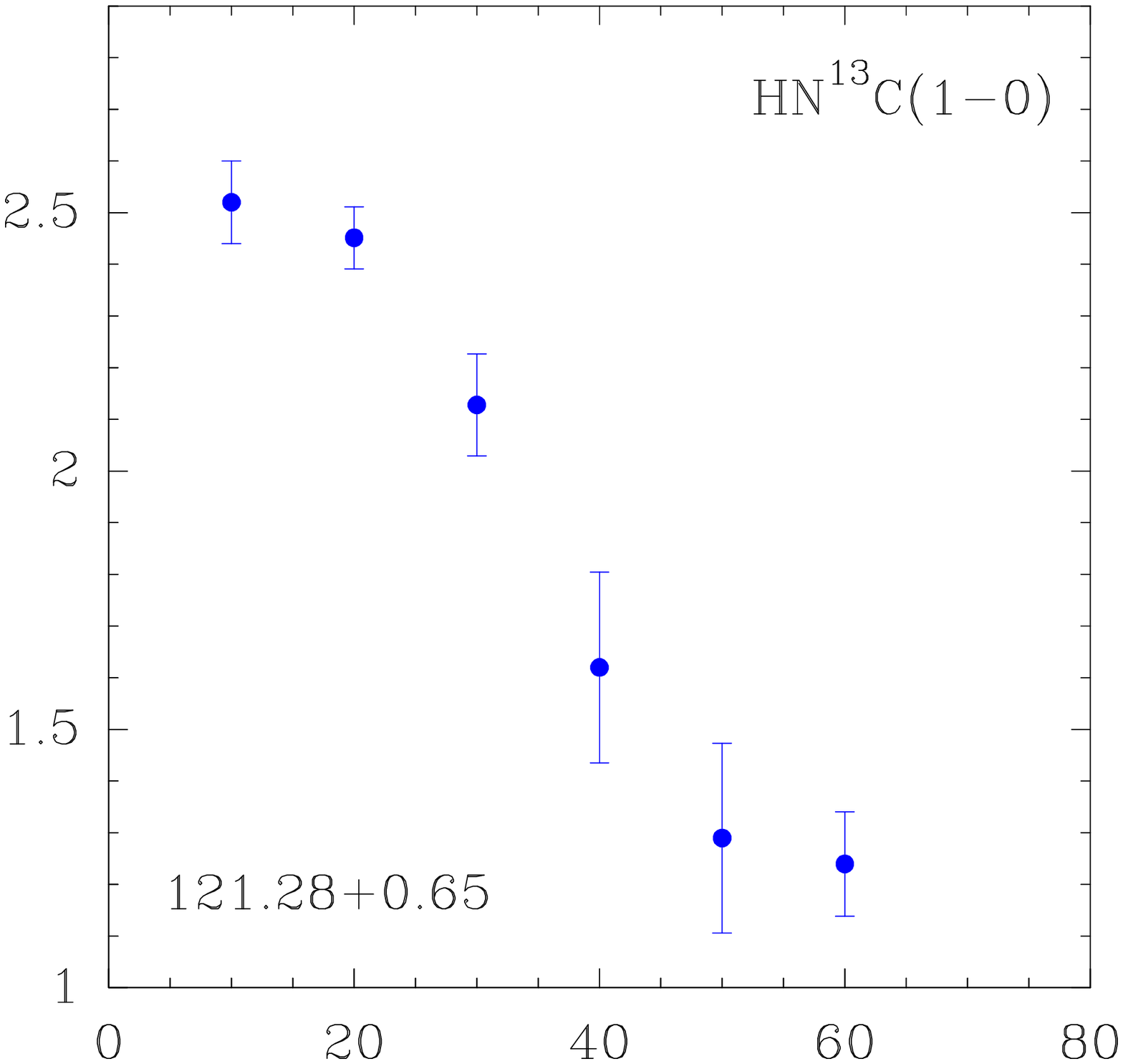}
\end{minipage}

\vskip 2mm

\begin{minipage}[b]{0.24\textwidth}
    \includegraphics[width=\textwidth]{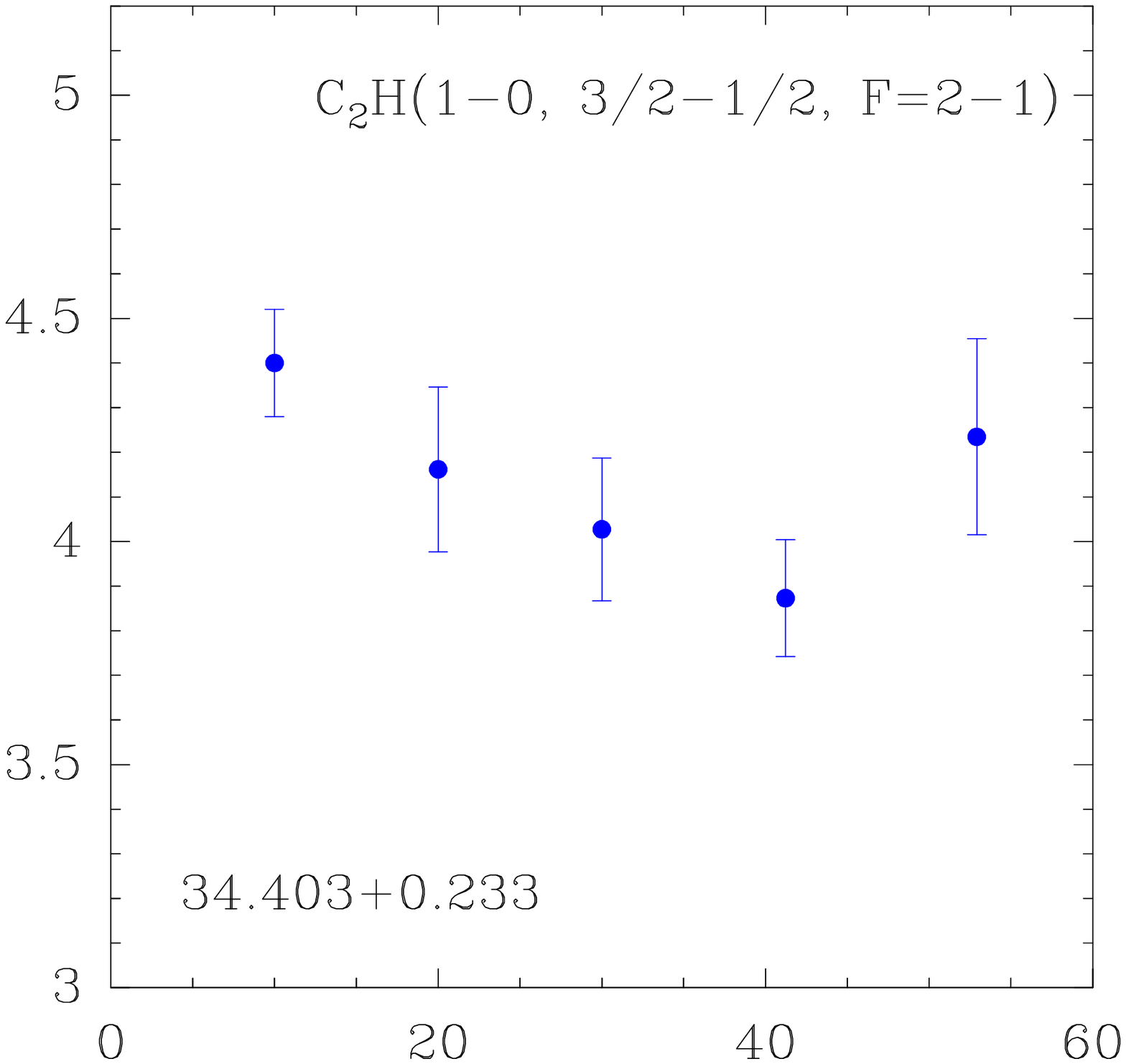}
\end{minipage}
\begin{minipage}[b]{0.24\textwidth}
    \includegraphics[width=\textwidth]{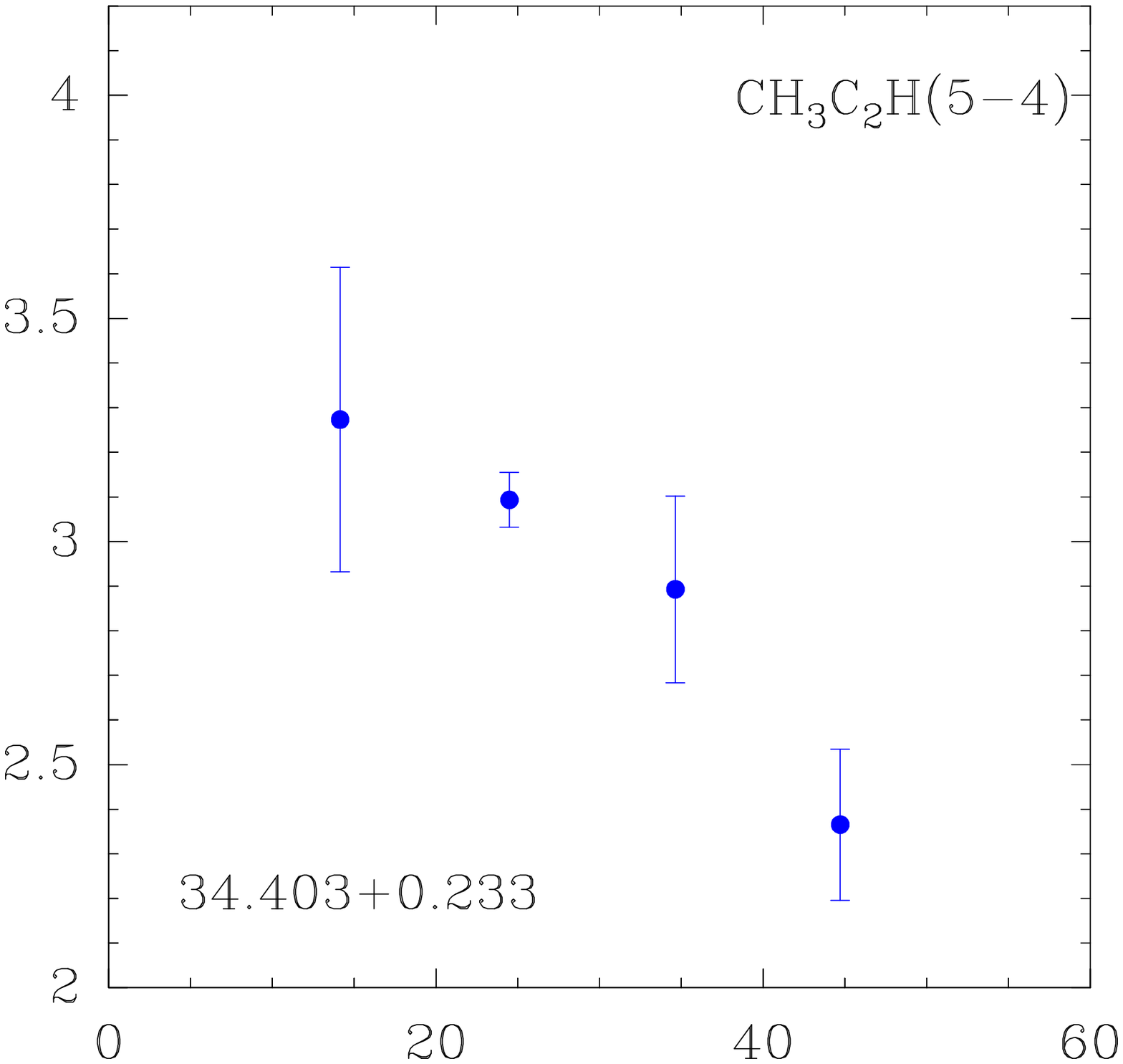}
\end{minipage}
\begin{minipage}[b]{0.24\textwidth}
    \includegraphics[width=\textwidth]{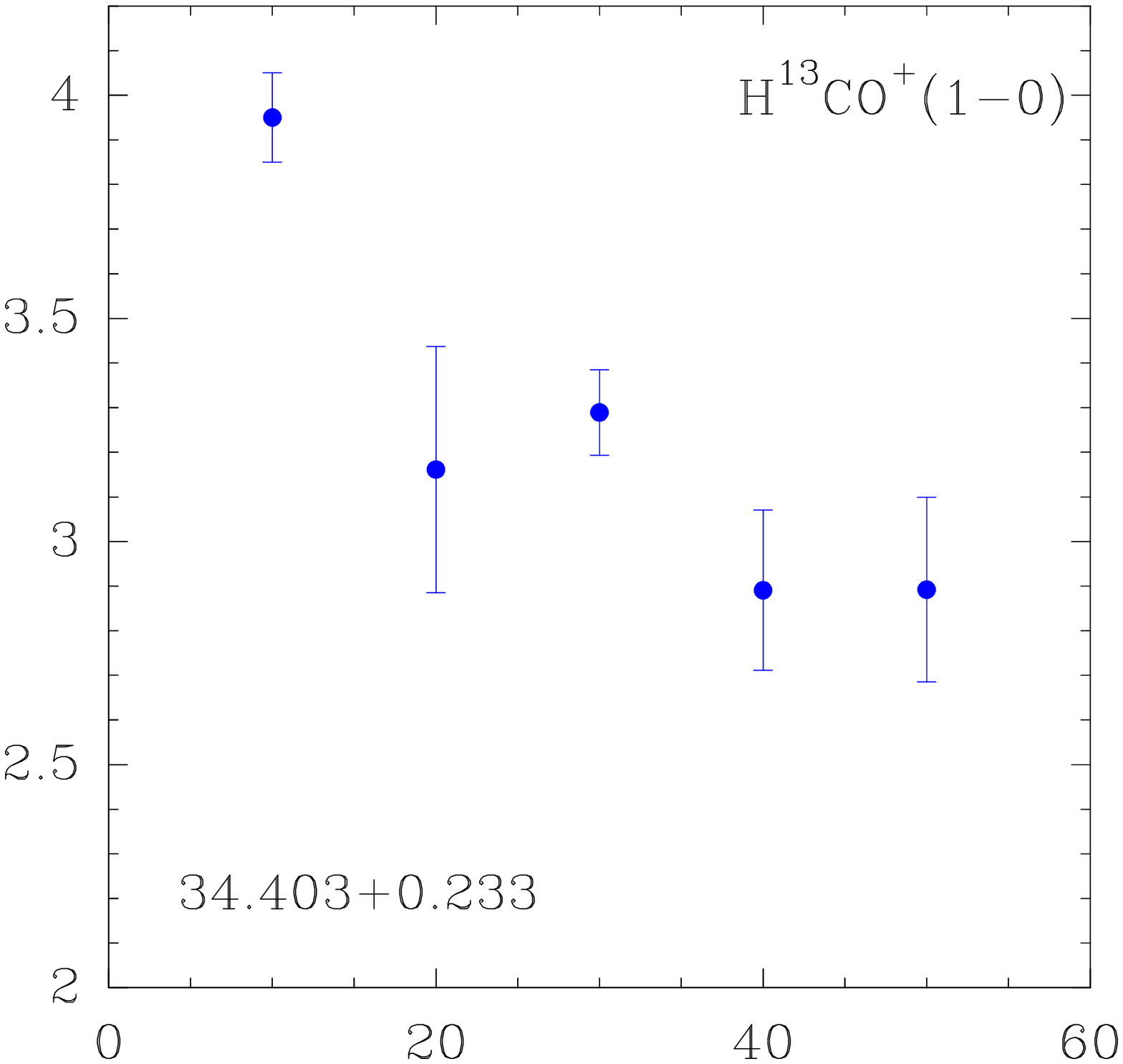}
\end{minipage}
\begin{minipage}[b]{0.24\textwidth}
    \includegraphics[width=\textwidth]{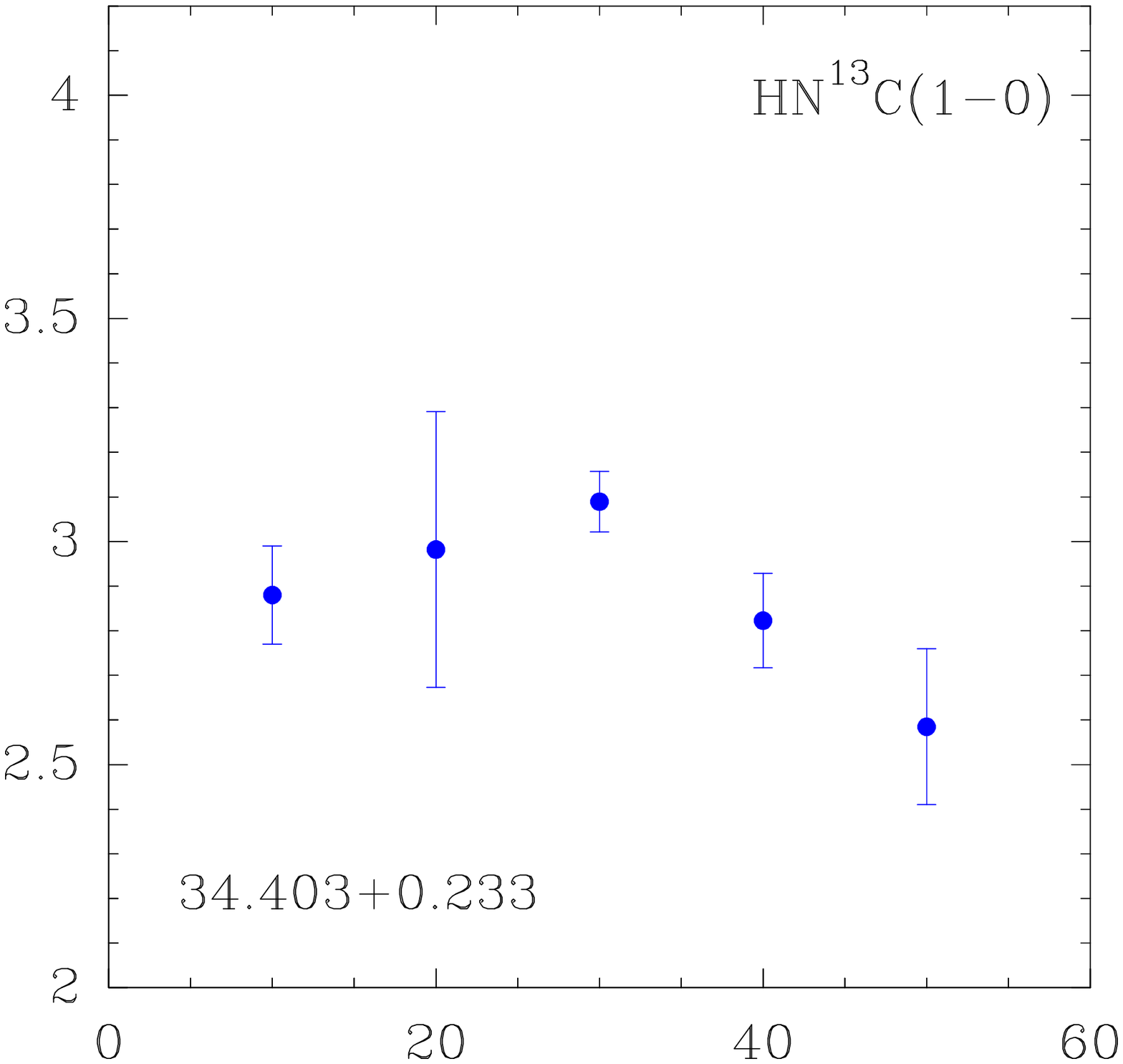}
\end{minipage}

\vskip 2mm

\begin{minipage}[b]{0.24\textwidth}
    \includegraphics[width=\textwidth]{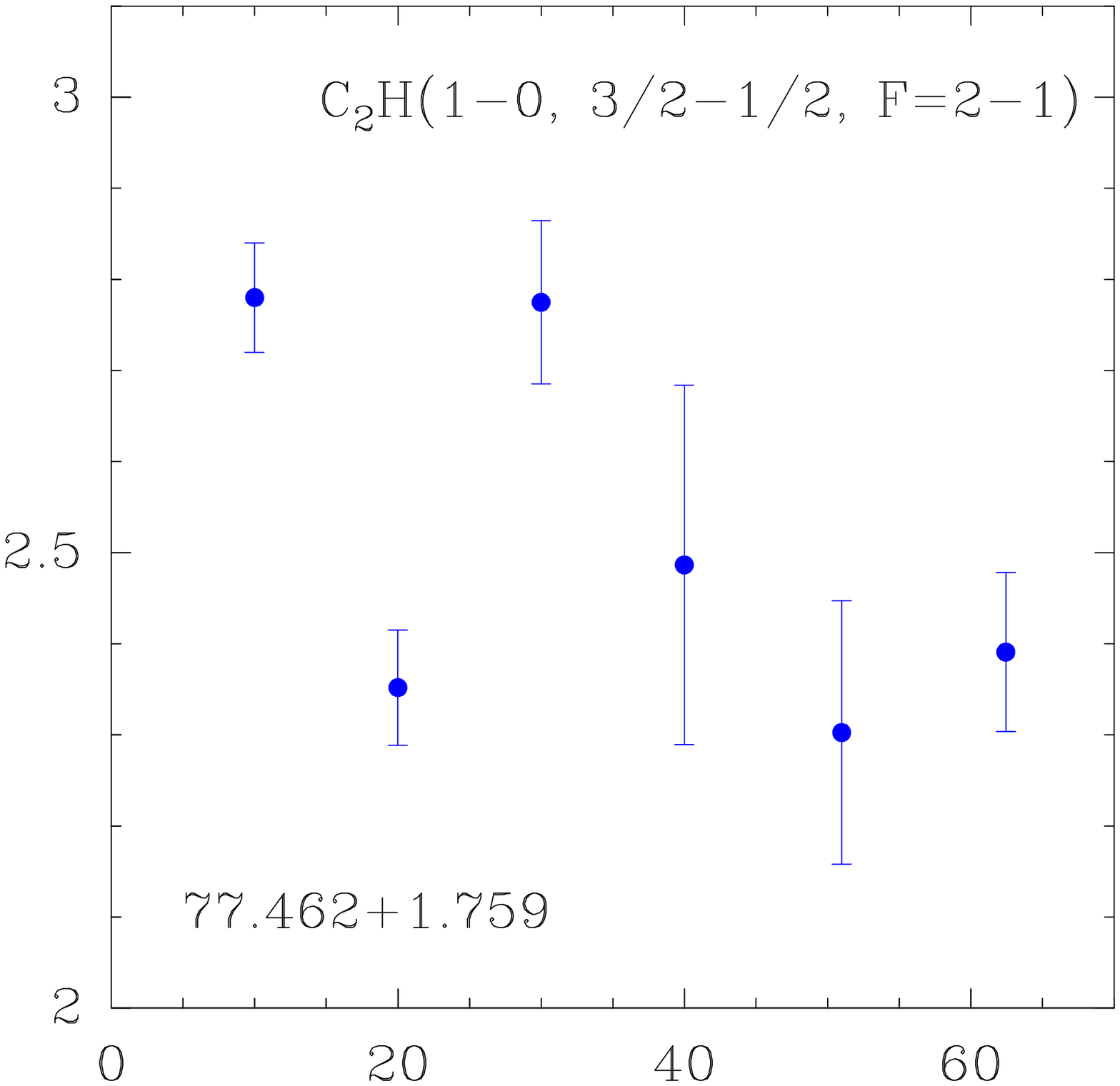}
\end{minipage}
\begin{minipage}[b]{0.24\textwidth}
    \includegraphics[width=\textwidth]{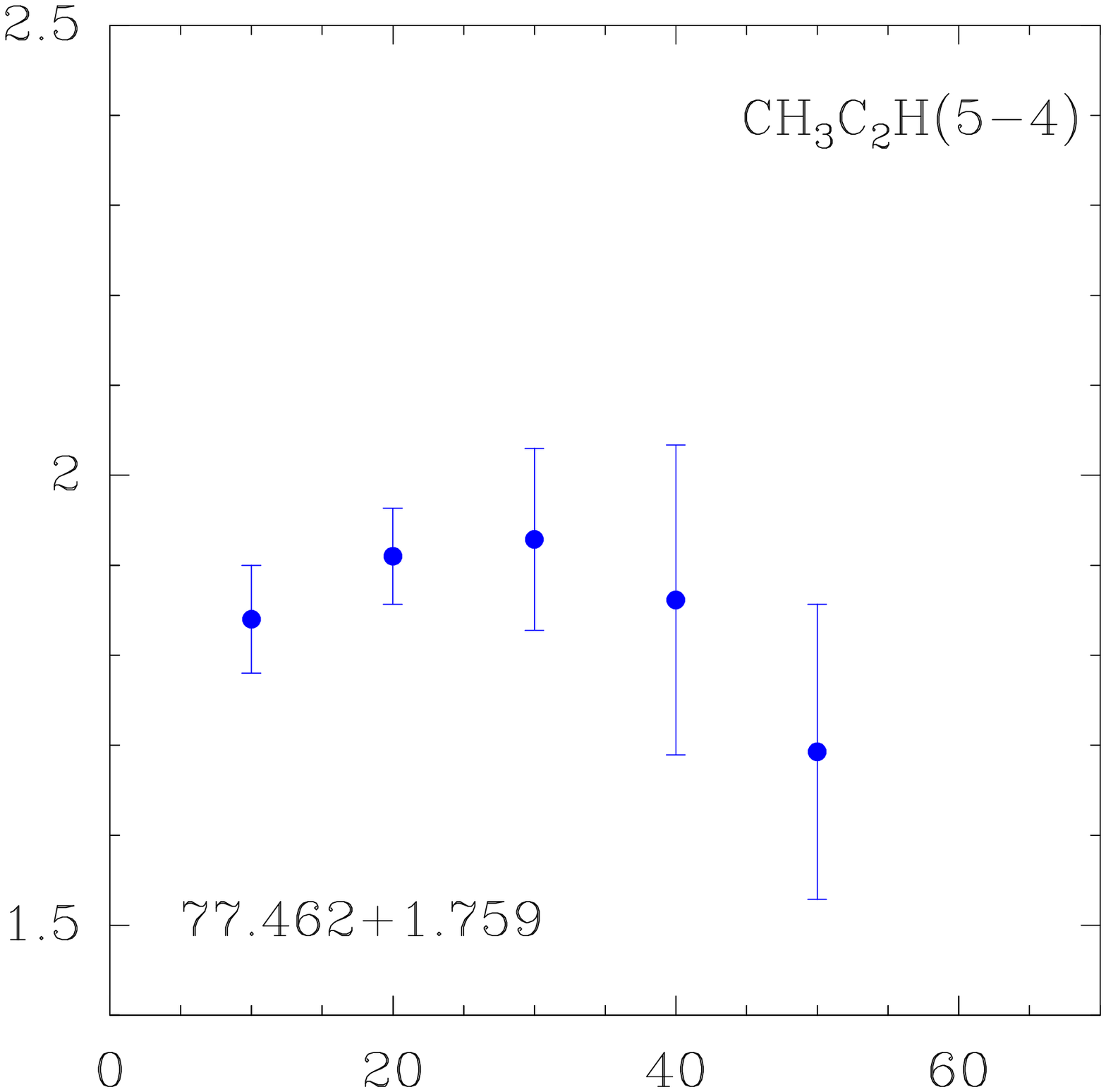}
\end{minipage}
\begin{minipage}[b]{0.24\textwidth}
    \includegraphics[width=\textwidth]{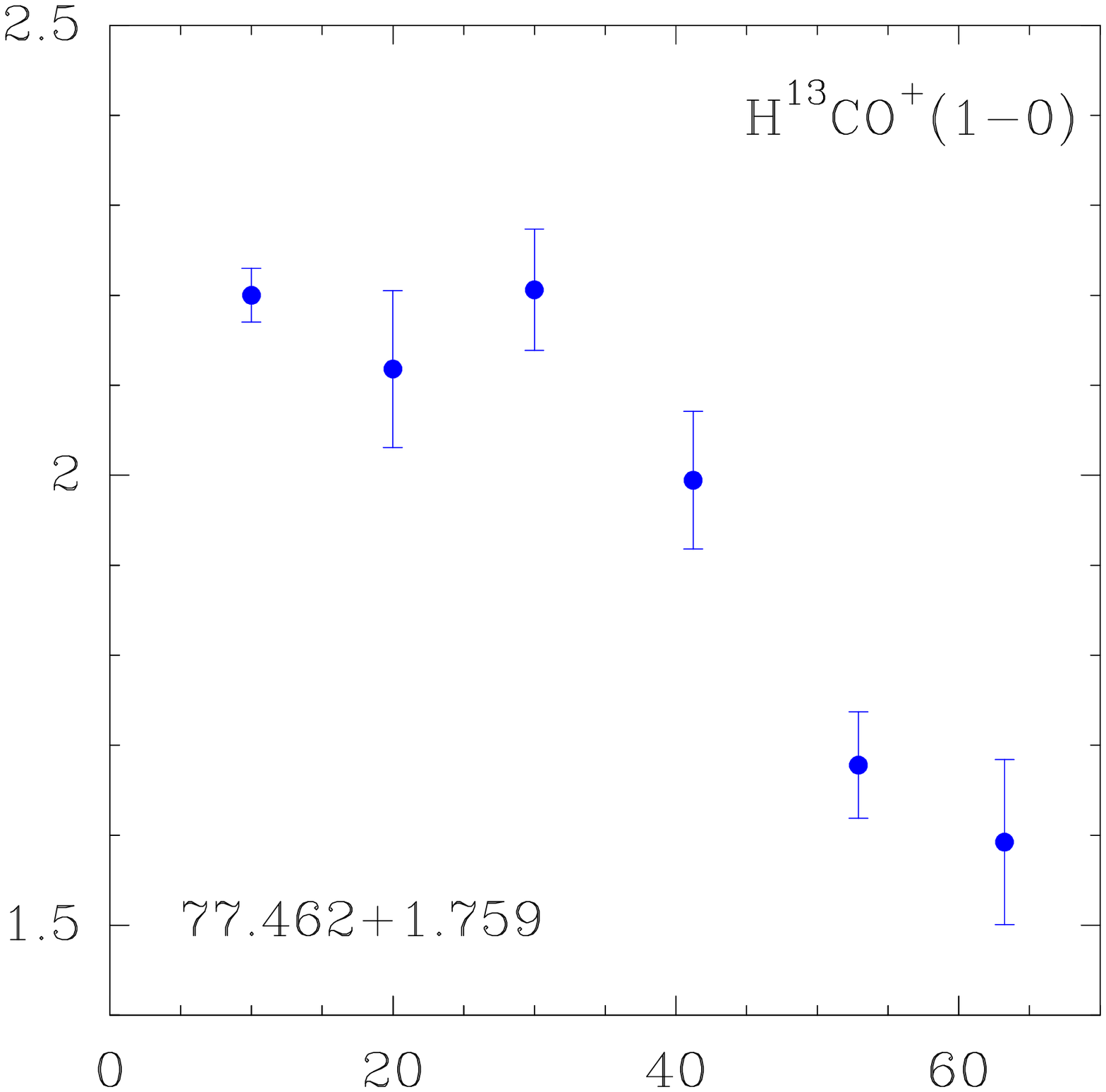}
\end{minipage}
\begin{minipage}[b]{0.24\textwidth}
    \includegraphics[width=\textwidth]{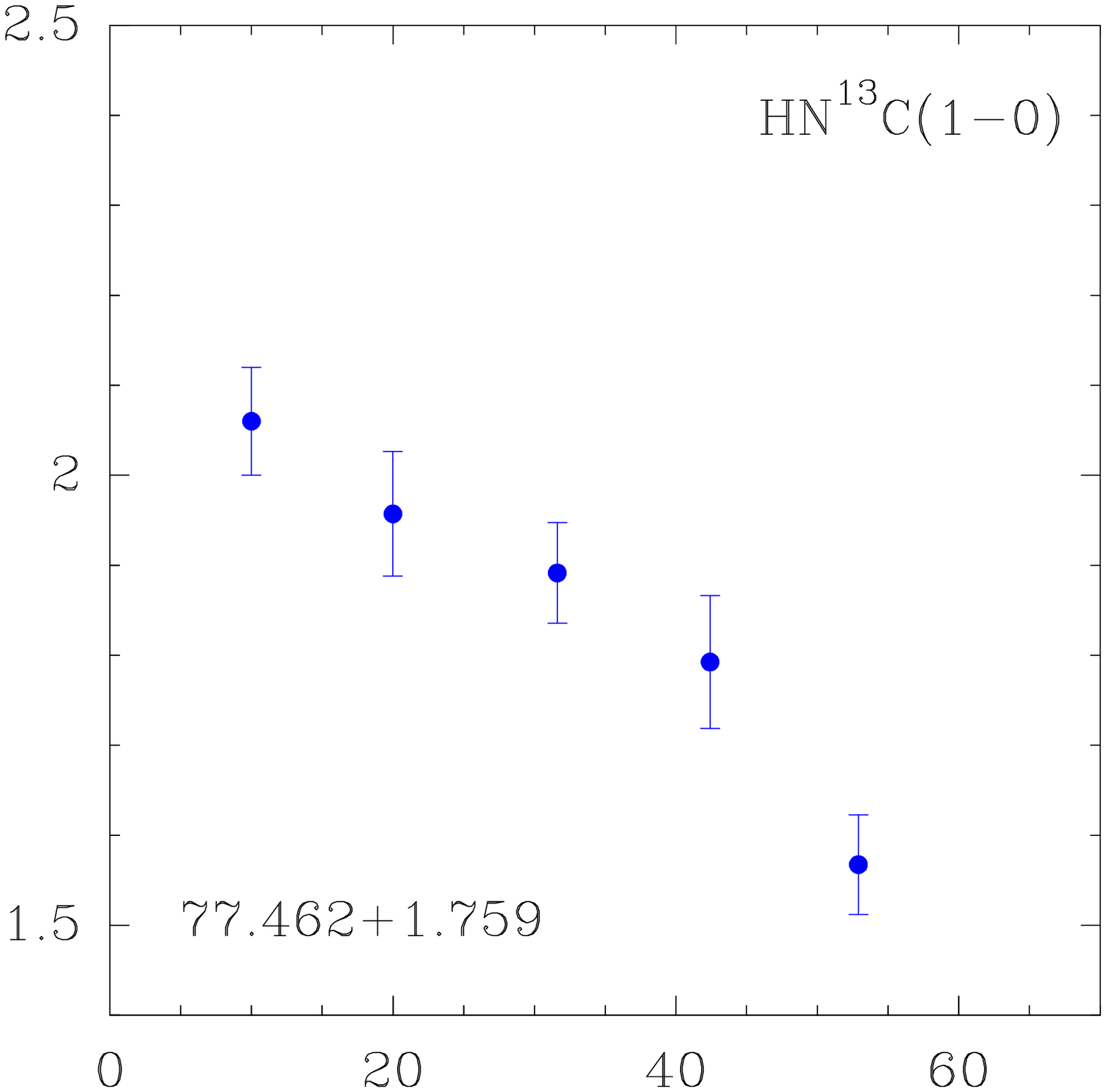}
\end{minipage}

\vskip 2mm

\begin{minipage}[b]{0.24\textwidth}
    \includegraphics[width=\textwidth]{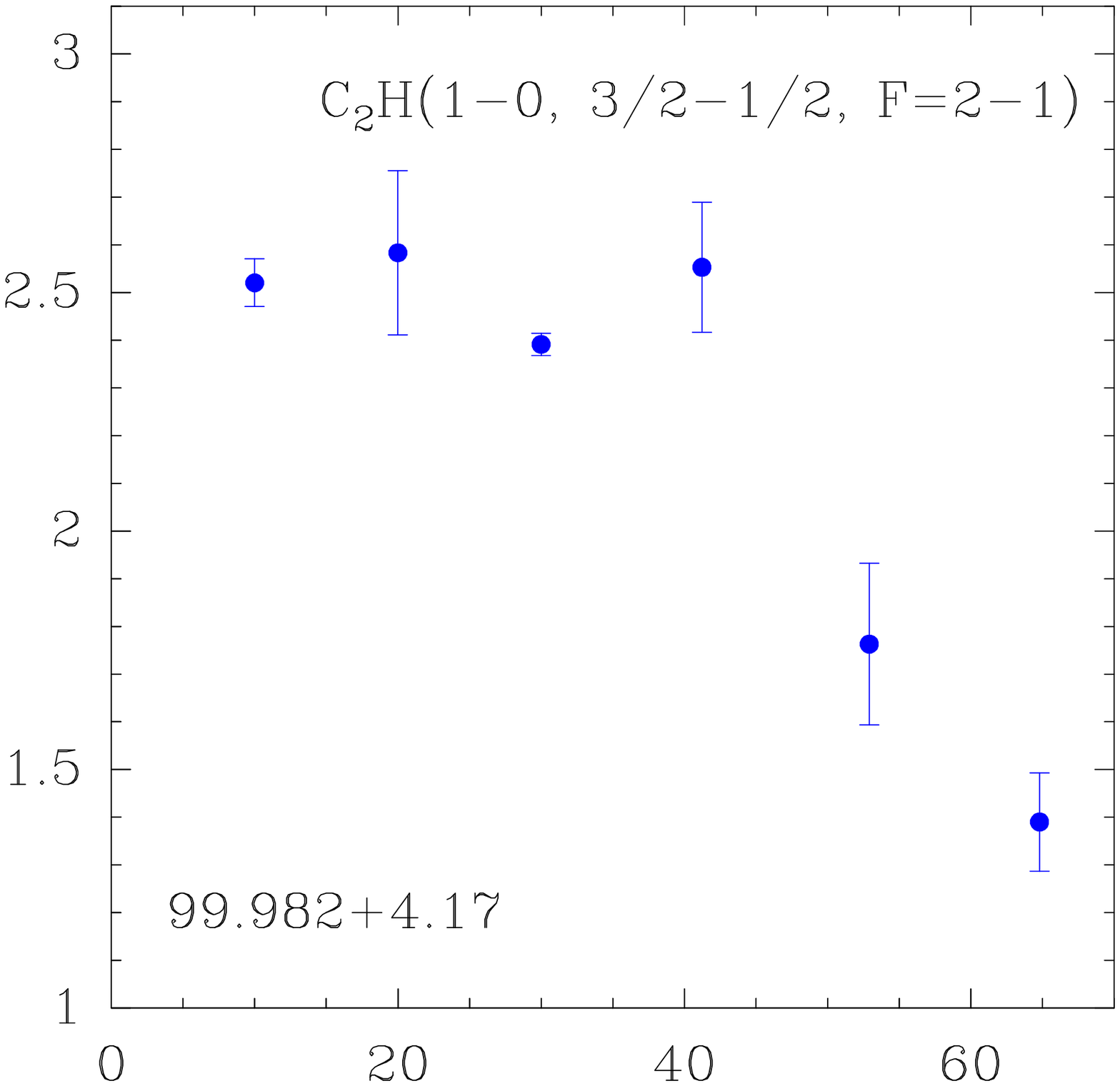}
\end{minipage}
\begin{minipage}[b]{0.24\textwidth}
    \includegraphics[width=\textwidth]{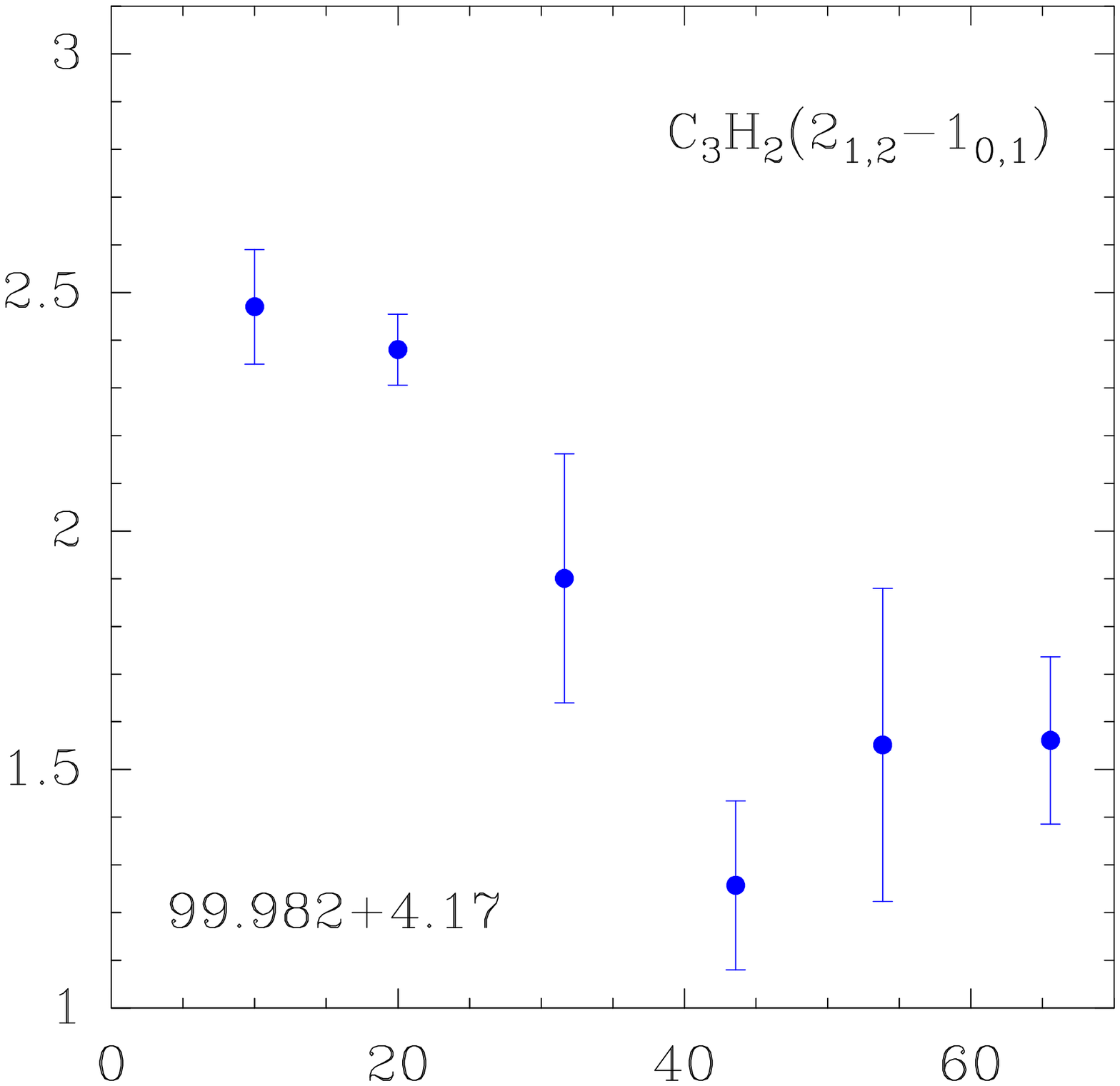}
\end{minipage}
\begin{minipage}[b]{0.24\textwidth}
    \includegraphics[width=\textwidth]{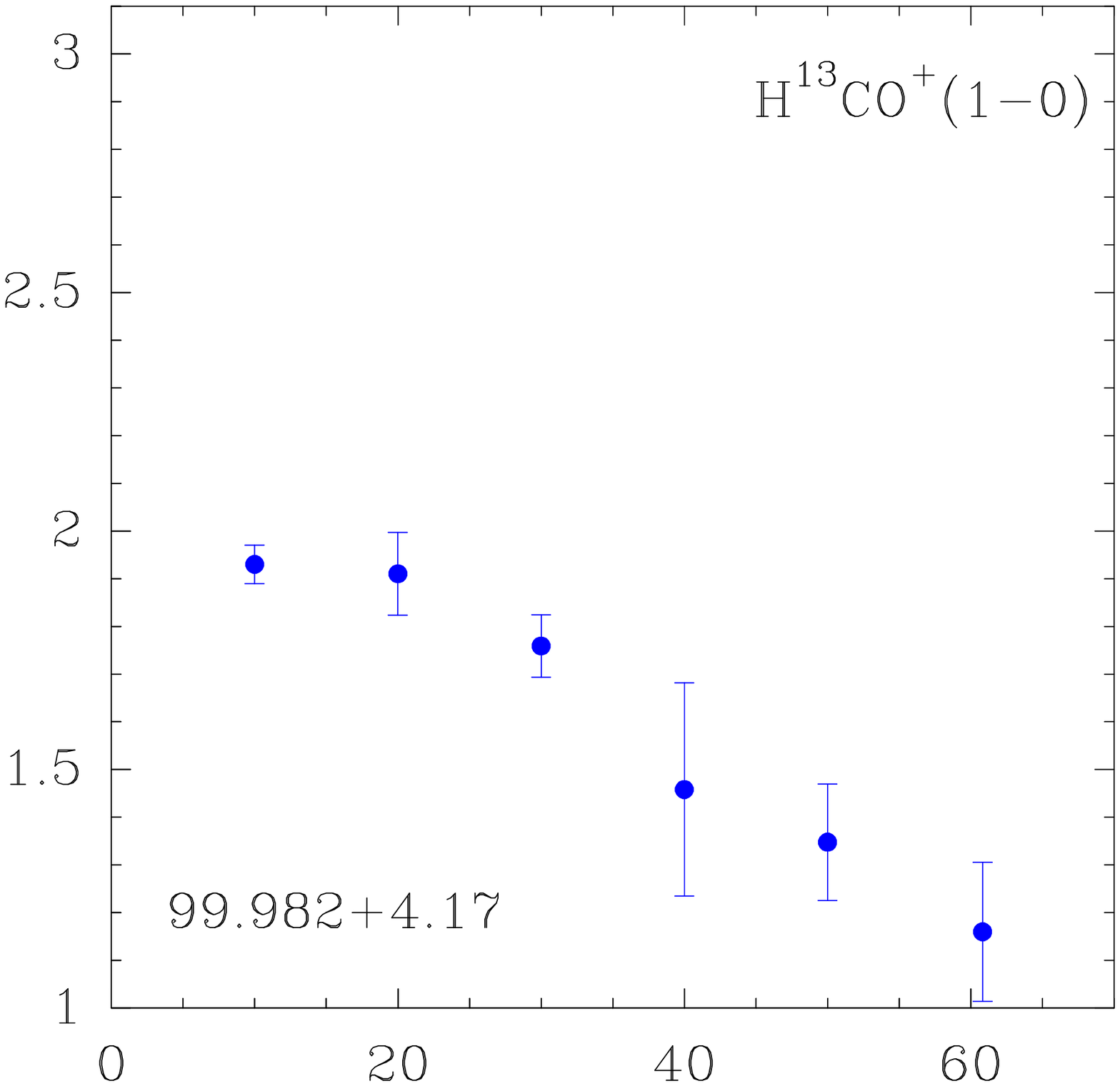}
\end{minipage}
\begin{minipage}[b]{0.24\textwidth}
    \includegraphics[width=\textwidth]{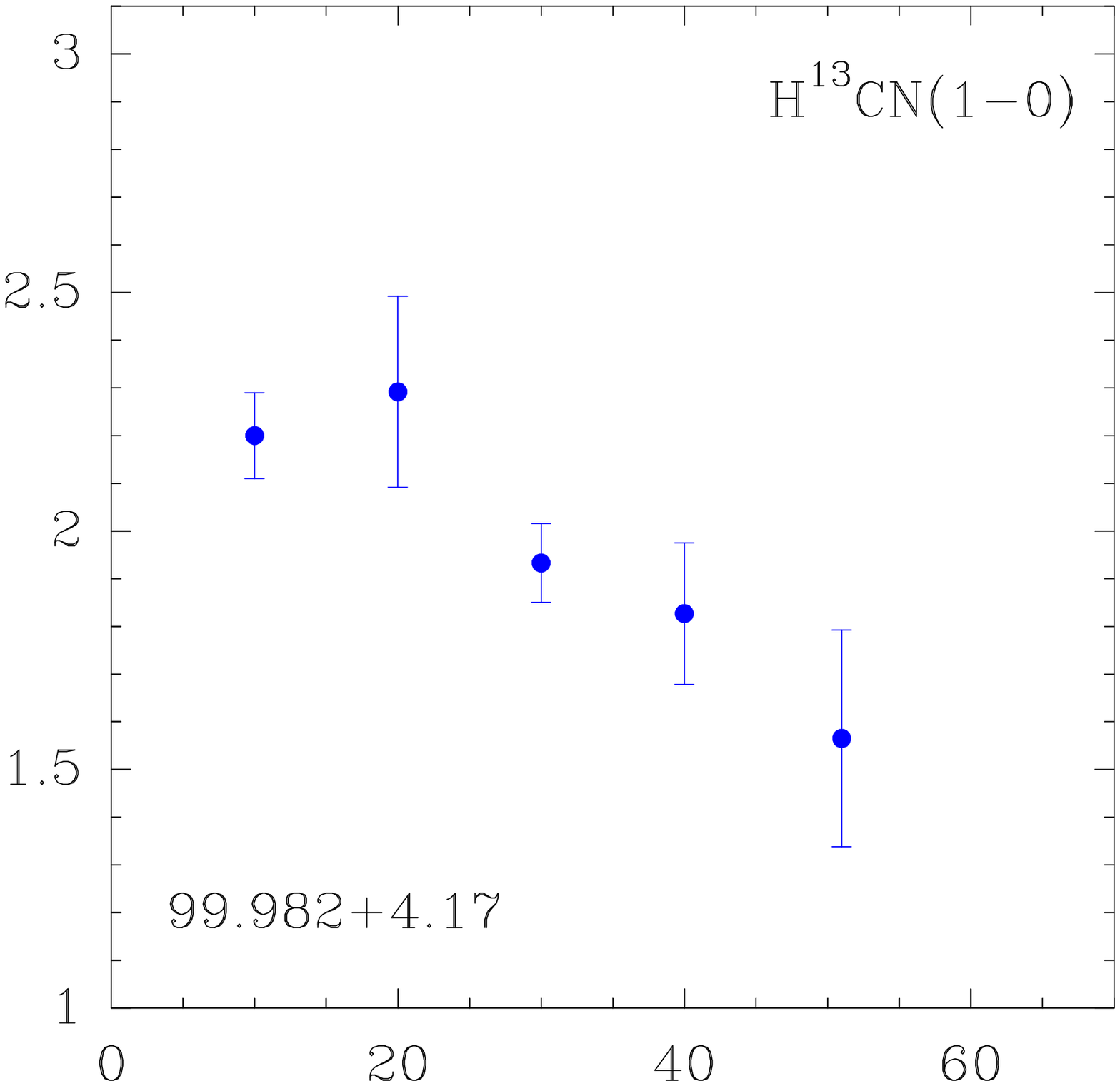}
\end{minipage}

\caption{
Profiles of velocity dispersions for different molecular lines in four regions. The vertical axis plots the averaged line widths ($\Delta V$) in km~s$^{-1}$ and the horizontal axis the impact parameters ($b$) in angular seconds.
}
\label{dvprof}
\end{figure}

\newpage

\begin{figure}[t!]
\setcaptionmargin{5mm}
\onelinecaptionsfalse
\captionstyle{flushleft}

\begin{minipage}[b]{0.48\textwidth}
    \includegraphics[width=\textwidth]{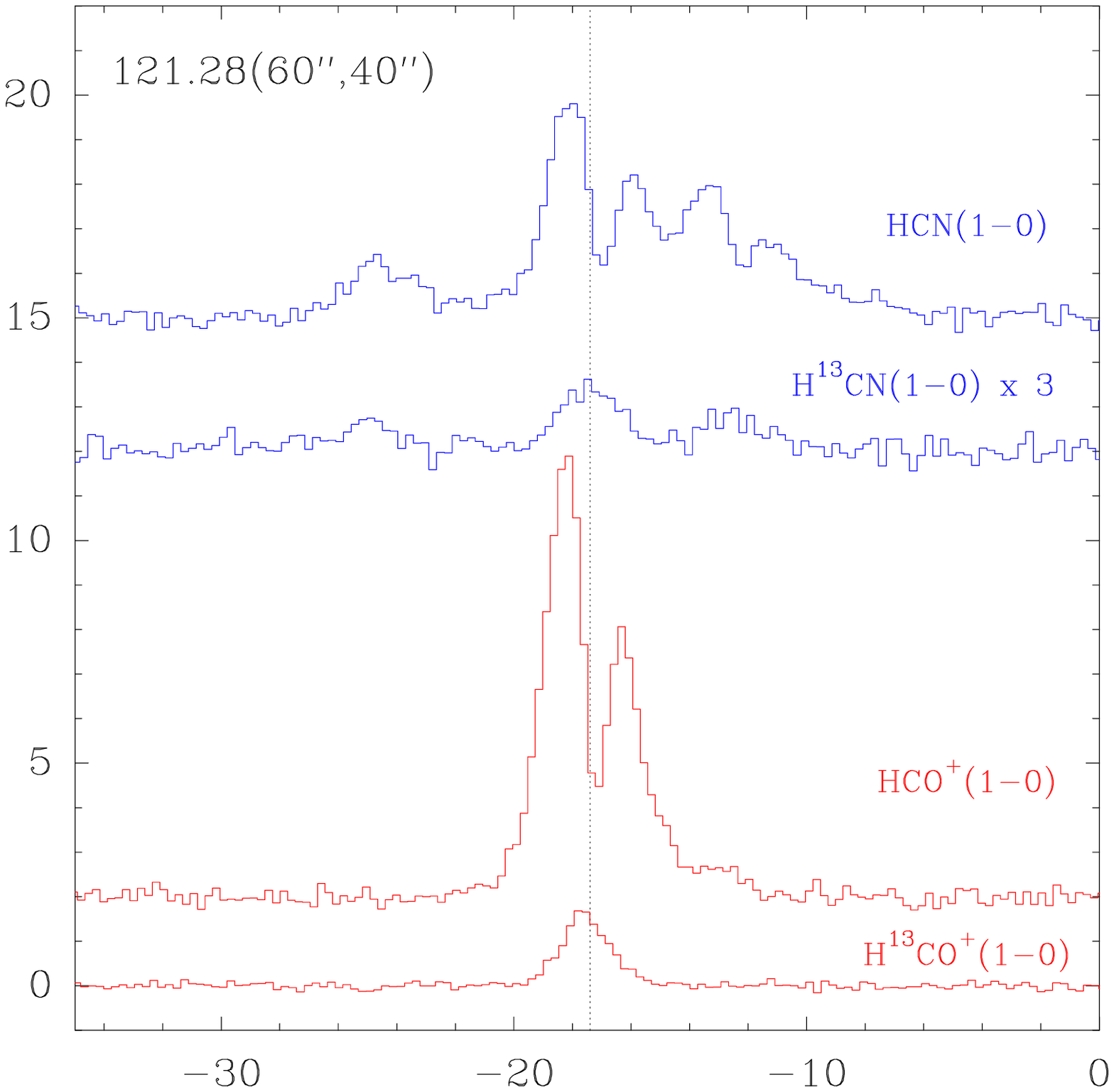}
\end{minipage}
\begin{minipage}[b]{0.485\textwidth}
    \includegraphics[width=\textwidth]{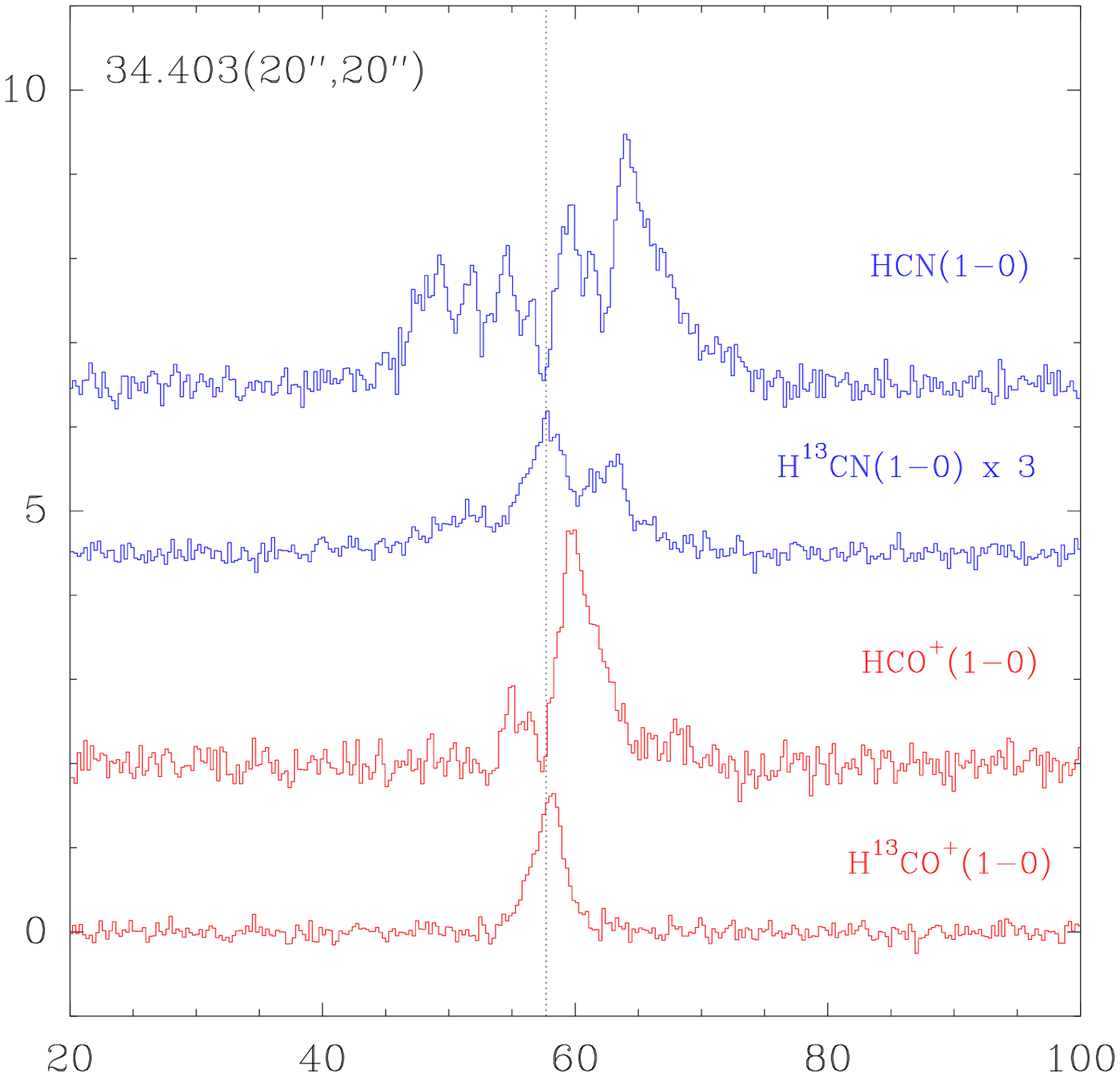}
\end{minipage}

\vskip 2mm
\begin{minipage}[b]{0.48\textwidth}
    \includegraphics[width=\textwidth]{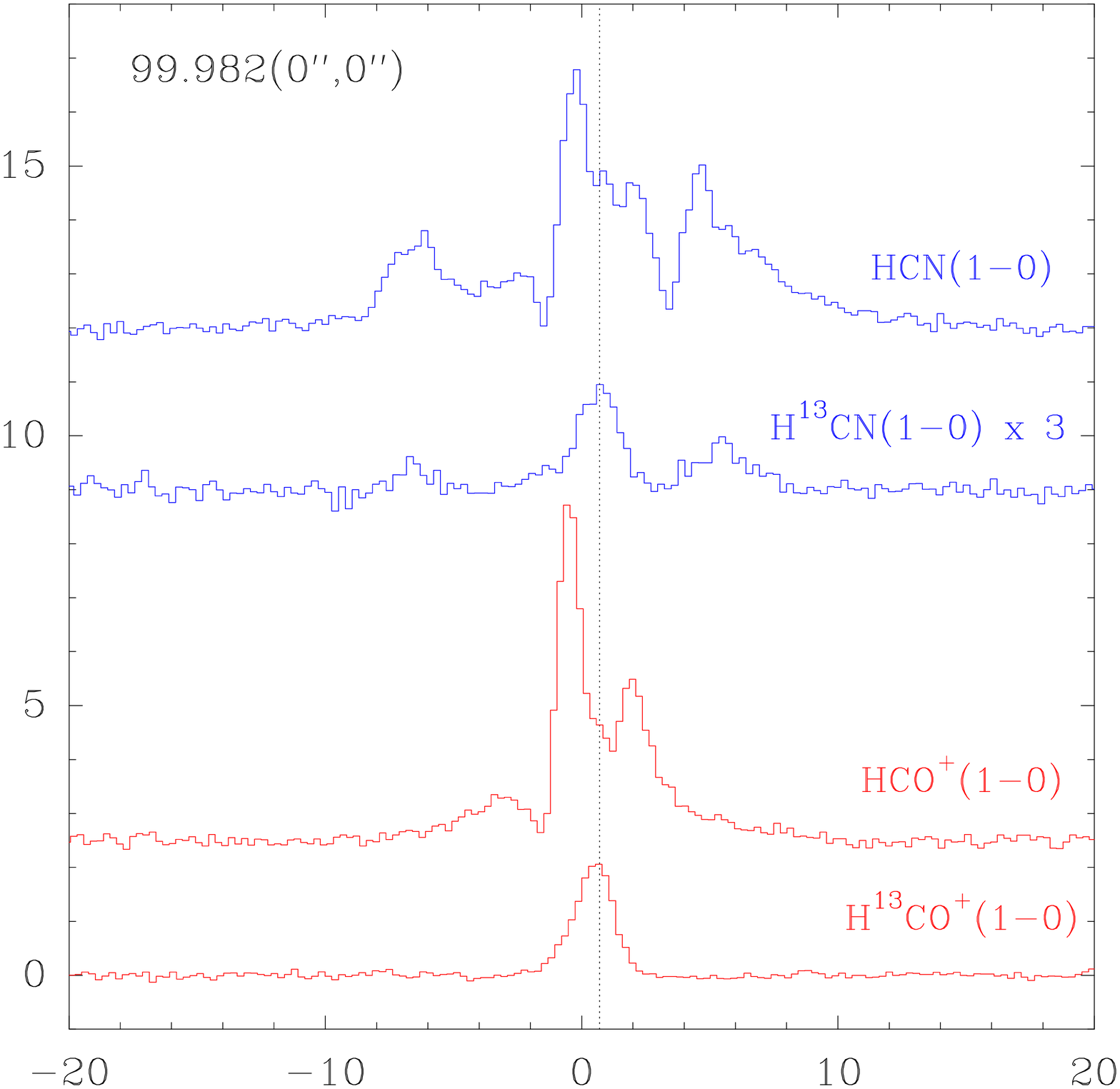}
\end{minipage}
\begin{minipage}[b]{0.48\textwidth}
    \includegraphics[width=\textwidth]{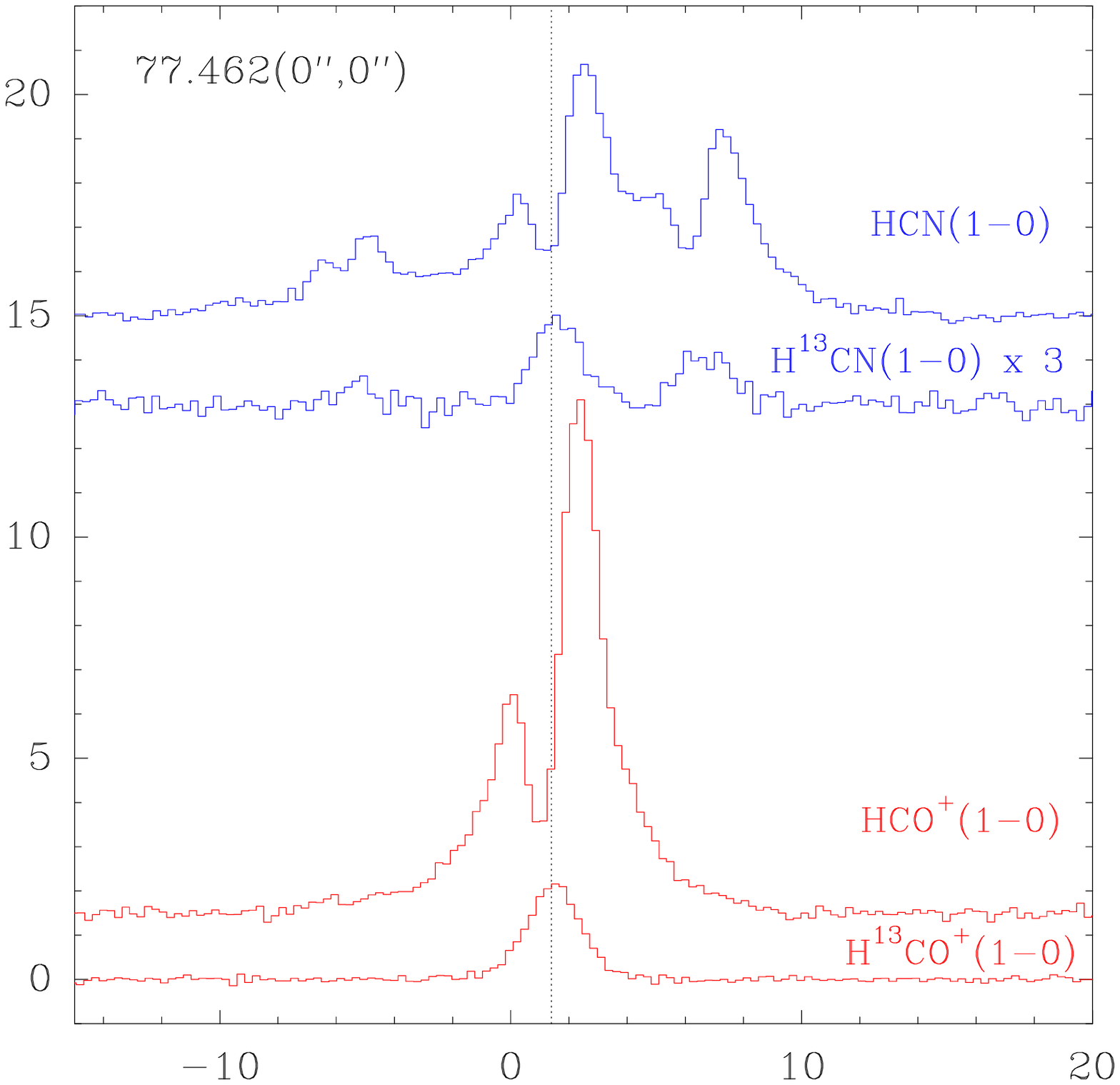}
\end{minipage}

\vskip 2mm
\centering
\begin{minipage}[b]{0.485\textwidth}
    \includegraphics[width=\textwidth]{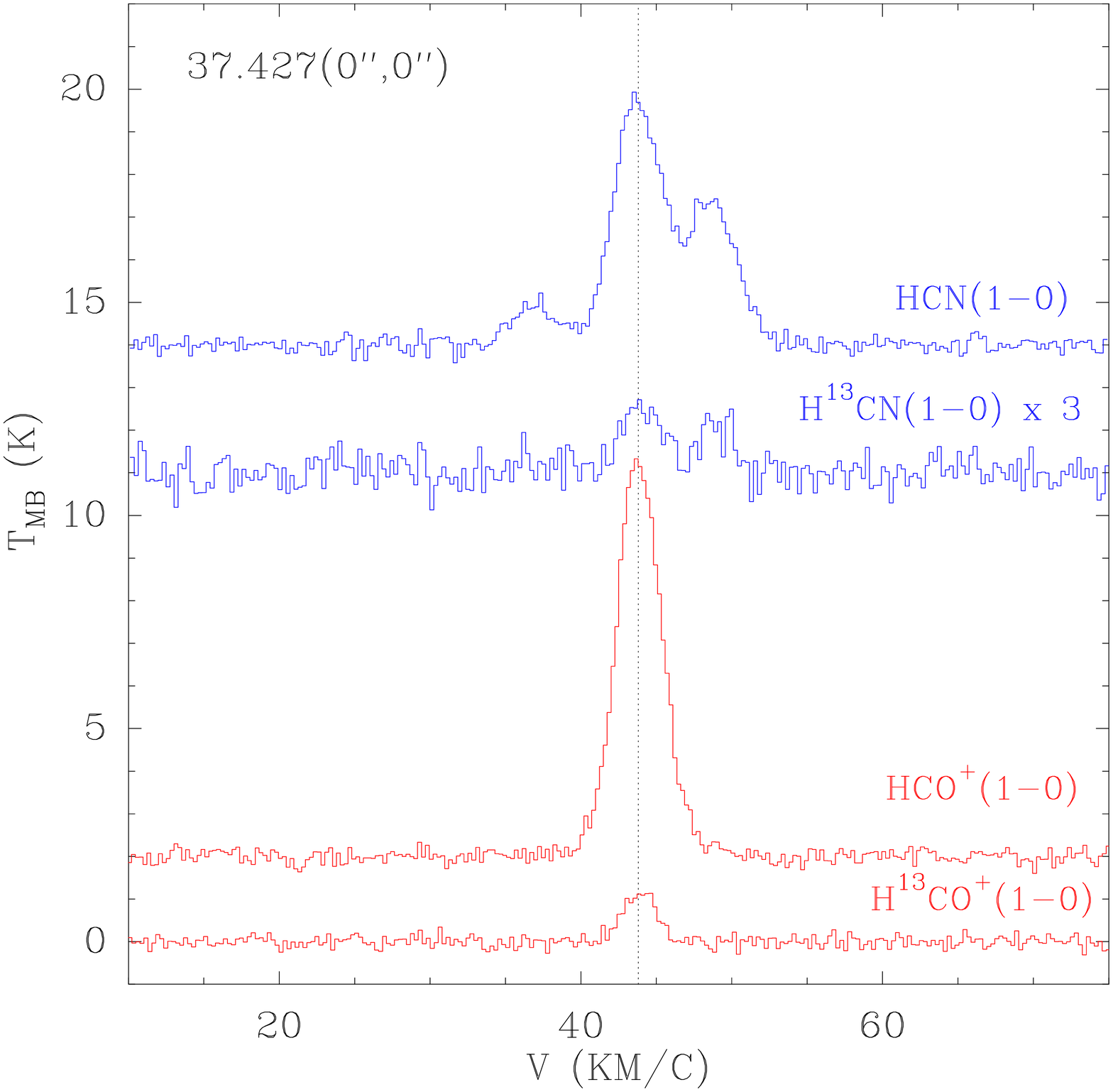}
\end{minipage}

\caption{\scriptsize 
Profiles of the HCN(1--0), HCO$^+$(1--0), H$^{13}$CN(1--0), and H$^{13}$CO$^+$(1--0) lines toward positions close to the emission peaks of these lines. The vertical dotted lines coincide with the peaks on the profiles of optically thin lines and close to the dips in the profiles of optically thick lines.
}
\label{hcn_hco+}
\end{figure}


\begin{thebibliography}{99}

\bibitem{Tan}
J.~C.~Tan, M.~Beltr\'an, P.~Caselli, F.~Fontani, A.~Fuente, M.~R.~Krunholz, C.~F.~McKee, A.~Stolte,
Protostars and Planets VI, H.~Beuther, R.~S.~Klessen, C.~P.~Dullemond, Th.~Henning (eds.), University of Arizona Press, Tucson, 149 (2014)
\bibitem{Sandell83}
G.~Sandell, B.~H\"oglund, A.~G.~Kislyakov, Astron. and Astrophys. \textbf{118}, 306 (1983)
\bibitem{Harju89}
J.~Harju, Astron. and Astrophys. \textbf{219}, 293 (1989)
\bibitem{Pir99}
L.~Pirogov,  Astron. and Astrophys. \textbf{348}, 600 (1999)
\bibitem{Bayandina}
O. S. Bayandina, I. E. Val'tts, and G. M. Larionov, Astron. Rep. \textbf{56}, 553 (2012)
\bibitem{Wu14}
Y. ~W.~Wu, M.~Sato, M.~J.~Reid, L.~Moscadelli, B.~Zhang, Y.~Xu, A.~Brunthaler, K.~M.~Menten,
T.~M.~Dame, X.~W.~Zheng, Astron. and. Astrophys. \textbf{566}, A17 (2014)
\bibitem{Varricatt}
W.~P.~Varricatt, C.~J.~Davis, S.~Ramsay, S.~P.~Todd, Mon.~Not.~RAS \textbf{404}, 661 (2010)
\bibitem{Xu06}
Y. ~Xu, Z.-Q.~Shen, J.~Yang, X.~W.~Zheng, A.~Miyazaki, K.~Sunada, H.~J.~Ma, J.~J.~Li, J.~X.~Sun, C.~C.~Pei, Astron.~J. \textbf{132}, 20 (2006)
\bibitem{Quanz}
S.~P.~Quanz, Th.~Henning, J.~Bouwman, H.~Linz, F.~Lahuis, Astrophys.~J. \textbf{658}, 487 (2007)
\bibitem{Pir07}
L.~Pirogov, I.~Zinchenko, P.~Caselli, L.~E.~B.~Johansson, Astron. and Astrophys. \textbf{461}, 523 (2007)
\bibitem{Rygl}
K.~L.~J.~Rygl, A.~Brunthaler, M.~J.~Reid, K.~M.~Menten, H.~J.~van~Langevelde, Y.~Xu, Astron. and Astrophys. \textbf{511}, A2 (2010)
\bibitem{Kurayama}
T.~Turayama, A.~Nakagawa, S.~Sawada-Satoh, K.~Sato, M.~Honma, K.~Sunada, T.~Hirota, H.~Imai, Publ.~ Astron.~Soc.~Japan \textbf{63}, 513 (2011)
\bibitem{Lumsden}
S.~L.~Lumsden, M.~G.~Hoare, J.~S.~Urquhart, R.~D.~Oudmaijer, B.~Davies, J.~C.~Mottram,
H.~D.~B.~Cooper, T.~J.~T.~Moore, Astrophys.~J. Suppl. \textbf{208}, 11 (2013)
\bibitem{Molinari}
S.~Molinari, J.~Brand, R.~Cesaroni, F.~Palla, Astron. and Astrophys. \textbf{308}, 573 (1996)
\bibitem{Belitsky}
V.~Belitsky, I.~Lapkin, M.~Fredrixon, E.~Sundin, L.~Helldner, L.~Pettersson, S.-E.~Ferm, M.~Pantaleev,
B.~Billade, P.~Bergman, A.~O.~H.~Olofsson, M.~S.~Lerner, M.~Strandberg, M.~Whale, A.~Pavolotsky, J.~Flygare, H.~Olofsson, J.~Conway, Astron. and Astrophys. \textbf{580}, A29 (2015)
\bibitem{Rathborne08}
J.~M.~Rathborne, J.~M.~Jackson, Q.~Zhang, R.~Simon, Astrophys.~J. \textbf{689}, 1141 (2008)
\bibitem{Pir03}
L.~Pirogov, I.~Zinchenko, P.~Caselli, L.~E.~B.~Johansson, P.~C.~Myers, Astron. and Astrophys.  \textbf{405}, 639 (2003)
\bibitem{ST}
G. T. Smirnov and A. P. Tsivilev, Sov. Astron. \textbf{26}, 616 (1982)
\bibitem{Press}
W.~H.~Press, S.~A.~Teukolsky, W.~T.~Vetterling, B.~P.~Flannery B. P.,
Numerical Recipes in Fortran 77, Cambridge Univ. Press (1992)
\bibitem{Mal}
S. Yu. Malafeev, I. I. Zinchenko, L. E. Pirogov, and L. E. B. Johansson, Astron. Lett. \textbf{31}, 239 (2005).
\bibitem{MangumShirley}
J.~G.~Mangum, Y.~L.~Shirley, Publ. Astron. Society Pacific \textbf{127}, 266 (2015)
\bibitem{Reitblat}
A. A. Reitblat, Sov. Astron. Lett. \textbf{6}, 406 (1980).
\bibitem{Busquet}
G.~Busquet, A.~Palau, R.~Estalella, J.~M.~Girart, A.~S\'anchez-Monge, S.~Viti, P.~T.~P.~Ho, Q.~Zhang, Astron. and Astrophys. \textbf{517}, L6 (2010)
\bibitem{Bester}
M.~Bester, S.~Urban, K.~Yamada, G.~Winnewisser, Astron. and Astrophys. \textbf{121}, L13 (1983)
\bibitem{Zin98}
I.~Zinchenko, L.~Pirogov, M.~Toriseva, Astron. and Astrophys. Suppl. \textbf{133}, 337 (1998)
\bibitem{McCutcheon}
W.~H.~McCutcheon, T.~Sato, C.~R.~Purton, H.~E.~Matthews, P.~E.~Dewdney, Astron. J. \textbf{110}, 1762 (1995)
\bibitem{Sanhueza}
P.~Sanhueza, J.~M.~Jackson, J.~B.~Foster, G.~Garay, A.~Silva, S.~C.~Finn, Astrophys.~J. \textbf{756}, 60 (2012)
\bibitem{Rathborne06}
J.~M.~Rathborne, J.~M.~Jackson, R.~Simon, Astrophys.~J. \textbf{641}, 389 (2006)
\bibitem{Beuther}
H.~Beuther, P.~Schilke, K.~M.~Menten, F.~Motte, T.~K.~Sridharan, F.~Wyrowski, Astrophys. J. \textbf{566}, 945 (2002)
\bibitem{Faundez}
S.~Fa\'undez, L.~Bronfman, G.~Garay, R.~Chini, L.-A.~Nyman, J.~May, Astron. and Astrophys. \textbf{426}, 97 (2004)
\bibitem{Sugitani}
K.~Sugitani, H.~Matsuo, M.~Nakano, M.~Tamura, K.~Ogura, Astron.~J. \textbf{119}, 323 (2000)
\bibitem{HW}
A.~Heske, H.~J.~Wendker, Astron. and Astrophys. \textbf{149}, 199 (2005)
\bibitem{Walker}
K.~C.~Walker, F.~C.~Adams, C.~J.~Lada, Astrophys.~J. \textbf{349}, 515 (1990)
\bibitem{Harju98}
J.~Harju, K.~Lehtinen, R.~S.~Booth, I.~Zinchenko, Astron. and Astrophys. Suppl. \textbf{132}, 211 (1998)
\bibitem{Pir09}
L. E. Pirogov, Astron. Rep. \textbf{53}, 1127 (2009).
\bibitem{Caselli02}
P.~Caselli, P.~J.~Benson, P.~C.~Myers, M.~Tafalla, Astrophys.~J. \textbf{572}, 238 (2002)
\bibitem{Phillips}
T.~G.~Phillips, P.~J.~Huggins, P.~G.~Wannier,  N.~Z.~Scoville, Astrophys.~J. \textbf{231}, 720 (1979)
\bibitem{Evans}
N.~J.~Evans~II, Ann. Rev. Astron. and Astrophys. \textbf{37}, 311 (1999)
\bibitem{Myers}
P.~C.~Myers, D.~Mardones, M.~Tafalla, J.~P.~Williams, D.~J.~Wilner,  Astrophys.~J. \textbf{465}, L133 (1996)
\bibitem{KW}
P.~D.~Klaassen, C.~D.~Wilson, Astrophys.~J. \textbf{663}, 1092 (2007)
\bibitem{Fuller}
G.~A.~Fuller, S.~J.~Williams, T.~K.~Sridharan, Astron. and Astrophys. \textbf{442}, 949 (2005)
\bibitem{Alakoz}
A. V. Alakoz, S. V. Kalenskii, V. G. Promyslov, L. E. B. Johansson, and A. Winnberg, Astron. Rep.
\textbf{46}, 551 (2002).
\bibitem{Sanchez}
A.~S\'anchez-Monge, A.~L\'opez-Sepulcre, R.~Cesaroni, C.~M.~Walmsley, C.~Codella, M.~T.~Beltr?n, M.~Pestalozzi, S.~Molinari, Astron. and Astrophys. \textbf{557}, A94 (2013)
\bibitem{Cyganowski}
C.~J.~Cyganowski, J.~Koda, E.~Rosolowsky, S.~Towers, J.~Donovan~Meyer, F.~Egusa, R.~Momose, T.~P.~Robitaille, Astrophys.~J. \textbf{764}, 61 (2013)
\bibitem{Zin97}
I.~Zinchenko, Th.~Henning, K.~Schreyer, Astron. and Astrophys. Suppl. \textbf{124}, 385 (1997)
\bibitem{Wu10}
J.~Wu, N.~J.~Evans, Y.~L.~Shirley, C.~Knez, Astrophys.~J.~Suppl. \textbf{188}, 313 (2010)
\bibitem{Gan}
C.-G.~Gan, X.~Chen, Z.-Q.~Shen, Y.~Xu, B.-G.~Ju, Astrophys.~J. \textbf{763}, 2 (2013)
\bibitem{Pestalozzi}
M.~R.~Pestalozzi, V.~Minier, R.~S.~Booth, Astron. and Astrophys. \textbf{432}, 737 (2005)
\bibitem{Valdettaro}
R.~Valdettaro, F.~Palla, J.~Brand, R.~Cesaroni, G.~Comoretto, S.~Di~Franco, M.~Felli, E.~Natale,  F.~Palagi, D.~Panella, G.~Tofani, Astron. and Astrophys. \textbf{368}, 845 (2001)
\bibitem{Sandell01}
G.~Sandell,  D.~A.~Weintraub, Astrophys.~J.~Suppl. \textbf{134}, 115 (2001)
\bibitem{Shepherd}
D.~S.~Shepherd, M.~S.~Povich, B.~A.~Whitney, T.~P.~Robitaille, D.~E.~A.~N\''urnberger, L.~Bronfman, D.~P.~Stark, R.~Indebetouw, M.~R.~Meade, B.~L.~Babler, Astrophys.~J., \textbf{669}, 464 (2007)
\bibitem{Xu03}
Y.~Xu, X.-W.~Zheng, D.-R.~Jiang, Chin. J. Astron. and Astrophys., \textbf{3}, 49 (2003)
\bibitem{Szymczak}
M.~Szymczak, A.~J.~Kus, G.~Hrynek, A.~Kepa, E.~Pazderski, Astron. and Astrophys., \textbf{392}, 277 (2002)
\bibitem{Chen11}
X.~Chen, S.~P.~Ellingsen, Z.-Q.~Shen, A.~Titmarsh, C.-G.~Gan, Astrophys. J. Suppl., \textbf{196}, 9 (2011)
\bibitem{Blaszkiewicz}
L.~Blaszkiewicz, A.~J.~Kus, Astron. and Astrophys., \textbf{413}, 233 (2004)
\bibitem{Fontani}
F.~Fontani, R.~Cesaroni, R.~S.~Furuya, Astron. and Astrophys. \textbf{517}, A56, (2010)
\bibitem{Palla}
F.~Palla, J.~Brand, G.~Comoretto, M.~Felli, R.~Cesaroni, Astron. and Astrophys. \textbf{246}, 249 (1991)
\bibitem{Urquhart}
J.~S.~Urquhart, M.~G.~Hoare, C.~R.~Purcell, S.~L.~Lumsden, R.~D.~Oudmaijer, T.~J.~T.~Moore, A.~L.~Busfield, J.~C.~Mottram, B.~Davies, Astron. and Astrophys., \textbf{501}, 539 (2009)
\bibitem{Kurtz}
S.~Kurtz, P.~Hofner,  C.~V.~Alvarez, Astrophys. J. Suppl., \textbf{155}, 149 (2004)
\bibitem{Slysh}
V.~I.~Slysh, I.~E.~Val'tts, V.~Migenes, E.~Fomalont, H.~Hirabayashi, M.~Inoue, T.~Umemoto, Astrophys.~J., \textbf{526}, 236 (1999)
\bibitem{Pollanen}
M.~D.~Pollanen, P.~A.~Feldman, Publ. Astron. Soc. of the Pacific, \textbf{107}, 617 (1995)
\bibitem{Chauhan}
N.~Chauhan,  A.~K.~Pandey,  K.~Ogura, D.~K.~Ojha, B.~C.~Bhatt, S.~K.~Ghosh, P.~S.~Rawat, Mon.~Not.~RAS, \textbf{396}, 964 (2009)
\bibitem{Choudhury}
R.~Choudhury, B.~Mookerjea, H.~C.~Bhatt, Astrophys.~J., \textbf{717}, 1067(2010)
\bibitem{DiFrancesco}
J.~Di~Francesco, D.~Johnstone, H.~Kirk, T.~MacKenzie, E.~Ledwosinska, Astrophys.~J.~Suppl., \textbf{175}, 277 (2008)

\end{thebibliography}
\end{document}